\documentclass[journal]{IEEEtran}
\usepackage{amsmath}
 
\usepackage{amssymb} 
\usepackage{amsthm}
\usepackage{mathbbol}
\usepackage{latexsym}
\usepackage{ascmac}
\usepackage{bm}
\usepackage{cite}
\usepackage{subfigure}
\usepackage{amsfonts}
\usepackage{accents}
\usepackage{ifpdf}
\usepackage{color}
\usepackage{lscape}
\usepackage{color}
\usepackage{hyperref}
\usepackage{booktabs}
\usepackage{tabularx}
\usepackage{fancyhdr}

\newcommand{\transp}{{\sf T}}

\ifCLASSINFOpdf
  \usepackage[pdftex]{graphicx}
\else
  \usepackage[dvips]{graphicx}
\fi
\usepackage{epstopdf}

\usepackage{algorithmic}
\usepackage{algorithm}

\usepackage{alphalph}

\usepackage{multirow}
\usepackage{multicol}
\usepackage{url}

\newcommand{\argmin}{\mathop{\rm arg\,min}\limits}

\begin{document}
\title{Time-Varying Graph Learning \\ with Constraints on Graph Temporal Variation}

\author{Haruki~Yokota, \IEEEmembership{Student Member,~IEEE},
        Koki~Yamada, \IEEEmembership{Member,~IEEE},
        Yuichi~Tanaka, \IEEEmembership{Senior Member,~IEEE},
        and~Antonio~Ortega,~\IEEEmembership{Fellow,~IEEE}% <-this % stops a space
\thanks{The work of Haruki Yokota was supported in part by JSPS KAKENHI under Grant 25KJ1755.
The work of Yuichi Tanaka was supported in part by JSPS KAKENHI under Grant 26H02536, and JST AdCORP under Grant JPMJKB2307.
The work of Koki Yamada was supported in part by JSPS KAKENHI under Grant 24K20788. (Corresponding author: Haruki Yokota.)}
\thanks{H. Yokota and K. Yamada contributed equally to this work.}
\thanks{H. Yokota and Y. Tanaka are with the Graduate School of Engineering, The University of Osaka, Osaka 565-0871, Japan (e-mail: h.yokota@sip.comm.eng.osaka-u.ac.jp and ytanaka@comm.eng.osaka-u.ac.jp).}% <-this % stops a space
\thanks{K. Yamada is with the Graduate School of Engineering, Tokyo University of Agriculture and Technology, Tokyo 184-8588, Japan (e-mail: k-yamada@go.tuat.ac.jp).}% <-this % stops a space
\thanks{A. Ortega is with the Department of Electrical and Computer Engineering, University of Southern California, Los Angeles, CA 90089 USA (aortega@usc.edu).}% <-this % stops a space
}

\maketitle

\thispagestyle{fancy}
\fancyhf{}
\setlength{\footskip}{15pt} 
\cfoot{\small © 2026 IEEE. Personal use of this material is permitted. Permission from IEEE must be obtained for all other uses, in any current or future media, including reprinting/republishing this material for advertising or promotional purposes, creating new collective works, for resale or redistribution to servers or lists, or reuse of any copyrighted component of this work in other works.}

\begin{abstract}
We propose a novel framework for learning time-varying graphs from spatiotemporal measurements.
Given an appropriate prior on the temporal behavior of signals, our proposed method can estimate time-varying graphs from a small number of available measurements.
To achieve this, we introduce three regularization terms in convex optimization problems that constrain the sparseness of temporal variations of the time-varying networks.
Moreover, a computationally scalable algorithm is introduced to solve the optimization problem efficiently.
The experimental results with synthetic and real datasets (point cloud, temperature, and EEG data) demonstrate that our proposed method outperforms state-of-the-art methods.
\end{abstract}


% \fi
\IEEEpeerreviewmaketitle

\section{Introduction}
\label{sec:intro}

\IEEEPARstart{S}{ignals} often have underlying network structures, e.g., in sensor, traffic, brain, and social networks.
Graphs, consisting of sets of nodes and edges, are fundamental tools for describing the relationship among entities.
Graph edges and the corresponding edge weights can capture the similarity between nodes (where a higher positive weight indicates greater similarity).
Introducing a graph representation enables us to efficiently analyze signals on networks in many practical applications, such as epidemics \cite{nowzari2016, ganesh2005}, transportation networks \cite{colizza2006}, and social networks \cite{zhang2014}.

Even when data is not associated with an actual network, graphs are efficient tools to represent latent structures in the data.
For example, a principal component analysis can be improved by imposing a prior based on graphs \cite{shahid2016, shang2012, zhang2013}. 
Graph-based image processing enables us to improve performance in several image processing tasks \cite{gadde2013, wang2014, onuki2016, elmoataz2008, couprie2013, ono2015}.  
However, graphs are often not known {\em a priori}.

{\it Graph learning} methods aim at identifying graphs from observed data \cite{dong2019, mateos2019, giannakis2018, kadambari2020a, liu2019b, lodhi2020, chen2010}. 
Each observation is a vector, with each entry in the vector corresponding to an observation at one node. 
The goal of graph learning is to obtain the weights of all the edges connecting those nodes.   
Most graph learning methods identify a single static graph from all the observations \cite{kalofolias2016, dong2016, egilmez2017, pasdeloup2018, thanou2017, mei2017, chepuri2017,  pavez2018, egilmez2018, pu2021, li2023,tamaru2024}.
These {\it static graph learning} methods assume that the node relationships obtained from the observations do not change during the measurement process.

In many applications, observations are obtained over a period of time, and a time-varying (TV) graph can provide a better model.
Examples of such applications include estimation of the time-varying brain functional connectivity from EEG or fMRI data \cite{preti2017}, identification of temporal transit of biological networks such as protein, RNA, and DNA \cite{kim2014}, inference of relationships among companies from historical stock price data \cite{hallac2017}, dynamic point cloud processing, and analysis of sensor network data.

A straightforward approach to estimating a TV graph consists of aggregating temporal observations into non-overlapping windows and then using an existing static graph learning method to estimate a graph for each time window. However, such an approach has some drawbacks.
First, this method estimates a graph {\em independently} for each temporal interval; thus, it ignores temporal relations that may exist in TV graphs.
Second, time-varying graph learning may require estimating graphs from time windows containing only a small fraction of observations, given the trade-off between the choice of window length and temporal resolution. 
Specifically, if we choose a short window to adapt to fast temporal graph changes, we may not have enough data to learn a graph within each window since existing static graph learning methods cannot successfully infer graphs from a small number of observations.
On the other hand, if we use a larger window, we may be unable to track relatively fast temporal changes.

\begin{figure}
    \centering
    \includegraphics[width=0.8\linewidth]{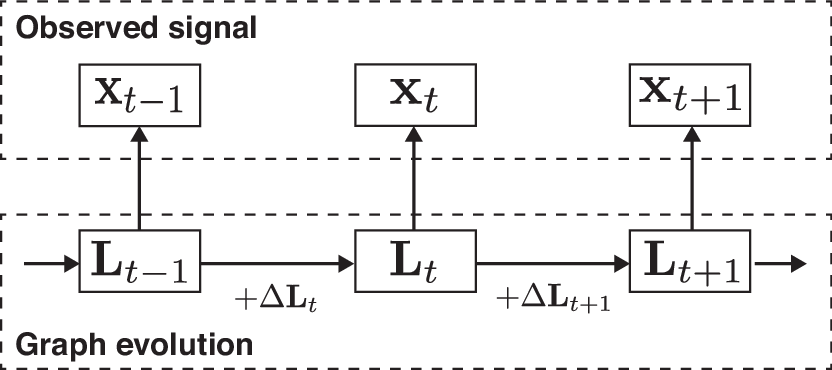}
    \caption{An overview of TV graph factor analysis. ${\bf L}_{t}$ and $\Delta {\bf L}_{t}$ represent the graph Laplacian at the $t$th time slot and the graph temporal variation. This study focuses on learning a TV graph, which is the sequence of the graph Laplacian, from the observed signal ${\bf x}$.}
    \label{fig:tgfa}
\end{figure}

This paper presents a {\it time-varying graph learning} method based on time-varying graph factor analysis (TGFA), which is an extension of its static counterpart, static graph factor analysis (SGFA)  \cite{dong2016}.
We propose TGFA-based methods to estimate TV graphs from a collection of spatiotemporal measurements.
SGFA assumes a signal generation model based on a graph signal processing (GSP) perspective, with the observed signals having specific spectral properties with respect to the graph Fourier transform (GFT) of the graph to be learned. 
For example, under a multivariate Gaussian model, we assume that observed signals are generated from a Gaussian distribution whose inverse covariance matrix is given by the graph Laplacian of the underlying graph \cite{dong2016, dong2019}. 
Unlike SGFA, TGFA considers the graph evolution, which is represented by a sequence of graph Laplacians as illustrated in \autoref{fig:tgfa}.
% The graph evolution can be represented by a sequence of graph Laplacians and their corresponding temporal variations.

\begin{table*}[tb]
    \centering
    \caption{Comparison of Graph Learning Methods}

    \begin{tabular}{c|c|c|c|c|c|c|c}
        \hline
        \textbf{Setting} & \textbf{Author} & \textbf{Model} & \textbf{Memory} & \textbf{Target Graphs} & \textbf{Regularization} & \textbf{Processing} & \textbf{Bidirectional}  \\ \hline
        \multirow{7}{*}{Time-varying} & Proposed & Smoothness & \textbf{$1\text{--} L$} & $2\text{--} T$ & Multiple & Batch & Yes \\ 
         & Yamada et al.~\cite{yamada2019} & Smoothness & $1$  & $2\text{--} T$ & L1 & Batch & No \\ 
         & Kalofolias et al.~\cite{kalofolias2017} & Smoothness  & 1 & $2\text{--} T$ & Tikhonov & Batch & No  \\ 
         & Hallac et al.~\cite{hallac2017} & Graphical & $1$  & $2\text{--} T$ & Multiple & Batch / Streaming & No \\ 
         & Zhang et al.~\cite{zhang2022} &  Smoothness & $1\text{--} T-1$ & $2\text{--} T$ & L1 & Batch & Yes \\ 
         & Ye et al.~\cite{ye2024} & Smoothness  & $1$  & $2\text{--} T$ & L2  & Batch & No  \\
         & Cardoso and Palomar \cite{cardoso2020learning} & Graphical & 1 & $2\text{--}T$ & Tikhonov & Batch & No \\ \hline
        \multirow{3}{*}{Online} & Natali et al.~\cite{Natali2022} &  Multiple & $1$ & $1$ & - & Streaming   & No \\ 
         & Bagheri et al.~\cite{Bagheri2024} & Stationarity & $1$  & $1$ & Low-rank & Batch & No \\
         & Shen et al. \cite{shen2017tensor} & Structural Equation & $1\text{--}T$ & $T$ & - & Streaming & No \\\hline
        \multirow{2}{*}{Multiple} & Saboksayr et al.~\cite{Saboksayr2021} & Smoothness & $1$ & C & - &Batch / Streaming & - \\ 
         & Rey et al.~\cite{rey2022} &  Stationarity & $1$  & C  & L1  &  Batch & -  \\ \hline
    \end{tabular}
    \label{tab:graph_comparison}
\end{table*}

This study focuses on three separate TV graph models, each  corresponding to one of the following properties:

\textbf{1) Temporal homogeneity (TH)}: 
We assume that most edges and their weights in the TV graph model remain unchanged over a short-term time horizon, i.e., only a few graph edges change at any given time. 
TV graphs in many applications satisfy this property.
For example, consider a sensor network where the nodes represent sensor locations. Often, the edges are chosen to capture information about correlations among sensor measurements. 
If the sensors record the temperature in a building, various factors, such as air conditioning, sunlight, and the number of people in the room, locally affect the correlations among the sensor measurements.
However, these factors vary smoothly over time. 
As a result, data on this sensor network can be modeled by a TV graph where the connectivity itself does not change frequently, but the edge weights can change smoothly over time.
TV graphs in fMRI and various biological networks can also be seen as having this property \cite{monti2014, kim2014}.

\textbf{2) Switching behavior (SB)}: We assume both edges and weights remain unchanged most of the time; however, %some of them 
they may change within a few time slots. 
% This TV graph type appears when some factors cause 
This model reflects sudden changes in graph topologies. 
Prominent examples include brain networks, where epileptic seizures can lead to sudden changes in graph model topology \cite{wang2014a}. 

\textbf{3) Local topological changes (LTC)}: We assume that most edges and weights remain unchanged over time, %as in \textbf{SB}, 
while the connectivity of a small portion of nodes changes for each time slot. An example of this behavior can be seen in mobile sensor networks, where some sensors may relocate themselves over time to form a different local neighborhood \cite{velmani2015}. 

\renewcommand{\arraystretch}{1}{
\begin{table}[tb]
	\centering
	\caption{List of notation}
	{\small
	\begin{tabular}{l l} \hline\hline
		$N$                & Number of nodes   \\
        $E$ & Number of edges \\
		$K$                & Number of data chunk   \\
		$T$             &  Number of time frames \\
            $L$             & Number of temporal filter taps \\
		${a}_{i}, ({\bf a})_{i},  a[i]$       & $i$th entry of a vector \\
		${A}_{ij}, ({\bf A})_{ij}$      & $(i,j)$ entry of a matrix\\
%		${\bf A}(i,:)$      & subvector of $\bf A$ at $i$-th row \\
		$(\mathbf{A})_{i}$      & $i$th column of $\mathbf{A}$ \\
		$\mathbf{A}^{\dagger}$ & Moore-Penrose pseudoinverse of $\mathbf{A}$ \\
		$ \circ$            & Hadamard product \\
		$ \| \mathbf{a} \|_{2}^{2}, \| \mathbf{A} \|_{F}^{2} $ & Sum of squared values of all elements \\
		$ \| \mathbf{a} \|_{1}, \| \mathbf{A} \|_{1} $ & Sum of absolute values of all elements \\
            $\| \mathbf{A} \|_{2,1}=\sum_{i}\|(\mathbf{A})_i\|_2$ \\
		$\mathrm{Tr}(\mathbf{A})$  & Trace of a matrix \\
		$\mathrm{diag}(\mathbf{A})$  & Vector formed by diagonal elements\\
		 \hline\hline	
	\end{tabular}
	\label{tb:notation}
	}
\end{table}
}

In this paper, we design an algorithm to estimate these three types of TV graphs with a consistent objective function.
For this purpose, we formulate the graph learning problem as a convex optimization with regularization of temporal variation derived from TGFA.
To solve the convex optimization problem, we utilize a primal-dual splitting algorithm \cite{condat2013}, which enables us to estimate TV graphs more successfully than static graph learning methods. 
Furthermore, while the model shown in \autoref{fig:tgfa} is a \textit{single-step memory} model where learning a new graph includes a constraint based on the most recently learned graph, it is also possible to consider \textit{multi-step memory} cases where the constraints are based on the previous $L$ learned graphs. We show that our formulation can be easily adapted to such $L$-step memory cases.

In experiments with synthetic datasets, our proposed method outperforms existing methods, especially when only small amounts of data are available.
We also evaluate the performance of our methods for tasks involving real data.  
Our results for dynamic point cloud denoising show that estimating graph topology information with our method allows us to improve the denoising performance.
In a meteorological and temperature data application, we show that our method can learn reasonable TV graphs that capture the geographical characteristics \textit{without using geographical information}. In an experiment with electroencephalogram (EEG) data, the graphs learned from EEGs of epileptic seizures with our method can illustrate the structural variance during the seizure, which is consistent with clinical observations.

Our published preliminary work\cite{yamada2019}
proposed a TV graph learning framework for \textbf{TH} under the single-step memory setting.
In our preprint \cite{yamada2020arxiv}, we introduced the \textbf{SB} model and proposed a solution.
This paper includes the \textbf{TH} model from our peer-reviewed publication \cite{yamada2019}, and extends it with an updated version of the \textbf{SB} model from \cite{yamada2020arxiv}  and the newly introduced \textbf{LTC} model along with the $L$-step memory case.
We also propose a new unified methodology to formulate and solve all three problems and evaluate our approach under different conditions through extensive experiments on synthetic and real-world datasets, comparing its computational complexity with competing approaches.

\subsection{Related Works}
Recent overviews of graph learning are given in \cite{dong2019, mateos2019}. 
Typical approaches for learning TV graphs solve problems that consist of two terms, one quantifying the smoothness of signals on the graph 
and another that provides regularization to characterize graph evolution over time.

Many approaches, including our method, assume (spatial) signal smoothness on the TV graphs. The key difference between the methods lies in the choice of the regularization terms. 
In \cite{kalofolias2017}, the smoothness of edge weights between adjacent time windows is considered in an $\ell_2$ sense, using Tikhonov regularization. In \cite{zhang2022}, total variation regularization is applied to capture sparse changes between pairs of time-varying graphs. However, the optimization proposed in \cite{kalofolias2017} lacks the flexibility for other types of temporal changes, such as \textbf{TH}, and the model in \cite{zhang2022} requires prior knowledge about how the graphs evolve.
In contrast, we demonstrate that our algorithm is suited for learning TV graphs with the properties \textbf{TH}, \textbf{SB}, and \textbf{LTC} and can accommodate other models.

The TV graph learning method in \cite{hallac2017} uses TV graphical Lasso, which combines graphical Lasso with a temporal regularization and finds the solution using the alternating direction method of multipliers (ADMM).
Note that graphs estimated with this approach often have negative edge weights.
In contrast, our proposed method is constrained to have non-negative edges in the estimated graph, which are often desired for many applications \cite{dong2019, mateos2019}.
Furthermore, our method has a lower computation complexity: TV graphical Lasso requires eigendecompositions of target matrices to calculate a proximal operator of a logarithm determinant term, whereas our approach is eigendecomposition-free.

The problem of identifying a TV graph to capture causal relationships in the evolution of network contagion has been studied in prior works \cite{hallac2017, cardoso2020learning}.
The model in \cite{hallac2017} focuses on the spread process on networks (e.g., infectious diseases and fake news) and aims to identify the links propagating information over nodes at each time.The one in \cite{cardoso2020learning} focuses on tracking the causal relationship in the financial data. Another work proposes a dynamic network tracking based on tensor decomposition \cite{shen2017tensor}.
However, only directed graphs can be estimated by these approaches.
In contrast, our method estimates undirected graphs whose connections change over time. This is often desired
for data that do not have causal relationships, such as data acquired by physical sensors (e.g.,  point cloud coordinates and temperature readings).

The above-mentioned models, including ours, are \textit{offline} approaches that compute a set of TV graphs from the observed data.
%In contrast
Conversely, several \textit{online} TV graph learning methods have been proposed in recent research.
% Other works on time-varying graph learning often typically estimate the graphs sequentially. 
For example, \cite{Bagheri2024} proposes a joint graph learning and signal interpolation method that assumes low-rank temporal variation, updating the graph at $t$ based on the one at $t-1$. Other online methods \cite{Natali2022,Saboksayr2021} also follow this sequential approach. \cite{Natali2022} estimates a single successor graph on the fly with new available observations, and \cite{Saboksayr2021} sequentially estimates stationary batches of graphs.
We provide a summary of state-of-the-art TV graph learning methods in \autoref{tab:graph_comparison}.

The remainder of this paper is organized as follows.
 \autoref{notations} presents preliminaries about graphs and proximal operators. 
 \autoref{sec:tvgl_form} introduces the problem formulation of time-varying graph learning.
 \autoref{sec:proposed} defines regularization terms for temporal graph variation and the corresponding graph learning optimization problem, and proposes an algorithm to find a solution.
 \autoref{exp} and \autoref{application} provide experimental results with synthetic and real data, respectively.
Finally, the conclusions and future work are presented in  \autoref{conclusion}.

\section{Preliminaries}
\label{notations}

The notation used in this paper is summarized in \autoref{tb:notation}.
Vectors and matrices are written in bold lowercase and uppercase letters, respectively.
The calligraphic capital letters, namely, $\mathcal{V}$ and $ \mathcal{W}_{m}$, denote sets.
$\mathcal{O}(\cdot)$ and $\Omega(\cdot)$ are the big-O and big-Omega notations used in complexity theory.

\subsection{Basic GSP Definitions}
An undirected weighted graph is represented as $\mathcal{G} = (\mathcal{V}, \mathcal{E}, {\bf W})$, where  $\mathcal{V}$ is a set of nodes, $\mathcal{E}=\{(i,j)\}_{i,j\in \mathcal{V}}$ is a set of edges, and ${\bf W}$ is a weighted adjacency matrix. 
The number of nodes is given by $N =|\mathcal{V}|$, and the number of edges is given by $E = |\mathcal{E}|\leq N(N-1)/2$.
Each element of the weighted adjacency matrix is defined by
\begin{equation}
\label{ }
({\bf W})_{mn} = \begin{cases}
W_{mn}  & \text{if nodes } m \text{ and } n \text{ are connected}, \\
0      & \text{otherwise},
\end{cases}
\end{equation}
where $W_{mn} \geq 0$ is the edge weight between nodes $m$ and $n$.
The degree matrix ${\bf D}$ is a diagonal matrix whose diagonal element is $D_{mm} = \sum_n W_{mn}$.
The graph Laplacian is given by ${\bf L} = {\bf D} - {\bf W}$.
Let ${\bf f} \in \mathbb{R}^{N}$ be a signal on the graph, then the Laplacian quadratic form is given by 
\begin{equation}
	{\bf f}^{\transp} {\bf L} {\bf f} = \frac{1}{2} \sum_{m,n} W_{mn} ({f}_{m} - {f}_{n})^{2},
	\label{eq:smooth}
\end{equation}
where $f_{m}$ and $f_{n}$ denote the signal values at nodes $m$ and $n$, respectively.
Note that \eqref{eq:smooth} provides a measure to quantify the smoothness of $\mathbf{f}$ \cite{shuman2013,ortega2022introduction}.

Because $\bf L$ is a real symmetric matrix, it has orthogonal eigenvectors and can be decomposed into ${\bf L} = {\bf U}{\bf \Lambda} {\bf U}^{\transp}$, where ${\bf U} = [{\bf u}_{0}, {\bf u}_{1}, \ldots , {\bf u}_{N-1}]$ is a matrix whose $i$-th column is the eigenvector ${\bf u}_{i}$ and ${\bf \Lambda} = {\rm diag}(\lambda_{0}, \lambda_{1}, \ldots \lambda_{N-1})$ is a diagonal eigenvalue matrix.
The graph Fourier transform (GFT) 
% and the inverse GFT are 
is defined as: $\hat{f}[i] = \langle {\bf u}_{i}, {\bf f} \rangle = \sum_{n=0}^{N-1} u_{i}[n]f[n]$.
% \begin{align}
% 	\hat{f}[i] &= \langle {\bf u}_{i}, {\bf f} \rangle = \sum_{n=0}^{N-1} u_{i}[n]f[n], \label{eq:gft}  \\
% 	{f}[n] &= \sum_{i=0}^{N-1} \hat f[i] u_{i}[n].
% 	\label{eq:igft}
% \end{align}
Eigenvalues of the graph Laplacian correspond to frequencies in the classical Fourier transform and, thus, are often called {\it graph frequencies}.

\subsection{Proximal Operator and Its Properties}
Let $f : \mathbb{R}^{n} \to \mathbb{R} \cup \{ \infty \} $ be a proper lower semicontinuous convex function. 
The proximal operator $\text{prox}_{\gamma f} : \mathbb{R}^{n} \to \mathbb{R}^{n}$ of $f$ with a parameter $\gamma > 0$ is defined by
\begin{equation}
	\text{prox}_{\gamma f}({\bf x}) = \argmin_{\bf y} f({\bf y}) + \frac{1}{2 \gamma} \| {\bf y - x} \|_{2}^{2}.
\end{equation}
If a proximal operator $\text{prox}_{\gamma f}$ can be computed efficiently, the function $f$ is called proximable. 

Proximal operators have the following properties \cite{parikh2014}:
\begin{itemize}
	\item If a function $f$ is separable across variables, i.e., $f({\bf x}) = f_{1}({\bf x}_{1}) + f_{2}({\bf x}_{2})$ with ${\bf x} = [{\bf x}_{1}^{\transp} \ {\bf x}_{2}^{\transp}]^{\transp}$, then
	\begin{equation}
		\text{prox}_{\gamma f}({\bf x}) = [(\text{prox}_{\gamma f_{1}}({\bf x}_{1}))^{\transp} \ (\text{prox}_{\gamma f_{2}}({\bf x}_{2}))^{\transp}]^{\transp},
		\label{eq:pprox1}
	\end{equation}
	where ${\bf x} = [{\bf x}_{1}^{\transp} \ {\bf x}_{2}^{\transp}]^{\transp}$.
	Thus, the computation in the proximal operator of separable functions reduces to the computation of the proximal operator for each variable block.
	
	\item If a function $f$ is fully separable to its elements, i.e., $f({\bf x}) = \sum_{i}f_{i}(x_{i})$, then
	\begin{equation}
		(\text{prox}_{\gamma f}({\bf x}))_{i} = \text{prox}_{\gamma f_{i}}(x_{i}).	
	\end{equation}
	Therefore, in this case, the computation of the proximal operator can be reduced to the element-wise computation. 
\end{itemize}

\section{Static and Time-Varying Graph Factor Analysis}
\label{sec:tvgl_form}
We first describe SGFA \cite{dong2016}, which introduces a model for the generation of graph signals. 
Subsequently, we introduce TGFA, which extends SGFA to formulate the time-varying graph learning problems studied in this paper.

\subsection{Static Graph Factor Analysis (SGFA)}
\label{sec:fags}
The observation model of SGFA is defined \cite{dong2016}:
\begin{equation}
	{\bf x} = {\bf U}{\bf h} + {\bm \epsilon},
	\label{eq:factora}
\end{equation}
where ${\bf x} \in \mathbb{R}^{N}$ is an observed signal, ${\bf h} \in \mathbb{R}^{N}$ is a latent variable represented in the graph frequency domain, ${\bf U}$ is the GFT matrix and ${\bm \epsilon} \sim \mathcal{N}(0, \sigma_{\epsilon}^2 {\bf I})$ is an additive white Gaussian noise. 
From \eqref{eq:factora}, signals 
%possessing graph structures 
can be represented by the inverse GFT of latent variables in the spectral domain of the underlying graph. 
SGFA assumes that the latent variable ${\bf h}$ follows the multivariate Gaussian distribution 
\begin{equation}
	p({\bf h}) = \mathcal{N}(0, {\bf \Lambda}^{\dagger}).
	\label{eq:hprob}
\end{equation}
This corresponds to the assumption that signal energy tends to be concentrated in the low frequencies and thus encourages graph signal smoothness.
  \eqref{eq:factora} and \eqref{eq:hprob} lead to the conditional probability of ${\bf x}$ given ${\bf h}$:
\begin{equation}
	p({\bf x} | {\bf h} ) = \mathcal{N}({\bf Uh}, \sigma_{\epsilon}^2 {\bf I}).
	\label{eq:condx}
\end{equation}
From \eqref{eq:hprob} and \eqref{eq:condx}, the marginal distribution of ${\bf x}$ is given by:
\begin{equation}
	p({\bf x}) = \mathcal{N}(0, {\bf L}^{\dagger} + \sigma_{\epsilon}^2 {\bf I}).
	\label{eq:margix}
\end{equation}
Note that the fact ${\bf L}^{\dagger} = {\bf U \Lambda^{\dagger} U^{\transp}}$ is used in the derivation of \eqref{eq:margix}.
The marginal distribution in \eqref{eq:margix} indicates that a graph signal ${\bf x}$ can be generated by the multivariate Gaussian distribution whose precision matrix is the graph Laplacian of the underlying graph.
Signals generated from the distribution \eqref{eq:margix} satisfy graph stationarity \cite{perraudin2017} because their covariance and graph Laplacian are jointly diagonalizable by the eigenvectors of the graph Laplacian.
The maximum a posteriori estimation of ${\bf h}$ leads to the following problem \cite{dong2016}:
\begin{equation}
	\min_{{\bf L, y}} \ \| {\bf x} - {\bf y} \|^{2}_{2} + \alpha {\bf y}^{\transp}{\bf L}{\bf y},
	\label{eq:dongf}
\end{equation}
where ${\bf y} = {\bf Uh}$ and ${\bf y}$ is regarded as a noiseless signal.
This problem is solved by an optimization that alternates denoising and graph learning steps \cite{dong2016}.
However, this optimization cannot guarantee the convergence to the optimal solution in general.
To avoid this, by assuming $\mathbf{y} = \mathbf{x}$, \eqref{eq:dongf} is reformulated as the problem of finding the graph Laplacian that minimizes the smoothness measure $\mathbf{x}^{\transp}\mathbf{Lx}$ under some constraints \cite{kalofolias2016, kalofolias2017, chepuri2017}.

\subsection{Time-varying Graph Factor Analysis (TGFA)}
Suppose that we have a multivariate time series data divided into non-overlapping time windows\footnote{With slight modifications, overlapping sliding windows can also be used.} $\{\mathbf{X}_{1}, \ldots , \mathbf{X}_{T}\}$, where $\mathbf{X}_{t} = [\mathbf{x}_{1}^{(t)}, \dots, \mathbf{x}_{K}^{(t)}] \in \mathbb{R}^{N \times K}$ contains the $K$ observations corresponding to time window $t$, and the graph corresponding to observations within a time window is invariant. 
%The choice of $K$ depends on a sampling frequency in measurements.
Our goal is to learn a sequence of graph Laplacians ${\bf L}_{1}, \ldots, {\bf L}_{T}$, where ${\bf L}_t$ corresponds to ${\bf X}_t$ .

We extend SGFA by incorporating a graph evolution process. We define the TGFA model as follows:
\begin{align}
    {\bf x}^{(t)} &\sim \mathcal{N}(0, {\bf L}_{t}^{\dagger} + \sigma_{\epsilon}^2 {\bf I}), \label{eq:tgfa_sig} \\
    {\bf L}_{t} &= \begin{cases} \label{eq:delta_L}
	{\bf 0}  & t =0 \\
	{\bf L}_{t-1} + \Delta{\bf L}_{t} &  t \geq 1,
\end{cases}
\end{align}
where ${\bf x}^{(t)}$, ${\bf L}_{t}$, and $\Delta{\bf L}_{t}$ are a signal, the graph Laplacian of the underlying graph and a graph update, representing the graph variation at a given time $t$, respectively. 
The probability distribution of $\mathbf{x}^{(t)}$ in \eqref{eq:tgfa_sig} indicates that time-varying graph learning requires the TV observations to be smooth on the corresponding learned graphs.
Introducing $\Delta \mathbf{L}_t$ in \eqref{eq:delta_L} allows us to incorporate prior knowledge about temporal graph variation.

Given the TGFA, a time-varying graph learning problem consists of estimating a set of graph Laplacians minimizing $\sum_t \mathbf{x}^{(t)}\mathbf{L}_t\mathbf{x}^{(t)}$.
The temporal graph variation is constrained via regularization term(s) corresponding to the prior knowledge on the graph temporal variation $\Delta\mathbf{L}_t$.

The generalized time-varying graph learning problem leads to the following optimization:
\begin{equation}
    \min_{{\bf L}_{t} \in \mathcal{L}} \sum_{t=1}^{T} \mathrm{Tr}({\bf X}_{t}^{\transp}{\bf L}_{t}{\bf X}_{t}) + f({\bf L}_{t}) + \eta \sum_{t=2}^{T} R(\Delta {\bf L}_{t} \circ {\bf H}), \\
\label{eq:tgl_lform}
\end{equation}
where $\mathcal{L}$ is the valid set of graph Laplacians and is given by
\begin{equation}
\begin{split}
    \mathcal{L} = \{ \ {\bf L} \in \mathbb{R}^{N \times N} \ | {\bf L}={\bf L}^{\transp}, L_{ij} \leq 0 \ \ (i \neq j), \\
    L_{ii} = -\sum_{j \neq i} L_{ij} \},
	\label{eq:Lspace}
\end{split}
\end{equation}
in which ${\bf H} = {\bf 1}{\bf 1}^{\transp} - {\bf I}$.
In \eqref{eq:tgl_lform}, $R(\Delta {\bf L}_{t} \circ {\bf H})$ is a regularization term that characterizes the temporal change in graph edges, that is, $\Delta\mathbf{L}_t\circ\mathbf{H}=\mathbf{W}_{t}-\mathbf{W}_{t-1}$.
The first term in \eqref{eq:tgl_lform} corresponds to the smoothness term in the static case \eqref{eq:dongf} and quantifies the smoothness of the signals on TV graphs.
The function in the second term $f({\bf L}_{t})$ is a regularization to avoid obtaining trivial solutions. This function, examples of which will be given later, depends on the assumptions made about the graph model.
The parameter $\eta$ controls the regularization strength.

We first focus on the single-step memory case formulation, where the temporal variation is described by its direct predecessor only: $\Delta\mathbf{L} := \mathbf{L}_t-\mathbf{L}_{t-1}$. The same approach can be applied for the multi-step memory case by replacing $\Delta\mathbf{L}$ in \eqref{eq:tgl_lform} by a function involving multiple $\mathbf{L}_t$ (see  \autoref{susec:Lstep}).

To obtain a tractable formulation with simplified conditions for the valid set, we reformulate the problem \eqref{eq:tgl_lform} with constraint \eqref{eq:Lspace} by directly optimizing the time-varying sequence of weighted adjacency matrices $\{\mathbf{W}_t\}_{t=1}^{T}$ instead of the graph Laplacians. This leads to 
a problem equivalent to that of \eqref{eq:tgl_lform}:
\begin{equation}
\begin{split}
    \min_{{\bf W}_{t} \in \mathcal{W}_{m}} \sum_{t=1}^{T} \frac{1}{2} \| {\bf W}_{t} \circ {\bf Z}_{t} \|_1 + f({\bf W}_{t}) + \eta \sum_{t=2}^{T} R({\bf W}_{t} - {\bf W}_{t-1}), \\
\end{split}
\label{eq:tgl_wform}
\end{equation}
where the valid set of weighted adjacency matrices is 
\begin{equation}
    \mathcal{W}_{m} = \left \{  {\bf W} \in \mathbb{R}_{+}^{N \times N}  \ | \  {\bf W}={\bf W}^{\transp}, W_{ii} = 0 \right \}.
    \label{eq:Wspace}
\end{equation}
Here, $\mathbf{W}_1$ is a valid learned graph adjacency carrying no temporal penalty.
The first term of \eqref{eq:tgl_wform} is equal to $\mathrm{Tr}({\bf X}_{t}^{\transp}{\bf L}_{t}{\bf X}_{t})$, since \cite{kalofolias2016}:
\begin{equation}
	\mathrm{Tr}({\bf X}_{t}^{\transp}{\bf L}_{t}{\bf X}_{t}) = \frac{1}{2} \mathrm{Tr}({\bf W}_{t}{\bf Z}_{t}) = \frac{1}{2} \|{\bf W}_{t} \circ {\bf Z}_{t} \|_1,
	\label{eq:smoothness}
\end{equation}
where ${\bf Z}_{t}$ is the pairwise distance matrix defined by $({\bf Z}_{t})_{ij} = \sum_{k=1}^{K}\| ({\bf x}^{(t)}_{k})_{i}-({\bf x}^{(t)}_{k})_{j} \|^{2}$. 
% \begin{equation}
% 	({\bf Z}_{t})_{ij} = \sum_{k=1}^{K}\| ({\bf x}^{(t)}_{k})_{i}-({\bf x}^{(t)}_{k})_{j} \|^{2}.
% 	\label{eq:dist}
% \end{equation}

In this paper, we solve the optimization problem in \eqref{eq:tgl_wform} to learn a TV graph.
Throughout this paper, we set the regularization for the weighted adjacency matrix to be \cite{kalofolias2017}:
\begin{equation}
		f({\bf W}_{t})= - \alpha {\bf 1}^{\top} \mathrm{log}({\bf W}_{t}{\bf 1})
		+ \beta ||{\bf W}_{t}||_F^2,
	\label{eq:reg_W}
\end{equation}
where $\alpha$ and $\beta$ are the parameters.
The first term in \eqref{eq:reg_W} forces the degree on each node to be positive without preventing edge weights from becoming zero. 
The second term controls the magnitude of edge weights.

For learning TV graphs, the graph temporal variation term $R(\cdot)$ should be chosen appropriately.
In \cite{kalofolias2017}, a temporal variation regularizer $R (\cdot) = \| \cdot \|_{F}^{2} $ is selected to learn TV graphs such that the edge weights change smoothly over time. 
In the next section, we describe the regularization terms to learn TV graphs with models \textbf{TH}, \textbf{SB}, and \textbf{LTC}, followed by the methods to solve the optimization problem.

\section{Learning TV Graph with Temporal Variation Regularization}
\label{sec:proposed}
Specific regularization terms for graph temporal variation should be chosen based on the type of temporal graph evolution suitable for a given application.
In this paper, we consider three %new 
types of graph evolution terms, which appear in many applications, and present an algorithm to solve the corresponding optimizations. 

\subsection{Formulation}
\subsubsection{\textbf{Temporal Homogeneity (TH)}}
In the first model, we assume that most edges and their weights will likely remain unchanged over a short-term time horizon. 
Thus, as the regularizer $R(\cdot)$ in \eqref{eq:tgl_wform}, we aim to minimize the $\ell_1$ norm of the adjacency matrices between the time slots $t$ and $t-1$.
This leads to the following {\it fused Lasso} formulation  \cite{tibshirani2005}: 
\begin{equation}
    \min_{{\bf W}_{t} \in \mathcal{W}_{m}} \sum_{t=1}^{T} \frac{1}{2} \| {\bf W}_{t} \circ {\bf Z}_{t} \|_1 + f({\bf W}_{t}) + \eta \sum^{T}_{t=2} \|{\bf W}_{t}-{\bf W}_{t-1} \|_{1}.
\label{eq:tgl_W_flasso}
\end{equation}

To make the optimization problem tractable, we rewrite \eqref{eq:tgl_W_flasso} in vector form, with ${\bf W}_{t}$ and ${\bf Z}_{t}$  replaced by ${\bf w}_{t} \in \mathbb{R}_{+}^{N(N-1)/2}$ and  ${\bf z}_{t} \in \mathbb{R}_{+}^{N(N-1)/2}$ respectively. Only the upper-triangular parts of ${\bf W}_{t}$ and ${\bf Z}_{t}$ are considered, given that the graph is undirected.

Let us define ${\bf w} = [{\bf w}_{1}^{\transp} \  {\bf w}_{2}^{\transp} \ldots {\bf w}_{T}^{\transp}]^{\transp}$, ${\bf z} = [{\bf z}_{1}^{\transp} \  {\bf z}_{2}^{\transp} \ldots {\bf z}_{T}^{\transp}]^{\transp}$, and ${\bf d} = [{\bf d}_{1}^{\transp} \  {\bf d}_{2}^{\transp} \ldots {\bf d}_{T}^{\transp}]^{\transp}$ where ${\bf d}_{t} = {\bf W}_{t}{\bf 1}$.
We also introduce the linear operators $\bf S$ and $\bf \Phi$ that satisfy ${\bf S}{\bf w} = {\bf d}$ and ${\bf \Phi w} = {\bf w} - {\bf \hat{w}}$ respectively, where ${\bf \hat{w}} = [{\bf w}_{1}^{\transp} \ {\bf w}_{1}^{\transp} \  {\bf w}_{2}^{\transp} \ldots {\bf w}_{T-1}^{\transp}]^{\transp}$.
Then, we can rewrite \eqref{eq:tgl_W_flasso} and \eqref{eq:reg_W} as
\begin{equation}
	\min_{{\bf w} \in \mathcal{W}_{v}} \ {\bf z}^{\transp}{\bf w} - \alpha {\bf 1}^{\transp} \mathrm{log}({\bf S}{\bf w})+ \beta \|{\bf w} \|_2^2 + \eta \| {\bf \Phi w} \|_{1},
	\label{eq:vopt}
\end{equation}
where 
\begin{equation}
	\mathcal{W}_{v} = \left \{ {\bf w} \in \mathbb{R}^{TN(N-1)/2} \ | \ w_{i} \geq 0 \ (i=1,2, \ldots) \right \}
	\label{eq:vset}
\end{equation}
represents the non-negativity constraint. 
The symmetric and diagonal constraints can be simplified by expressing \eqref{eq:Wspace} in vector form.

\subsubsection{\textbf{Switching Behavior (SB)}}
The second model aims to learn graphs with switching behavior. In this case, the graphs remain almost constant for most time slots, but are allowed to undergo infrequent sudden changes at a small number of time slots.
To reflect this characteristic, we consider the Frobenius norm of the differences between the adjacency matrices at $t$ and $t-1$, i.e., $\|{\bf W}_{t}-{\bf W}_{t-1} \|_{F}$.
To maintain this term small, the TV graphs should be almost static over time.
Note that SB is designed to capture \emph{sudden} (few-time-slot) changes rather than \emph{rapid} (fast-varying) changes: the $\ell_1$ sum of Frobenius norms penalizes the \emph{number of time slots} at which changes occur, while the Frobenius norm itself permits a large change in topology at each of those time slots.
The resulting optimization problem is:
\begin{equation}
\min_{{\bf W}_{t} \in \mathcal{W}_{m}} \sum_{t=1}^{T} \frac{1}{2} \| {\bf W}_{t} \circ {\bf Z}_{t} \|_1 + f({\bf W}_{t}) + \eta \sum^{T}_{t=2} \|{\bf W}_{t}-{\bf W}_{t-1} \|_{F},
\label{eq:pmodel2}
\end{equation}
where only the last term differs from \eqref{eq:tgl_W_flasso}, allowing more edge weights to change, while changes remain infrequent.

To make the optimization problem tractable, we rewrite \eqref{eq:pmodel2} in the same manner as in the previous model as:
\begin{equation}
	\min_{{\bf w} \in \mathcal{W}_{v}} \ {\bf z}^{\transp}{\bf w} - \alpha {\bf 1}^{\transp} \mathrm{log}({\bf S}{\bf w})+ \beta \|{\bf w} \|_2^2 + \eta \sum _{j} \| [ {\bf \Phi w} ] _{j} \|_{2},
	\label{eq:optmodel2}
\end{equation}
where $ [ {\bf \Phi w} ] _{j} = {\bf w}^{\transp}_{j} - {\bf w}^{\transp}_{j-1}$, $j=2, \dots T$.

\subsubsection{\textbf{Local Topological Changes (LTC)}}
In the third model, we assume that the connectivity is nearly static, but local changes are possible.
That is, for a few nodes, several connections will change, while for most nodes, at most one connection change will occur. For instance, if the connectivity of a hub node changes, its neighbors are affected, and at least one of the neighboring nodes edges is altered. At each time, the set of nodes changing can be different. 
Mathematically, when this happens, the matrix of changes ${\bf W}_{t}-{\bf W}_{t-1}$ will exhibit row sparsity (only a few rows will contain non-zero entries), which can be characterized by the matrix $\ell_{2,1}$-norm regularization \cite{yang2011ell}. 
Thus, we consider the following formulation for LTC: 
\begin{equation}
    \min_{{\bf W}_{t} \in \mathcal{W}_{m}} \sum_{t=1}^{T} \frac{1}{2} \| {\bf W}_{t} \circ {\bf Z}_{t} \|_1 + f({\bf W}_{t}) + \eta \sum^{T}_{t=2} \|{\bf W}_{t}-{\bf W}_{t-1} \|_{2,1}.
\label{eq:pmodel3}
\end{equation}
The $\ell_{2,1}$ norm is also known as \textit{group Lasso} \cite{meier2008}. 
Let us consider a case of a \textit{concentrated} change in an unweighted graph, where a node changes its connection with $M$ nodes. In this case, the $\ell_{2,1}$ norm penalty will be $M+\sqrt{M}$, while the Frobenius norm penalty will be $\sqrt{2M}$. In contrast, if all these changes are \textit{distributed}, e.g., two nodes each change their connection with $M/2$ distinct nodes (no overlap between the two sets of newly connected or disconnected neighbors), the $\ell_{2,1}$ norm penalty will be $M+\sqrt{M/2}$ while the Frobenius norm penalty will be the same as the previous case. This indicates that the Frobenius norm regularization is insensitive to the structure of the change, while the $\ell_{2,1}$ norm regularization encourages the model to learn structurally sparse solutions.
To make the optimization problem tractable, we rewrite \eqref{eq:pmodel3} in the same manner as the previous model:
\begin{equation}
    \min_{{\bf w} \in \mathcal{W}_{v}} \ {\bf z}^{\transp}{\bf w} - \alpha {\bf 1}^{\transp} \mathrm{log}({\bf S}{\bf w})+ \beta \|{\bf w} \|_2^2 + \eta \sum_{j}\sum_{i}\| [ {\bf \Phi w} ] _{j}^{(i)} \|_{2},
\label{eq:optmodel3}
\end{equation}
where $[\mathbf{\Phi w}]_{j}^{(i)}=[\mathbf{w}^\top_j-\mathbf{w}^\top_{j-1}]^{(i)},\ i=1,\dots,N, j = 2,\dots,T$.
$[\mathbf{\Phi w}]_{j}^{(i)}$ represents a vector whose elements are $i$-th row of the temporal variation of the adjacency matrix.

Note that the regularization terms mentioned above are convex. Combining these may lead to a meaningful model, but some combinations could be redundant. For example, the combination of TH and the Tikhonov regularization from \cite{kalofolias2017} leads to an elastic-net \cite{zou2006elastic} regularization. These combined problems can still be solved via the same algorithm shown below, as long as the terms are additive and convex.

The choice of regularizer should follow domain knowledge about the expected
temporal dynamics: TH is appropriate when most edges change slowly and smoothly;
SB when abrupt topology switches occur at few time instants; LTC when connectivity
changes are localized to a few nodes per time step. Absent such prior knowledge,
cross-validation or model-selection criteria can guide the choice, which we leave
for future work.

\subsection{Optimization}

Since the proposed regularization terms are not differentiable, these optimization problems cannot be solved directly using the methods in \cite{kalofolias2017}. 
Thus, we solve \eqref{eq:vopt}, \eqref{eq:optmodel2} and \eqref{eq:optmodel3} with a primal-dual splitting (PDS) method \cite{condat2013}, which solves a problem in the form,
\begin{equation}
	\min_{{\bf w}} f_{1}({\bf w}) + f_{2}({\bf w}) + f_{3}({\bf Mw}),
	\label{eq:pds}
\end{equation}
where $f_{1}$ is a differentiable convex function with gradient $\nabla f_{1}$ and Lipschitz constant $\xi$; $f_{2}$ and $f_{3}$ are proper lower semi-continuous convex functions which are proximable; and ${\bf M}$ is a linear operator.

Many algorithms to solve \eqref{eq:pds} have been proposed \cite{condat2013, komodakis2015, daubechies2004}. 
We use a forward-backward-forward (FBF) approach \cite{combettes2012}, which allows parallel computation of the proximal operators of $f_{2}$ and $f_{3}$.

\subsubsection{Temporal Homogeneity (TGL-TH)}
By introducing the indicator function of $\mathcal{W}_{v}$ in \eqref{eq:vset}, we rewrite \eqref{eq:vopt} as follows:
\begin{equation}
\begin{split}
	\min_{{\bf w}} \ {\bf z}^{\transp}{\bf w} - \alpha {\bf 1}^{\transp} \mathrm{log}({\bf S}{\bf w})+ \beta \|{\bf w} \|_2^2
	+ \eta \| {\bf \Phi w} \|_{1} + \iota_{\mathcal{W}_{v}}({\bf w}),
\end{split}
\label{eq:vopt2}
\end{equation}
where the indicator function $\iota_{\mathcal{W}_{v}}$ is defined by
\begin{equation}
	\iota_{\mathcal{W}_{v}}({\bf w}) = \begin{cases}
	0  & w_{i} \geq 0 \\
	\infty      & \text{otherwise}.
\end{cases}
	\label{eq:indf}
\end{equation} 
The objective function \eqref{eq:vopt2} further reduces to the applicable form of the PDS as follows:
\begin{equation}
\begin{split}
	f_{1}({\bf w}) &= \beta \|{\bf w} \|_2^2 \ \ \text{with} \ \xi = 2\beta , \\
	f_{2}({\bf w}) &= {\bf z}^{\transp}{\bf w} + \iota_{\mathcal{W}_{v}}({\bf w}), \\
	f_{3}({\bf v}) &= - \alpha {\bf 1}^{\transp} \mathrm{log}({\bf v}_{1}) + \eta \| {\bf v}_{2} \|_{1}, \\
	{\bf M} &= 
		\begin{bmatrix}
				{\bf S} \\
				{\bf \Phi}
		\end{bmatrix},
\end{split}
\label{eq:model1ff}
\end{equation}
where the dual variable is ${\bf v} := {\bf Mw} = [{\bf v}_{1}^{\transp} \ {\bf v}_{2}^{\transp}]^{\transp}$.

The proximal operator of $f_{2}$ is given by
\begin{equation}
	\bigl(\text{prox}_{\gamma (\| \cdot \|_{1}  + \iota_{\mathcal{W}_{v}})} ({\bf x}) \bigr)_{i} = \begin{cases}
	0  & x_{i} \leq \gamma \\
	x_{i} - \gamma     & \text{otherwise}.
\end{cases}
\end{equation}
Because $f_{3}$ is separable across variables, the proximal operator can be computed separately for each term (see \eqref{eq:pprox1}).
The proximal operator for the log barrier function \cite{parikh2014} as the first term of $f_{3}$ is given by 
\begin{equation}
\label{eq:prox_l1}
	\bigl(\text{prox}_{\gamma (- {\bf 1}^{\transp} \log(\cdot))} ({\bf x}) \bigr)_{i} = \frac{x_{i} + \sqrt{x_{i}^{2} + 4 \gamma}}{2}.
\end{equation}
The proximal operator for the $\ell_{1}$-norm, i.e., the second term in $f_{3}$, is well known to be the element-wise soft-thresholding operation \cite{parikh2014}:
\begin{equation}
	\bigl(\text{prox}_{\gamma \| \cdot \|_{1}} ({\bf x}) \bigr)_{i} = \begin{cases}
	0  & |x_{i}| \leq \gamma \\
	\text{sgn}(x_{i})(|x_{i}| - \gamma)      & \text{otherwise}.
\end{cases}
\end{equation}
See \autoref{alg1} for the complete algorithm. 

\subsubsection{Switching Behavior (TGL-SB) and LTC (TGL-LTC)}
The objective functions \eqref{eq:optmodel2} and \eqref{eq:optmodel3} for \textbf{SB} and \textbf{LTC}, respectively, can be reduced to the PDS applicable form similar to the first model. For \eqref{eq:optmodel2}, we can replace $f_{3}$ with
\begin{equation}
	f_{3}({\bf v}) = - \alpha {\bf 1}^{\transp} \mathrm{log}({\bf v}_{1}) + \eta \sum_{j=2}^{T} \| [ {\bf v}_{2} ]_{j} \|_{2}, 
	\label{eq:model2ff}
\end{equation}
where $[ {\bf v}_{2} ] _{j} = [ {\bf \Phi w} ] _{j}.$

For \eqref{eq:optmodel3}, we can replace $f_3$ with
\begin{equation}
    f_{3}({\bf v}) = - \alpha {\bf 1}^{\transp} \mathrm{log}({\bf v}_{1}) + \eta \sum_{j=2}^{T}\sum_{i=1}^{N} \| [ {\bf v}_{2} ]_{j}^{(i)} \|_{2}, 
	\label{eq:model3ff}
\end{equation}
where $N$ is the number of nodes and $[(\mathbf{v}_2)]_{i}^{(i)}=[\mathbf{\Phi w}]_{i}^{(i)}$.

The proximal operator of the second term in the equations above is the block-wise soft-thresholding operation \cite{parikh2014}:
\begin{equation}
\label{eq:prox_l2}
	\text{prox}_{\gamma \| \cdot \|_{2}} ( [ {\bf x} ]_{j} ) = 
	\begin{cases}
	{\bf 0}  & \| [ {\bf x} ]_{j} \|_{2} \leq \gamma \\
	(1 - \gamma / \| [ {\bf x} ]_{j} \|_{2} )( [ {\bf x} ]_{j})      & \text{otherwise}.
	\end{cases}
\end{equation}
Because the proximal operator of the functions $f_{2}$ and $f_{3}$ can be computed efficiently, the second model can also be solved with the PDS algorithm.

\autoref{alg1} summarizes the above solvers.
The hyperparameter $\gamma$ in the PDS algorithm is determined such that the convergence condition \cite{komodakis2015} is satisfied.
Since our proposed optimization problem is convex and lower-semicontinuous, algorithm convergence is guaranteed. The initial iterates are set to zero vectors.

\begin{algorithm}
 \caption{Proposed PDS Algorithm Solver}
 \label{alg1}
 \begin{algorithmic}
 \REQUIRE ${\bf w}^{(0)}, {\bf v}^{(0)}_{1}, {\bf v}^{(0)}_{2}$
 \ENSURE ${\bf w}^{(i)}$
 \WHILE{ $\| {\bf w}^{(i+1)} - {\bf w}^{(i)}\| / \| {\bf w}^{(i)} \| > \epsilon $ }
 \STATE ${\bf y}^{(i)} = {\bf w}^{(i)} -\gamma ( 2 \beta {\bf w}^{(i)} + {\bf S}^{\transp}{\bf v}^{(i)}_{1} + {\bf \Phi}^{\transp}{\bf v}^{(i)}_{2}) $
 \STATE $\bar{\bf y}^{(i)}_{1} = {\bf v}^{(i)}_{1} + \gamma({\bf S} {\bf w}^{(i)})$
 \STATE $\bar{\bf y}^{(i)}_{2} = {\bf v}^{(i)}_{2} + \gamma({\bf \Phi} {\bf w}^{(i)})$
 \STATE ${\bf p}^{(i)} = \mbox{prox}_{\gamma (\| \cdot \|_{1}  + \iota_{\mathcal{W}_{v}}) } ({\bf y}^{(i)})$
 \STATE $\bar{{\bf p}}_{1}^{(i)} = \bar{\bf y}^{(i)}_{1} - \gamma \mbox{prox}_{\frac{1}{\gamma} (- {\bf 1}^{\transp} \log(\cdot))} \bigl( \frac{\bar{\bf y}^{(i)}_{1}}{\gamma} \bigl) $
 \STATE $\bar{{\bf p}}_{2}^{(i)} = \begin{cases}  \bar{\bf y}^{(i)}_{2} - \gamma \mbox{prox}_{\frac{1}{\gamma} \| \cdot \|_{1}} \bigl( \frac{\bar{\bf y}^{(i)}_{2}}{\gamma} \bigl) & \text{to solve \eqref{eq:vopt} } \\  \bar{\bf y}^{(i)}_{2} - \gamma \mbox{prox}_{\frac{1}{\gamma} \| \cdot \|_{2}} \bigl( \frac{\bar{\bf y}^{(i)}_{2}}{\gamma} \bigl) & \text{to solve \eqref{eq:optmodel2} and \eqref{eq:optmodel3} }\end{cases} $
\STATE ${\bf q}^{(i)} = {\bf p}^{(i)} - \gamma (2 \beta {\bf p}^{(i)} + {\bf S}^{\transp} \bar{{\bf p}}_{1}^{(i)} + {\bf \Phi}^{\transp} \bar{{\bf p}}_{2}^{(i)} )$
 \STATE $\bar{{\bf q}}_{1}^{(i)} = \bar{{\bf p}}_{1}^{(i)} + \gamma ({\bf S} {\bf p}^{(i)})$
 \STATE $\bar{{\bf q}}_{2}^{(i)} = \bar{{\bf p}}_{2}^{(i)} + \gamma ({\bf \Phi} {\bf p}^{(i)})$
 \STATE ${\bf w}^{(i+1)} = {\bf w}^{(i)} - {\bf y}^{(i)} + {\bf q}^{(i)}$
 \STATE ${\bf v}_{1}^{(i+1)} = {\bf v}_{1}^{(i)} - \bar{{\bf y}}_{1}^{(i)} + \bar{{\bf q}}_{1}^{(i)}$
 \STATE ${\bf v}_{2}^{(i+1)} = {\bf v}_{2}^{(i)} - \bar{{\bf y}}_{2}^{(i)} + \bar{{\bf q}}_{2}^{(i)}$
\ENDWHILE
\end{algorithmic}
\end{algorithm}

\renewcommand{\arraystretch}{1.4}
\begin{table*}[tb]
	\centering
	\caption{List of alternative and proposed methods}
	\label{tab:mlist}
	\begin{tabular}{ c | c | c | c | c } \hline
		& \bf{Method} & \bf{Approach} & \bf{Temporal Regularization} & \bf{Computational Complexity} \\ \hline
		\multirow{2}{*}{Static} & SGL-Glasso \cite{egilmez2017} & Graphical Lasso based model & - & $\mathcal O (T(N^{2} + \Omega(N^{3})))$  \\
		& SGL-Smooth \cite{kalofolias2016} & Signal smoothness model & - & $\mathcal O(TN^2)$\\ \hline
		\multirow{3}{*}{TV} & TGL-Tik \cite{kalofolias2017} & Signal smoothness + Regularization & $\sum_{t=2}^{T}\|\mathbf{W}_t-\mathbf{W}_{t-1}\|_F^2$ & $\mathcal O(TN^2)$ \\
		& TGL-TH (proposed)    & Signal smoothness + Regularization & $\sum_{t=2}^{T}\|\mathbf{W}_t-\mathbf{W}_{t-1}\|_1$ & $\mathcal O(TN^2)$ \\
		& TGL-SB (proposed)   & Signal smoothness + Regularization & $\sum_{t=2}^{T}\|\mathbf{W}_t-\mathbf{W}_{t-1}\|_F$ & $\mathcal O(TN^2)$ \\
		& TGL-LTC (proposed)   & Signal smoothness + Regularization & $\sum_{t=2}^{T}\|\mathbf{W}_t-\mathbf{W}_{t-1}\|_{2,1}$ & $\mathcal O(TN^2)$ \\
		\hline	
	\end{tabular}
\end{table*}
\renewcommand{\arraystretch}{1}

\subsection{Extension to \texorpdfstring{$L$}{L}-step Memory Cases}
\label{susec:Lstep}
In the model proposed in \eqref{eq:tgl_wform}, we define the temporal variation of the graph as a difference in edge weights in two consecutive time windows,
$\mathbf{W}_t - \mathbf{W}_{t-1}$.
The temporal difference operator $\mathbf{\Phi}\in\mathbb{R}^{ET\times ET}$ thus needs to satisfy% $\mathbf{\Phi}\mathbf{w} = [\mathbf{w}_2-\mathbf{w}_1, \mathbf{w}_3-\mathbf{w_2},\dots,\mathbf{w}_T-\mathbf{w}_{T-1}]$. 
\begin{equation}
\mathbf{\Phi}\mathbf{w} = \begin{bmatrix}\mathbf{w}_1-\mathbf{w}_1\\
\mathbf{w}_2-\mathbf{w}_1\\
% \mathbf{w}_3-\mathbf{w}_2\\
\vdots\\
\mathbf{w}_T-\mathbf{w}_{T-1}
\end{bmatrix}.
\end{equation}
As a result, $\mathbf{\Phi}$ is defined as:
\begin{equation}\label{eq:phi}
\mathbf{\Phi} = 
\begin{bmatrix}&\mathbf{0}_E& &\mathbf{0}_E& &\mathbf{0}_E& && &\mathbf{0}_E\\
&-\mathbf{I}_E& &\mathbf{I}_E& &\mathbf{0}_E& &\dots& &\mathbf{0}_E\\
&\mathbf{0}_E& &-\mathbf{I}_E& &\mathbf{I}_E& && &\mathbf{0}_E\\
&& &\vdots& && &\ddots& &\\
&\mathbf{0}_E& &\mathbf{0}_E& &\mathbf{0}_E& &-\mathbf{I}_E& &\mathbf{I}_E
\end{bmatrix},
\end{equation}
which is the first-order difference of the time-varying edge weights.

The choice of $\mathbf{\Phi}$ reflects an assumption about the expected type of time-varying behavior.
Other $\mathbf{\Phi}$ can be chosen to extend these ideas to the $L$-step memory setting for $L\ge 2$.
For example, $\mathbf{\Phi}$ for $L=2$ is defined as:
\begin{equation}\label{eq:phi2}
\mathbf{\Phi} = 
\begin{bmatrix}&\mathbf{0}_E& &\mathbf{0}_E& &\mathbf{0}_E& && &\mathbf{0}_E\\
&\mathbf{0}_E& &\mathbf{0}_E& &\mathbf{0}_E& &\dots& &\mathbf{0}_E\\
&-\mathbf{I}_E& &-\mathbf{I}_E& &2\mathbf{I}_E& && &\mathbf{0}_E\\
&& &\vdots& && &\ddots& &\\
&\mathbf{0}_E& &\mathbf{0}_E& &-\mathbf{I}_E& &-\mathbf{I}_E& &2\mathbf{I}_E
\end{bmatrix},
\end{equation}
For $L=2$,  we can also define $\mathbf{\Phi}$ as a cyclic operator
\begin{equation}
    \mathbf{\Phi} = 
\begin{bmatrix}&2\mathbf{I}_E& &-\mathbf{I}_E& &\mathbf{0}_E& && &-\mathbf{I}_E\\
&-\mathbf{I}_E& &2\mathbf{I}_E& &-\mathbf{I}_E& &\dots& &\mathbf{0}_E\\
&\mathbf{0}_E& &-\mathbf{I}_E& &2\mathbf{I}_E& && &\mathbf{0}_E\\
&& &\vdots& && &\ddots& &-\mathbf{I}_E\\
&-\mathbf{I}_E& &\mathbf{0}_E& &\mathbf{0}_E& &-\mathbf{I}_E& &2\mathbf{I}_E
\end{bmatrix},
\end{equation}
which is a bidirectional prediction of time-varying graphs.
Note that we can further parameterize the coefficient on each time slot, creating a trade-off between flexibility and computational efficiency. Also, $L$ can be set to be any number as long as $L \le T-1$.
Note that increasing the memory size will lead to a more static temporal behavior.

The $L$-step memory setting can also be found in a TV graph where the graph alternates between two states, such as dense to sparse, with some probability over time \cite{Lambiotte2014}.
In this case, we could design a TV graph learning problem with an appropriate $\mathbf{\Phi}$ by connecting the time slots in the same/similar state.
We can also design $\mathbf{\Phi}$ to take the difference of adjacency matrices among several time slots according to a (pre-defined) temporal connectivity as in \cite{zhang2022}. We compare the performance of the standard first-order temporal model, which only penalizes the deviations from the previous time step ($L=1$), against models incorporating longer-range dependencies, specifically with memory length of $L=2$ and $L=3$ in \autoref{subsec:memory_comp}.

Defining a ``good'' design of $\mathbf{\Phi}$ or the optimal choice of the memory length $L$ generally depends on the application and would require prior knowledge about the characteristics of the temporal evolution of the graph.
Therefore, it is beyond the scope of this paper and is left for future work.

\subsection{Computational Complexity}
In \autoref{tab:mlist}, we summarize the computational complexities of representative graph learning methods.
We compare static graph learning (SGL) methods, including graphical lasso (SGL-Glasso) \cite{egilmez2017} and smoothness prior (SGL-Smooth)\cite{kalofolias2016}, to time-varying graph learning (TGL) such as Tikhonov regularization (TGL-Tik) \cite{kalofolias2017}, to our three proposed approaches: 
temporal homogeneity regularization (TGL-TH), switching behavior regularization (TGL-SB), and local topological changes regularization (TGL-LTC). 

SGL-Glasso updates each row/column of the graph Laplacian at each time slot by solving non-negative quadratic programming, which requires at least $N^3$ order of computation. Therefore the computational complexity for SGL-Glasso is $\mathcal{O} (T(N^2 + \Omega(N^3)))$ per iteration. For SGL-Smooth, the upper triangular components of the adjacency matrix are updated per time slot, which leads to the complexity $\mathcal{O}(TN^2)$ per iteration.

The computational complexity of the TV graph learning algorithms mainly depends on the computation of the proximity operators (e.g., \eqref{eq:prox_l1} and \eqref{eq:prox_l2}). %of the temporal regularization.
For TGL-Tik and all three proposed learning problems, the per-iteration computational complexity is $\mathcal O (TN^{2})$ per iteration. In our experiments, PDS generally converged within several hundred iterations, although TGL-TH sometimes required more than one thousand iterations for larger problems. TGL-LTC has the same asymptotic complexity as TGL-SB. However, it performs per-node block-shrinkage operations per time slot. Consequently, TGL-LTC has larger computational and memory constants than TGL-SB. We compare the computational time in the following section.\\

\begin{figure*}[tb]
	\centering
	\subfigure[TV-RW]{\includegraphics[width=0.22\linewidth]{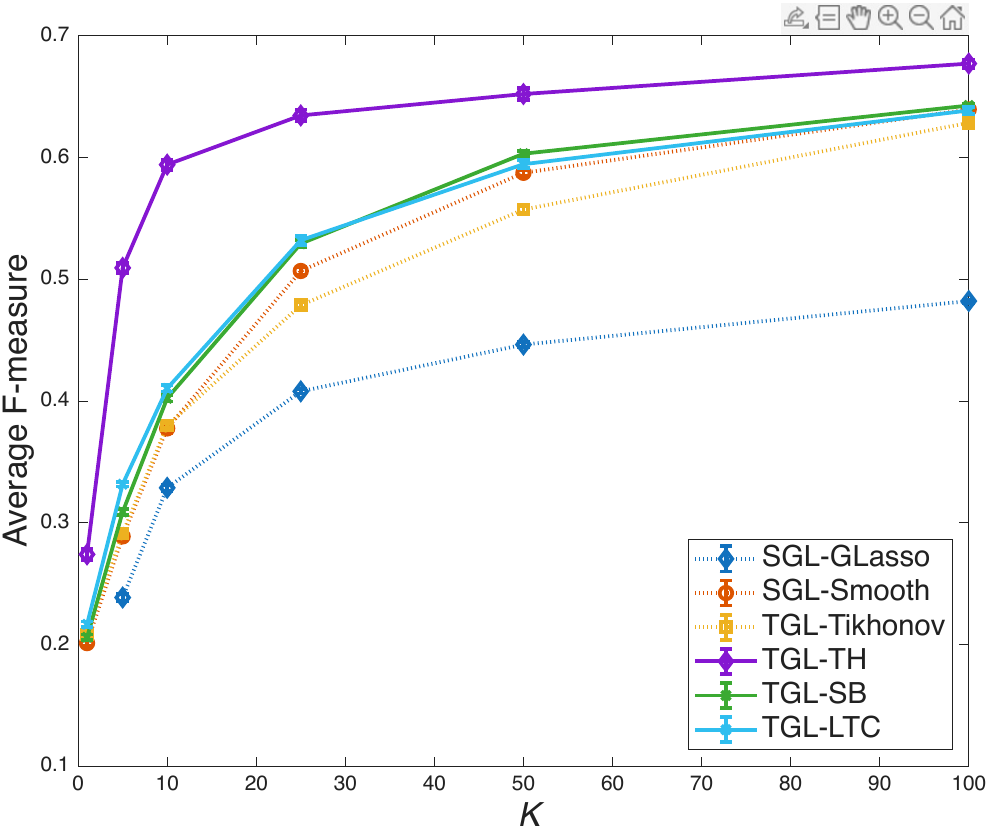} \label{fig:fm_RW}}\
	\subfigure[TH-ER]{\includegraphics[width=0.22\linewidth]{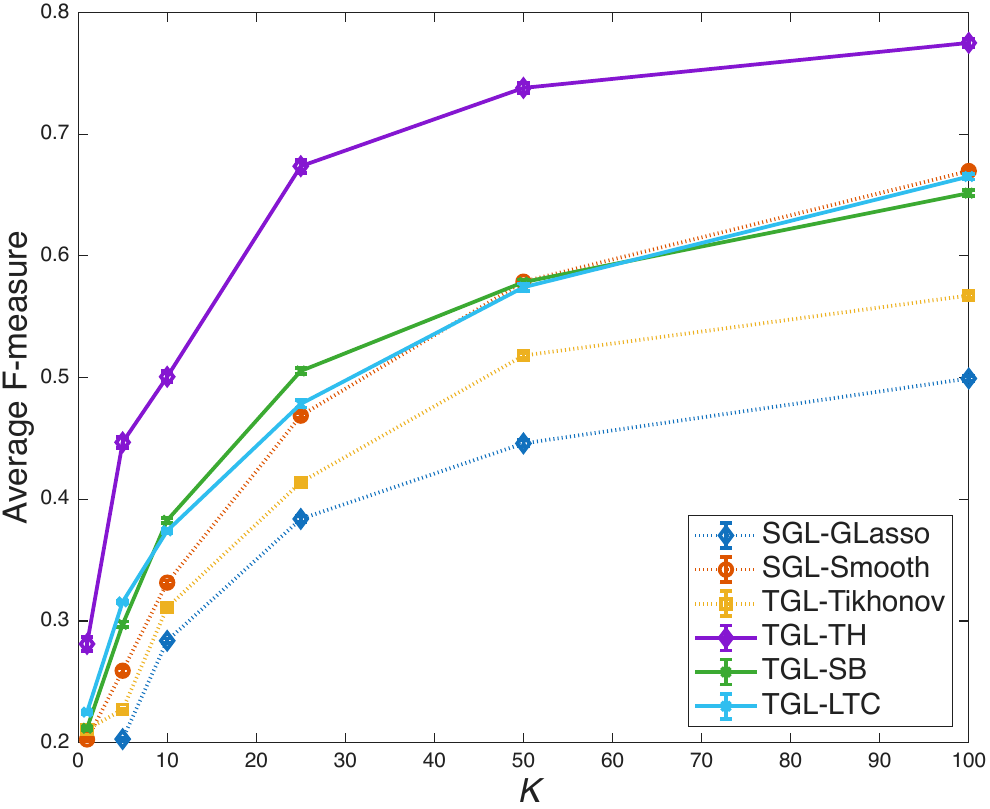} \label{fig:fm_ER}}\
    \subfigure[SB-ER]{\includegraphics[width=0.22\linewidth]{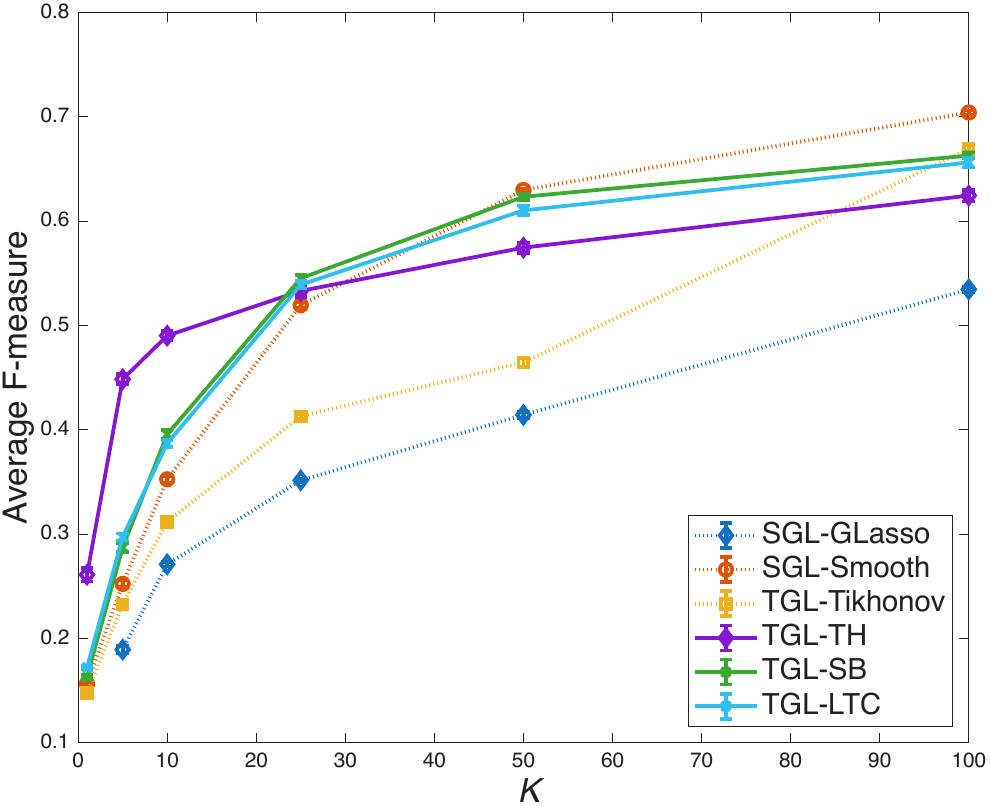} \label{fig:fm_SB}}\
    \subfigure[LTC-ER]{\includegraphics[width=0.22\linewidth]{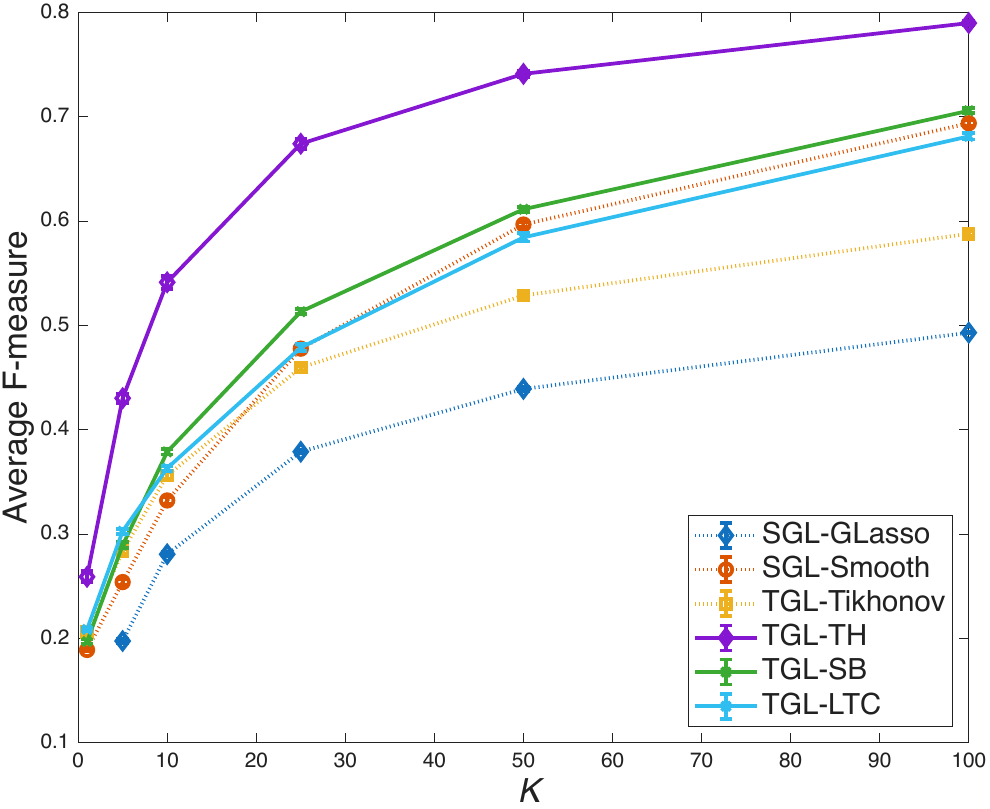} \label{fig:fm_NG}}\\
    \vspace{-3mm}
	\subfigure[TV-RW]{\includegraphics[width=0.22\linewidth]{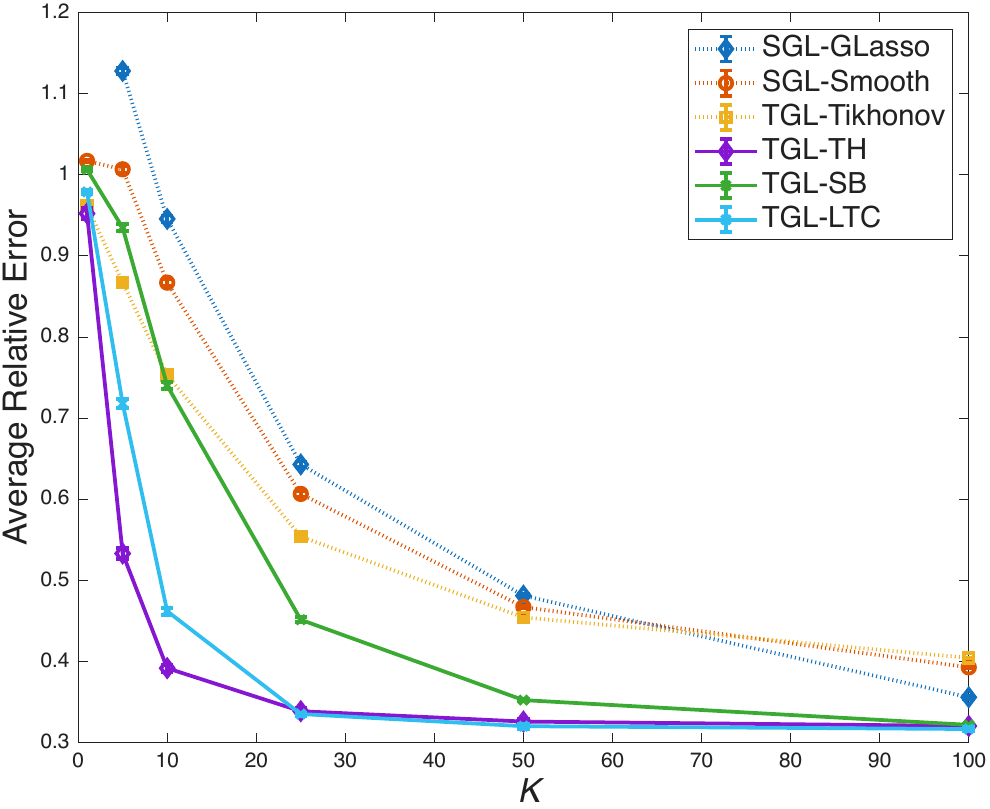} \label{fig:re_RW}}\
	\subfigure[TH-ER]{\includegraphics[width=0.22\linewidth]{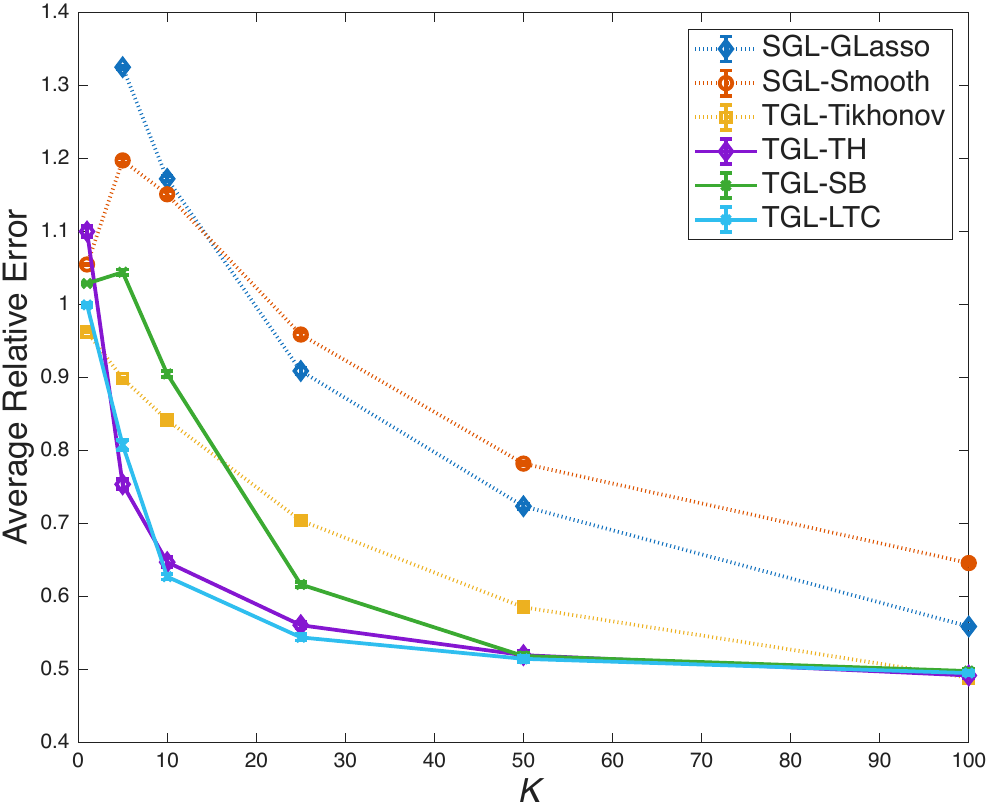} \label{fig:re_ER}}\
    \subfigure[SB-ER]{\includegraphics[width=0.22\linewidth]{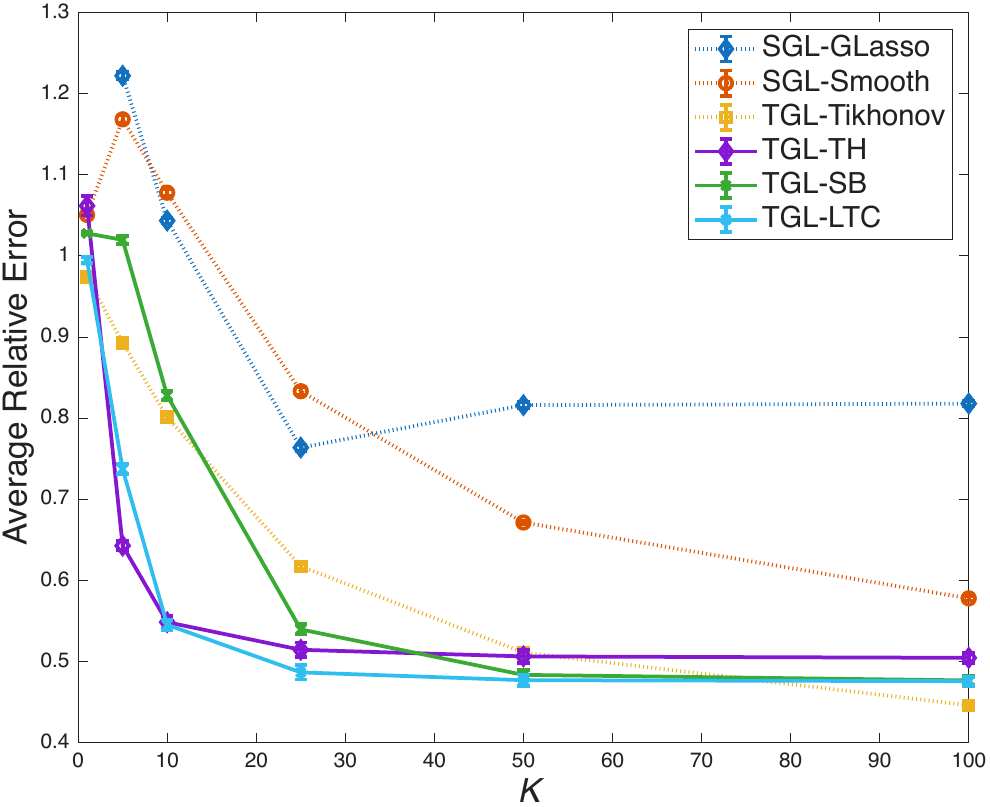} \label{fig:re_SB}}\
    \subfigure[LTC-ER]{\includegraphics[width=0.22\linewidth]{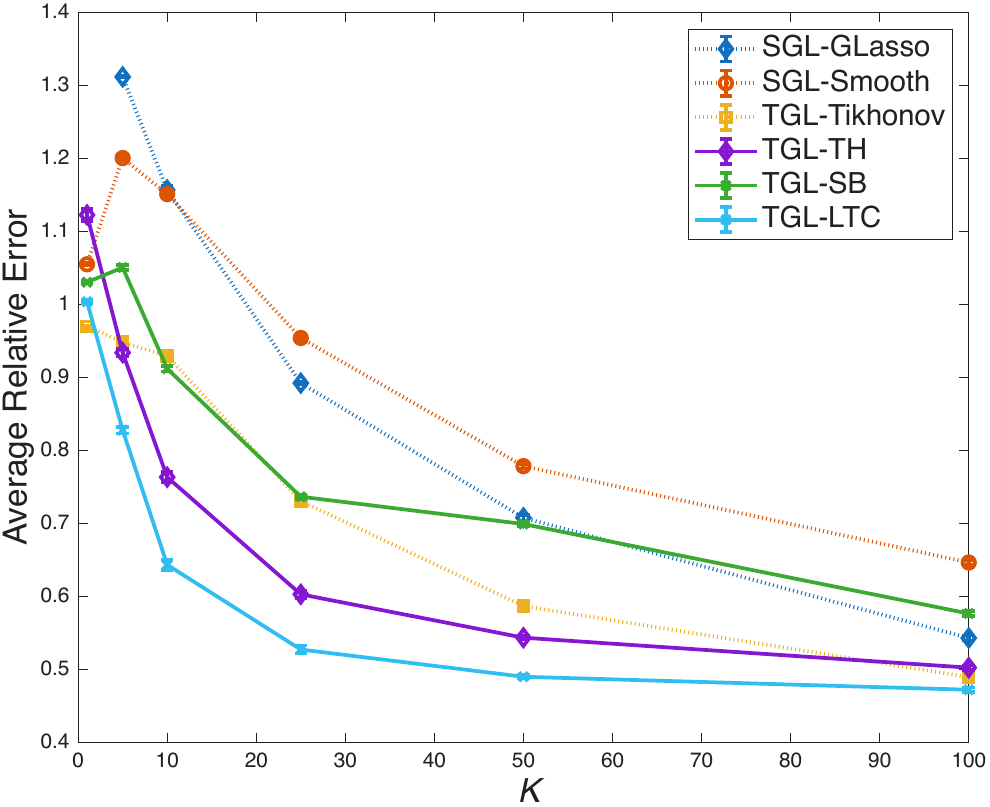} \label{fig:re_NG}}\\
    \caption{The performance of learning TV graphs for different numbers of data samples. The top row demonstrates the F-measure for the datasets based on TV-RW graph, TH-ER graph, SB-ER graph, and LTC-ER graph, respectively. The bottom row demonstrates the relative error for those datasets.}
    \label{fig:result_synth}\vspace{-10pt}
\end{figure*}

\section{Experimental Results with Synthetic Dataset}
\label{exp}

\begin{figure*}[tb]
	  \centering
	  \subfigure[Ground truth]{\includegraphics[width=0.2\linewidth]{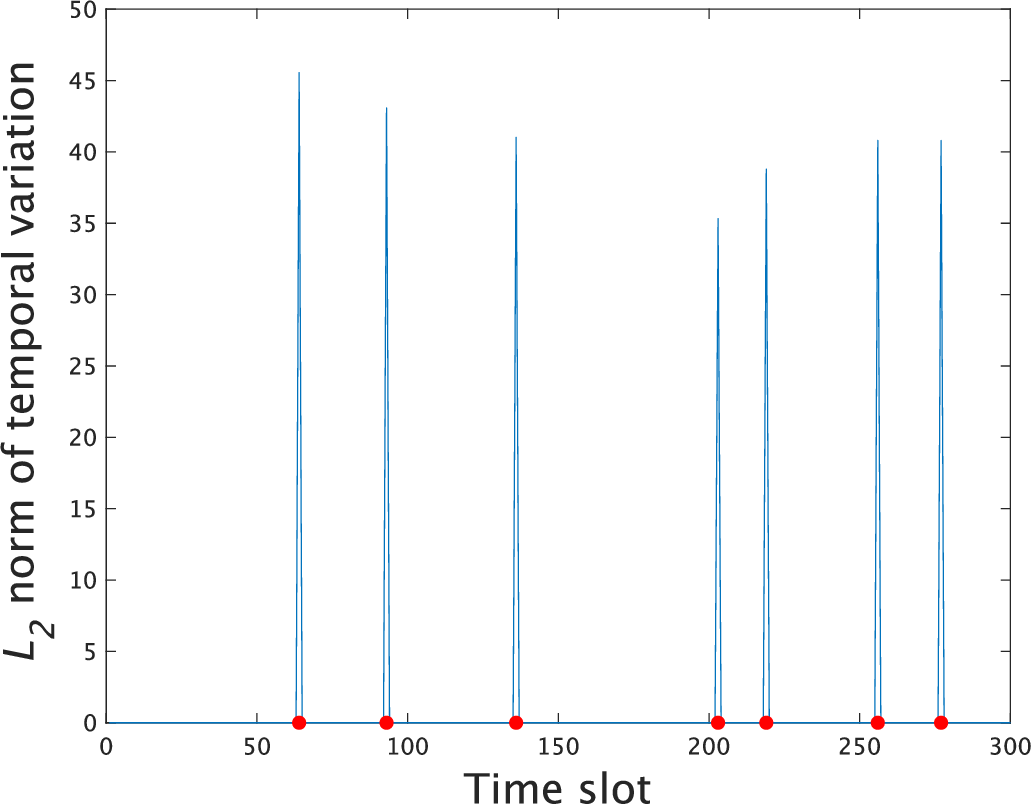}}\
      \subfigure[SGL-GLasso\cite{egilmez2017}]{\includegraphics[width=0.2\linewidth]{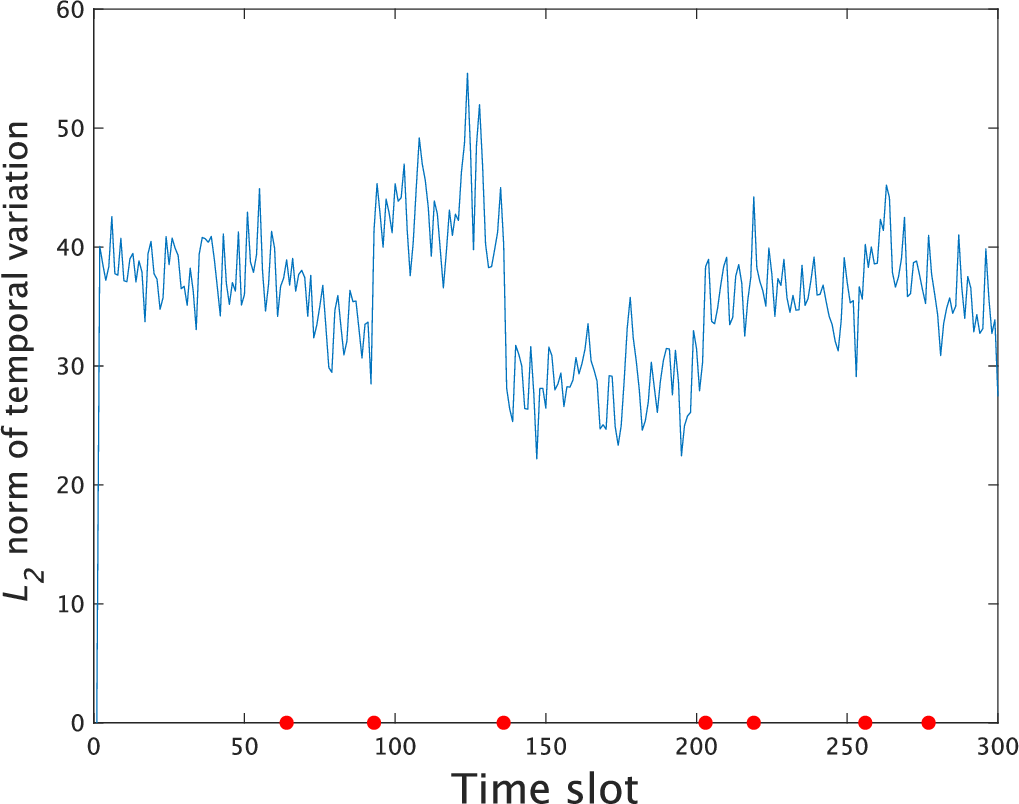}}\
      \subfigure[SGL-Smooth\cite{kalofolias2016}]{\includegraphics[width=0.2\linewidth]{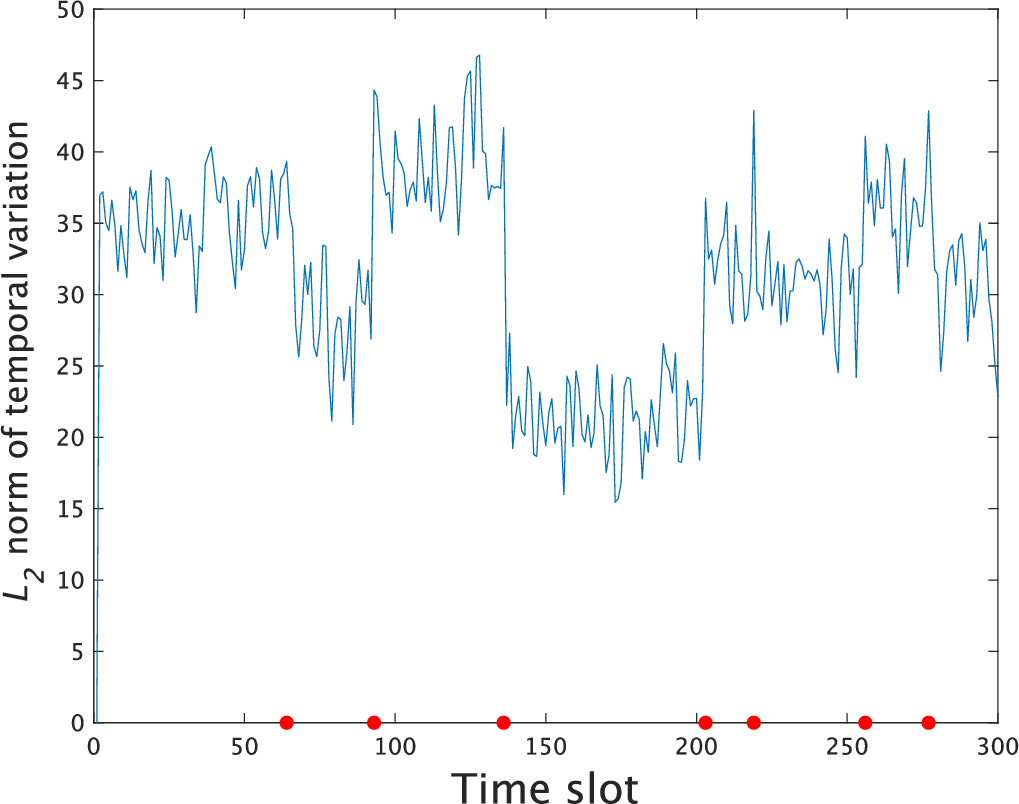}}\
      \subfigure[TGL-Tikhonov]{\includegraphics[width=0.2\linewidth]{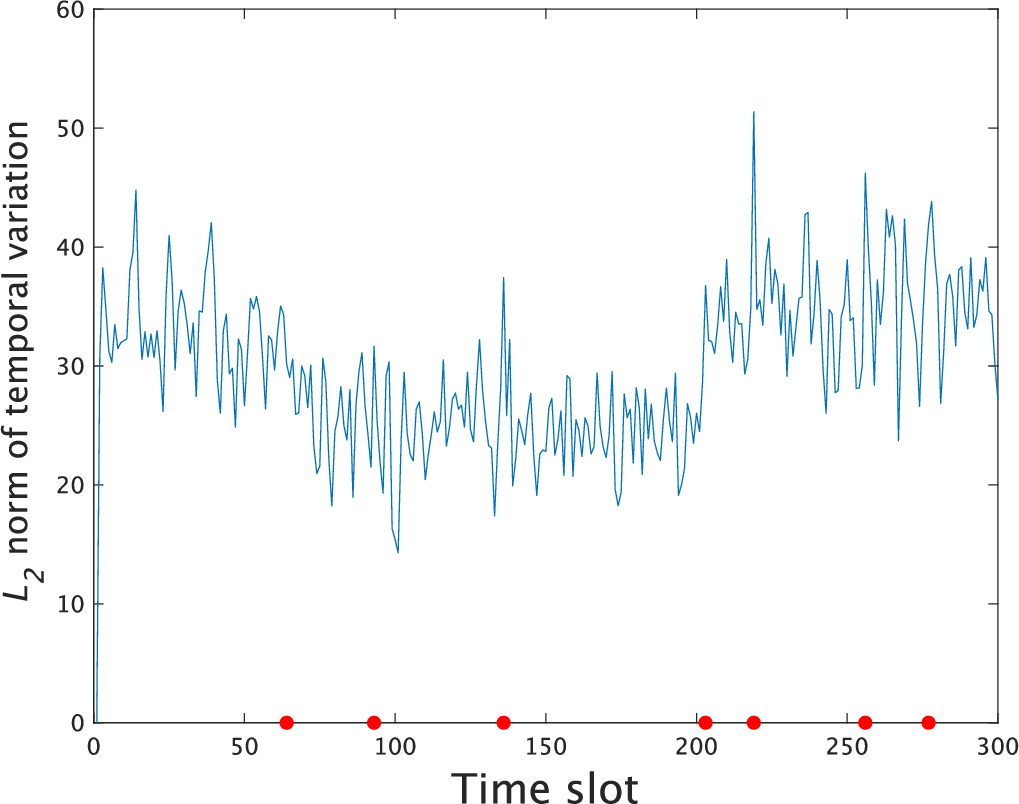}}\\
      \vspace{-3mm}
      \subfigure[TGL-TH (proposed)]{\includegraphics[width=0.2\linewidth]{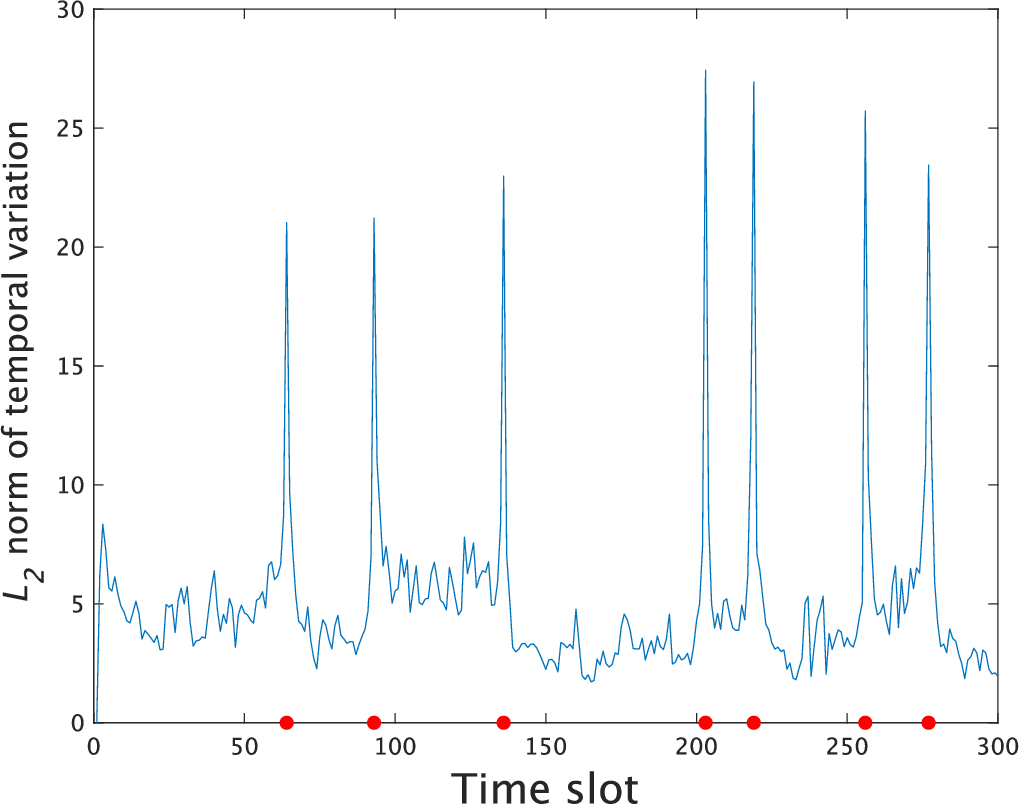}}\       
      \subfigure[TGL-SB (proposed)]{\includegraphics[width=0.2\linewidth]{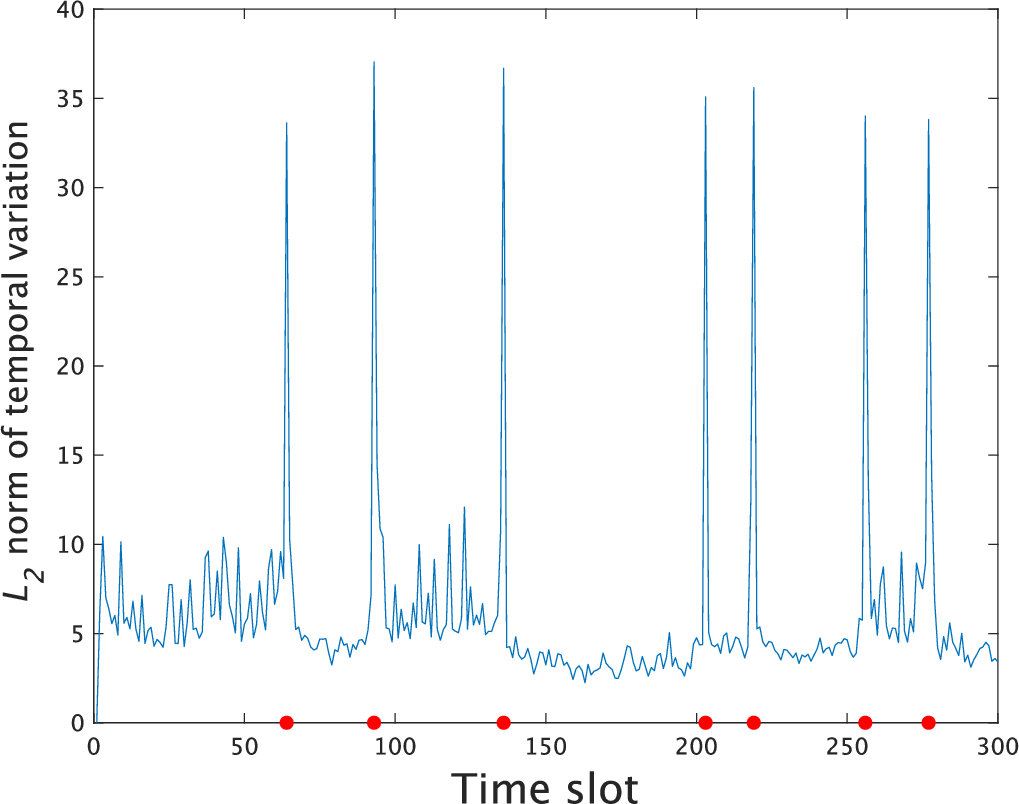}}\
      \subfigure[TGL-LTC (proposed)]{\includegraphics[width=0.2\linewidth]{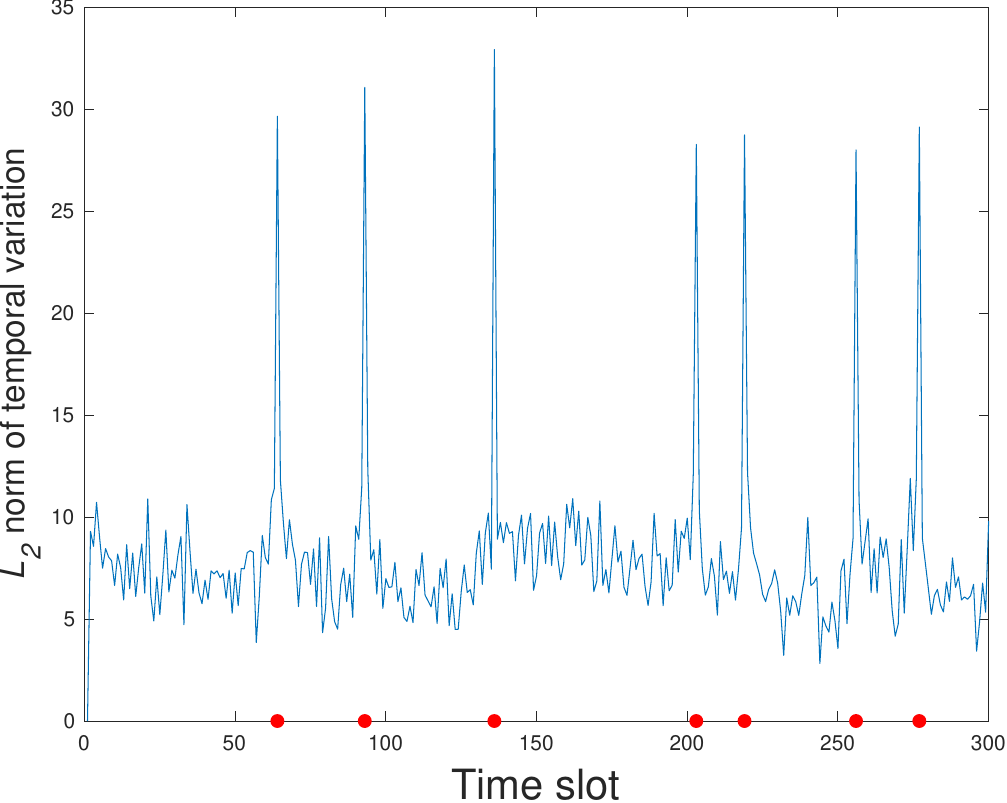}}\ 
  \caption{The visualization of the temporal variations in the TV graph learned from the dataset based on the graph in which large fluctuations occur at a few time slots. Red points in these figures represent time slots where the connectivity state changes.}
  \label{fig:l2norm}\vspace{-10pt}
\end{figure*}

The proposed TGL methods are performed on synthetic data in this section.
First, the details of the datasets and performance metrics are shown, and then the accuracies of graph learning methods are compared. All experiments, including the runtime comparison, were conducted on MATLAB2025b, Apple M2 chip with 24GB RAM.

\subsection{Synthetic Datasets}
%The dataset construction consists of two steps: 1) constructing a TV graph, and 2) generating TV graph signals from probability distributions based on graph Laplacians of the created TV graph.
We construct datasets in two steps: 1) building TV graphs, and 2) generating signals from a probability distribution based on the graph Laplacians of the created TV graph.
Four types of synthetic graphs are considered: 1) a TV graph based on the random waypoint model (abbreviated as TV-RW graph) \cite{johnson1996}, 2) a temporal homogeneity-based Erd\H{o}s--R\'enyi graph (TH-ER graph), 3) a
switching behavior-based ER graph 
% , where topologies change suddenly at a few time slots 
(SB-ER graph), and 4) an LTC-based ER graph 
% , where a small portion of nodes change connectivity at each time slot 
(LTC-ER graph).
Each TV-graph is created with $N=36$ nodes for 300 time slots.
The details of the four TV graphs are as follows:
% by referring to the desired properties \textbf{TH}, \textbf{SB}, and \textbf{LTC} described in \autoref{sec:intro}.
% \subsubsection{TV Graph Construction}
\begin{enumerate}
    \item TV-RW: TV graphs with smooth edge variation. A 3-nearest-neighbor graph was constructed based on the simulated movement of 36 sensors in a square space, with positions sampled over time.
    \item TH-ER: TV graphs with sparse edge variation. An initial ER graph was generated with edge connection probability $p=0.05$, and at each subsequent time step, $5\%$ of the edges were resampled. 
    \item SB-ER: TV graphs with infrequent, sudden changes. Six different static ER graphs were created with $p=0.50$. Starting from a random initial state at $t=1$, the graph at each time slot had a $5\%$ probability of switching to a different state. 
    \item LTC-ER: Localized, frequent connectivity changes. An initial ER graph was created with $p=0.30$. At each time step, $5\%$ of the nodes had their connectivity recomputed, while the rest of the graph remained stationary.
\end{enumerate}
The edge weights for TV-RW are determined based on the Euclidean distance $d_{ij}$ between nodes $i$ and $j$ as $W_{ij} = \exp(- d_{ij}/2\theta)$ where $\theta$ is a parameter. The edge weights for TH-ER, SB-ER, and LTC-ER are set to be uniformly distributed in the interval $[0, 1]$.
Using the graph Laplacian $\mathbf{L}_t$ at time $t$, a data sample ${\bf x}_{t}$ is generated from a Gaussian Markov random field model 
\begin{equation}
	p({\bf x}_{t}|{\bf L}^{(t)}) = \mathcal{N}({\bf x}_{t} | 0, {\bf L}_{t}^{\dagger}+\sigma ^2{\bf I}),
	\label{eq:gmf}
\end{equation}
where $\sigma^2$ is the covariance of the Gaussian noise. 
We set $\sigma=0.5$ in this experiment. Samples generated with \eqref{eq:gmf} have close signal values when two nodes are connected.
% Pairs of variables generated from this distribution have closer values to each other when the corresponding nodes connect with a larger edge weight.

\subsection{Performance Comparison}

\subsubsection{Experimental Conditions}
We evaluate the performance in terms of relative error and F-measure, each averaged over all the time slots.
The relative error is given by
\begin{equation}
	\text{Relative error} = \frac{\| \widehat{{\bf W}} - {\bf W}^{\ast} \|_{F} }{\| {\bf W}^{\ast} \|_{F} },
	\label{eq:re}
\end{equation}
where $\widehat{{\bf W}}$ is the estimated weighted adjacency matrix, and ${\bf W^{\ast}}$ is the ground truth.
It reflects the accuracy of edge weights on the estimated graph.
The F-measure is given by
\begin{equation}
	\text{F-measure} = \frac{2\text{tp}}{2\text{tp} + \text{fn} + \text{fp} },
	\label{eq:fmeasure}
\end{equation}
where the true positive (tp) is the number of edges that are included both in $\widehat{{\bf W}}$ and ${\bf W^{\ast}}$, the false positive (fp) is the number of edges that are not included in $\widehat{{\bf W}}$ but are included in ${\bf W^{\ast}}$, and the false negative (fn) is the number of edges that are included in $\widehat{{\bf W}}$ but are not included in ${\bf W^{\ast}}$.
The F-measure, the harmonic average of the precision and recall, represents the accuracy of the estimated graph topology and takes values between 0 and 1. The higher the F-measure is, the higher the performance of capturing the graph topology is.

% In this experiment, our goal is to compare all the methods based on their best achievable performance. 
We perform Monte-Carlo simulations for each method to find the parameters minimizing the relative error. Note that, as shown in the next section, fixed parameters still work well for our graph learning with real-world data.
For SGL-Smooth, TGL-Tik, and the proposed methods, $\alpha$ is selected by fine-tuning, and $\beta$ is fixed to $\beta=0.05$.
The parameter $\eta$ for all TGL methods is selected from 0.1 to 2 in steps of 0.1 to achieve the best relative error.
The parameter for SGL-GLasso is selected by the method described in \cite{egilmez2017}.
The tolerance value $\epsilon$ in \autoref{alg1} is set to $1.0 \times 10^{-3}$.
We evaluate the performance of each method using different data sample sizes, $K = \{1, 5, 10, 25, 50, 100\}$, and measure the average relative error and F-measure for each sample size across 10 independent trials.
Note that SGL-GLasso is not evaluated with $K=1$ because the sample covariance used in SGL-GLasso cannot be calculated.

\begin{table*}[t]
  \centering
  \caption{Empirical Running Time Comparison.}
  \label{tab:runtime}
  \setlength{\tabcolsep}{5pt}
  \begin{tabular}{l cc cc cc}
    \toprule
    \multicolumn{7}{c}{(a) Increasing graph size with $T=50$} \\
    \midrule
    & \multicolumn{2}{c}{$N=40$} & \multicolumn{2}{c}{$N=80$} & \multicolumn{2}{c}{$N=160$} \\
    \cmidrule(lr){2-3}\cmidrule(lr){4-5}\cmidrule(lr){6-7}
    Method & Time (s) & Iter. & Time (s) & Iter. & Time (s) & Iter. \\
    \midrule
    SGL-Smooth & \textbf{$0.295\pm0.019$} & \textbf{$374\pm5$}
               & \textbf{$1.579\pm0.101$} & \textbf{$483\pm21$}
               & \textbf{$8.092\pm0.880$} & \textbf{$606\pm61$} \\
    TGL-Tik    & $0.651\pm0.058$ & $555\pm25$
               & $3.139\pm0.323$ & $632\pm61$
               & $14.105\pm2.290$ & $702\pm109$ \\
    TGL-TH     & $1.037\pm0.160$ & $794\pm108$
               & $6.072\pm0.813$ & $1095\pm148$
               & $32.532\pm2.821$ & $1440\pm110$ \\
    TGL-SB     & $0.650\pm0.085$ & $507\pm60$
               & $3.384\pm0.785$ & $612\pm131$
               & $17.381\pm4.103$ & $776\pm191$ \\
    TGL-LTC    & $0.760\pm0.070$ & $464\pm43$
               & $3.874\pm0.571$ & $548\pm80$
               & $19.622\pm2.936$ & $690\pm108$ \\
    \midrule
    \multicolumn{7}{c}{(b) Increasing sequence length with $N=80$} \\
    \midrule
    & \multicolumn{2}{c}{$T=25$} & \multicolumn{2}{c}{$T=50$} & \multicolumn{2}{c}{$T=100$} \\
    \cmidrule(lr){2-3}\cmidrule(lr){4-5}\cmidrule(lr){6-7}
    Method & Time (s) & Iter. & Time (s) & Iter. & Time (s) & Iter. \\
    \midrule
    SGL-Smooth & \textbf{$0.814\pm0.080$} & \textbf{$489\pm12$}
               & \textbf{$1.579\pm0.101$} & \textbf{$483\pm21$}
               & \textbf{$3.127\pm0.195$} & \textbf{$475\pm29$} \\
    TGL-Tik    & $1.626\pm0.195$ & $645\pm48$
               & $3.139\pm0.323$ & $632\pm61$
               & $6.211\pm0.750$ & $621\pm75$ \\
    TGL-TH     & $2.647\pm0.228$ & $965\pm13$
               & $6.072\pm0.813$ & $1095\pm148$
               & $15.839\pm6.010$ & $1436\pm546$ \\
    TGL-SB     & $1.447\pm0.192$ & $528\pm32$
               & $3.384\pm0.785$ & $612\pm131$
               & $8.073\pm3.328$ & $740\pm304$ \\
    TGL-LTC    & $1.753\pm0.301$ & $493\pm18$
               & $3.874\pm0.571$ & $548\pm80$
               & $9.260\pm3.096$ & $661\pm223$ \\
    \bottomrule
  \end{tabular}

  \vspace{2pt}
  \parbox{0.96\textwidth}{\footnotesize The experiments used connected weighted Erd\H{o}s--R\'enyi graph sequences with an expected node degree of eight and five temporal evolution patterns: stationary graphs, smoothly varying edge weights, sparse edge changes, global graph switches, and node-local changes.}
\end{table*}

\begin{table*}[tb]
\centering
\caption{Relative Error (RE) and F-measure (FM) with Different Memory Length}
\label{tab:memory_comp}
\resizebox{\textwidth}{!}{%
\begin{tabular}{lcccccccccc}
\toprule
\textbf{Method} & \multicolumn{2}{c}{\textbf{SGL-smooth}} & \multicolumn{2}{c}{\textbf{TGL-Tik}} & \multicolumn{2}{c}{\textbf{TGL-LTC}} & \multicolumn{2}{c}{\textbf{TGL-TH}} & \multicolumn{2}{c}{\textbf{TGL-SB}}\\
\midrule
Metric & RE & FM & RE & FM & RE & FM & RE & FM & RE & FM \\\midrule
$L=0$ & 0.898 & 0.459 &   -   &   -   &   -   &   -   &   -   &   -   &   -   &   -   \\
$L=1$  &   -   &   -   & $0.712\pm0.005$ & $0.517\pm0.006$ & $0.539\pm0.006$ & $0.533\pm0.005$ & $0.492\pm0.005$ & $0.592\pm0.004$& $0.765\pm0.006$ & $0.493\pm0.006$ \\
$L=2$  &   -   &   -   & $0.674\pm0.005$ &$0.528\pm0.007$ & $0.412\pm0.005$ & $0.550\pm0.007$ & $0.417\pm0.006$ & $0.634\pm0.005$ & $0.683\pm0.007$ & $0.513\pm0.006$ \\
$L=3$  &   -   &   -   & $\bf0.661\pm0.005$ & $\bf0.529\pm0.007$ & $\bf0.382\pm0.006$ & $\bf0.557\pm0.007$ & $\bf0.411\pm0.007$ & $\bf0.653\pm0.006$ & $\bf0.610\pm0.007$ & $\bf0.527\pm0.006$ \\
\bottomrule
\end{tabular}%
}
\end{table*}

\subsubsection{Results}
As shown in \autoref{fig:result_synth}, our proposed time-varying (TV) graph learning methods outperform static approaches and the TGL-Tik baseline (Figs.~\ref{fig:fm_RW}, \ref{fig:re_RW}). Each method demonstrated strength in specific scenarios: TGL-TH's $\ell_1$ regularization worked well for graphs with smooth/sparse variations like TH-ER and LTC-ER (Figs.~\ref{fig:fm_RW}, \ref{fig:fm_NG}); TGL-SB was most effective for the sudden, infrequent changes of the SB-ER graph, especially with more observations ($K\geq50$) (Figs.~\ref{fig:fm_SB}, \ref{fig:re_SB}); and TGL-LTC showed robust, flexible performance across many models, especially when changes were localized as in the LTC-ER graph (Figs.~\ref{fig:re_RW}, \ref{fig:re_ER}, \ref{fig:re_NG}).
% As shown in \autoref{fig:result_synth}, TV graph learning methods generally outperform static methods, which estimate graphs independently at each time slot. For TV-RW graphs, which have smooth edge variations, all of the proposed TGL methods outperform TGL-Tik (Figs.~\ref{fig:fm_RW}, \ref{fig:re_RW}). 

% TGL-TH significantly outperforms the other methods in terms of F-measure on TH-ER and LTC-ER graphs since the regularization in \eqref{eq:tgl_W_flasso} is well-suited for the graphs with partially constant structure (Figs.~\ref{fig:fm_ER} and \ref{fig:fm_NG}). This advantage holds even for a small number of signals $K$.

% On SB-ER graphs, characterized by sudden but infrequent changes, TGL-SB demonstrates superior performance, particularly as the number of observations $K$ increases (Figs.~\ref{fig:fm_SB} and \ref{fig:re_SB}). Specifically, for $K\ge50$, it consistently yields the lowest relative error among all methods.

% TGL-LTC marks the best or the second-best performance in multiple graph models (Figs.~\ref{fig:re_RW}, \ref{fig:re_ER}, and \ref{fig:re_NG}). Focusing on Fig.~\ref{fig:re_NG}, TGL-LTC outperforms all other methods, especially in a small observation setting. This suggests that the regularization in \eqref{eq:pmodel3} flexibly captures temporal variations when changes are localized on a small number of nodes.

In addition, \autoref{fig:l2norm} shows the $L_{2}$-norm of the temporal difference of learned adjacency matrices in each time slot, i.e., $ \| {\bf W}_{t} - {\bf W}_{t-1} \|_{2}$. Clearly, TGL-TH, TGL-SB, and TGL-LTC can detect sudden changes in the graph in time slots, while SGL-GLasso and SGL-Smooth have several unclear jumps. On the other hand, TGL-Tik can suppress the temporal variation of graphs better than SGL-Smooth, but it obscures the change points in the TV graph. As a result, the proposed TGL methods are suitable for change detection. 

\subsection{Run-time Comparison}
Table~\ref{tab:runtime} reports the empirical runtime required to reach a common stopping tolerance of $10^{-4}$. Upper table (a) shows that the runtime increases with the number of nodes for all methods. SGL-Smooth is consistently the fastest because it does not include temporal regularization. Among the time-varying methods, TGL-Tik and TGL-SB have comparable runtimes for $N=40$, whereas TGL-Tik is the fastest for the two larger graph sizes. Although TGL-LTC requires fewer iterations than TGL-Tik for $N=40$ and $N=80$, its structured proximal operations increase the cost per iteration. Lower table (b) shows the effect of the sequence length. For SGL-Smooth and TGL-Tik, doubling $T$ approximately doubles the runtime while the iteration count remains nearly constant. The runtime of TGL-TH, TGL-SB, and TGL-LTC grows faster because their iteration counts also increase with $T$. Nevertheless, the measured fixed-iteration scaling exponents at $N=80$ range from $1.02$ to $1.06$, confirming that the per-iteration cost is approximately linear in $T$. All methods converged in every trial.

\subsection{Effect of Memory Size \texorpdfstring{$L$}{L}}
\label{subsec:memory_comp}
To verify the effect of memory size $L$, mentioned in Section \ref{susec:Lstep}, we perform TV graph learning for the TV-RW graph with perturbation on edge weights.
That is, data samples $\mathbf{x}_t$ is generated by \eqref{eq:gmf} but from the following \textit{noisy} edge weights:
\begin{equation}
\label{eq:ew_noise}
    \tilde{W}_{ij}=\begin{cases} W_{ij} + \epsilon & \text{if}\quad W_{ij} + \epsilon > 0\\
    0 & \text{otherwise}
    \end{cases}
\end{equation}
where $\epsilon\sim\mathcal{N}(0,0.1^2)$. 
We set $K=5$ and compare the average relative error and F-measure of $50$ independent runs.
We set $L=\{1,2,3\}$.
Note that $L=1$ is the baseline that only considers the relationship to the immediate previous adjacency matrix $\mathbf{W}_{t-1}$.
The temporal variation operator $\mathbf{\Phi}$ for the proposed algorithm with $L\in\{2,3\}$ is defined in \eqref{eq:phi2}. 
For comparison, we extend the TGL-Tik \cite{kalofolias2017} by generalizing its temporal regularizer $\sum_{t=2}^T\|\mathbf{W}_t-\mathbf{W}_{t-1}\|_F^2$, shown in  \autoref{tab:mlist}, for any memory length $L$ by: 
\begin{equation}
    \sum_{l=1}^{L} \sum_{t=l+1}^{T} \|\mathbf{W}_t - \mathbf{W}_{t-l}\|_F^2.    
\end{equation}
We apply this formulation for $L\in\{2,3\}$ in our experiments.

The results presented in \autoref{tab:memory_comp} show a consistent performance increase for all TGL algorithms as $L$ grows.
This implies that the robustness against the noise induced to edge weights in \eqref{eq:ew_noise} is improved by increasing the memory size $L$, as long as the edge weights vary smoothly over time.

We note that the ``Static'' row in \autoref{tab:memory_comp} refers to the SGL-Smooth applied \emph{independently to each window} ($K=5$ samples), which is data-starved. As $L$ grows, the proposed estimator approaches the pooled static graph, so in this slow-variation regime, monotone improvement is expected. A turning point may appear if the temporal variation among the underlying ground-truth graphs exceeds the variation due to perturbations (e.g., noise), such as the LTC-ER graphs.

\section{Experiments with Real-World Data}
\label{application}
\subsection{Dynamic Point Clouds}

\begin{table}[tb]
	\centering
	\caption{Denoising Results: SNR (dB)}
	\label{tab:denoise}
    \resizebox{0.5\textwidth}{!}{%
	\begin{tabular}{c c c c c} \toprule
	\bf{Method}	& {\it dog} & {\it handstand}  & {\it skirt}  & {\it wheel}  \\ \midrule
		noisy       & $10.31\pm0.03$ & $12.58\pm0.02$ & $12.62\pm0.02$ & $14.60\pm0.02$ \\
		$k$-NN      & $11.90\pm0.03$ & $14.30\pm0.01$ & $14.16\pm0.02$ & $16.18\pm0.02$ \\
		SGL-Smooth  & $11.74\pm0.04$ & $14.20\pm0.01$ & $14.21\pm0.02$ & $16.12\pm0.02$ \\
		TGL-Tik     & $12.08\pm0.03$ & $15.09\pm0.02$ & $15.05\pm0.02$ & $17.03\pm0.02 $ \\
		TGL-TH      & $\bf15.26\pm0.06$ & $20.36\pm0.03$ & $\bf21.34\pm0.03$ & $22.60\pm0.03$ \\
        TGL-SB      & $11.45\pm0.03$ & $15.93\pm0.03$ & $15.26\pm0.03$ & $17.60\pm0.02$ \\
        TGL-LTC     & $14.41\pm0.06$ & $\bf20.39\pm0.03$ & $21.34\pm0.03$ & $\bf22.62\pm0.03$\\
        \bottomrule	
	\end{tabular}%
    }
\end{table}

\begin{figure}[tp]
  \centering
    \subfigure[Ground truth]{\includegraphics[width=0.23\linewidth]{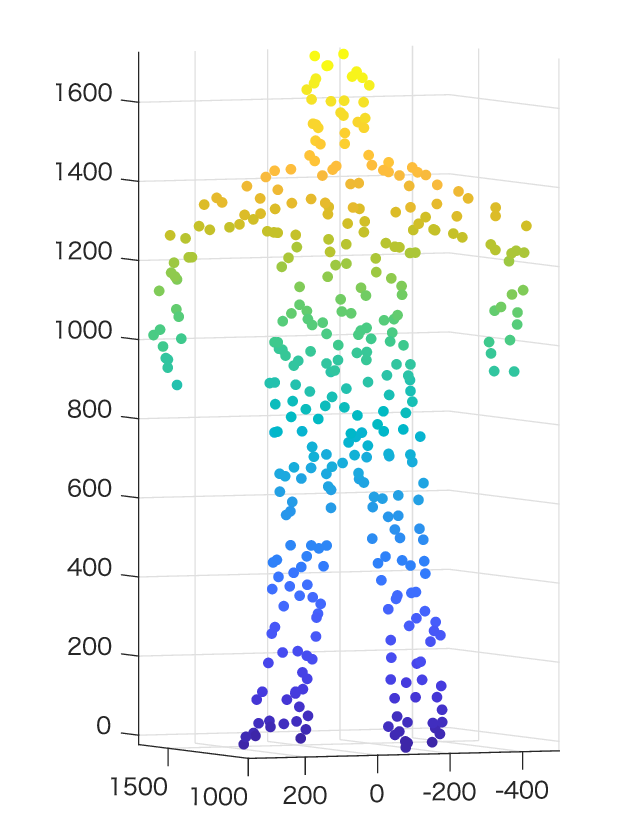}}\
    \subfigure[Noisy]{\includegraphics[width=0.23\linewidth]{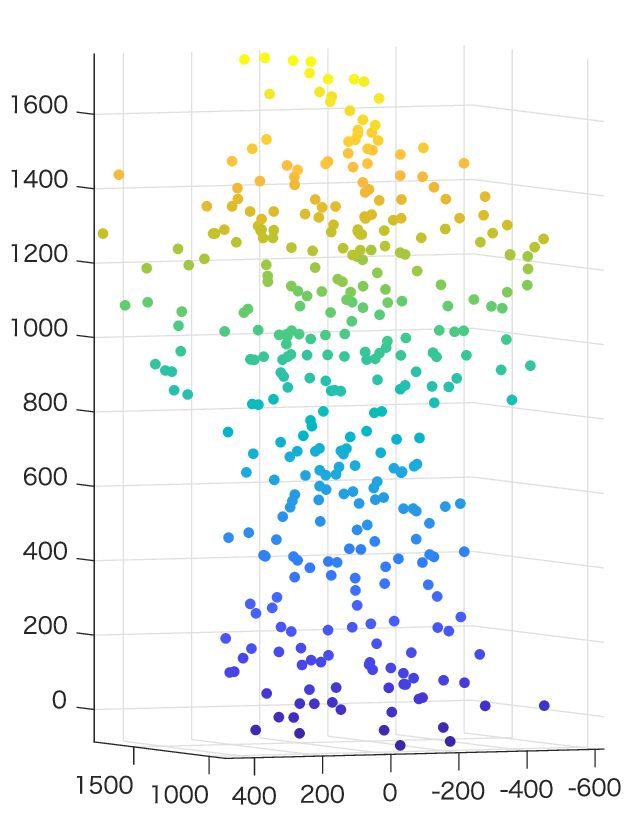}}\
    \subfigure[$k$-NN]{\includegraphics[width=0.23\linewidth]{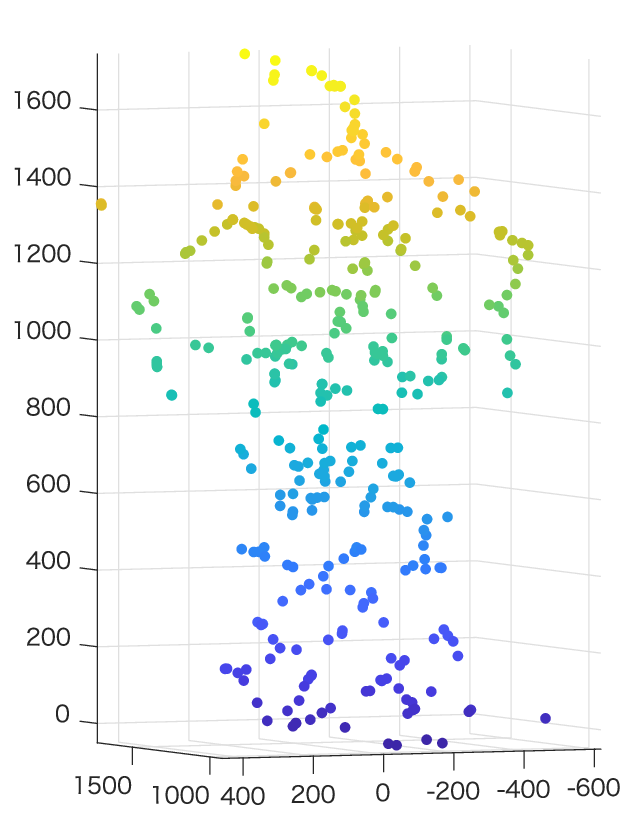}}\\
    \subfigure[SGL-Smooth]{\includegraphics[width=0.23\linewidth]{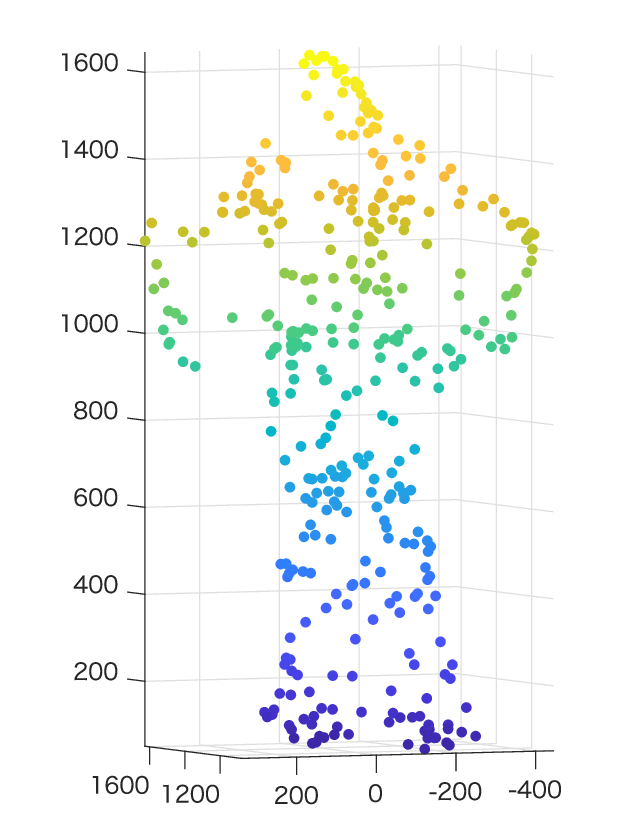}}\
    \subfigure[TGL-Tik]{\includegraphics[width=0.23\linewidth]{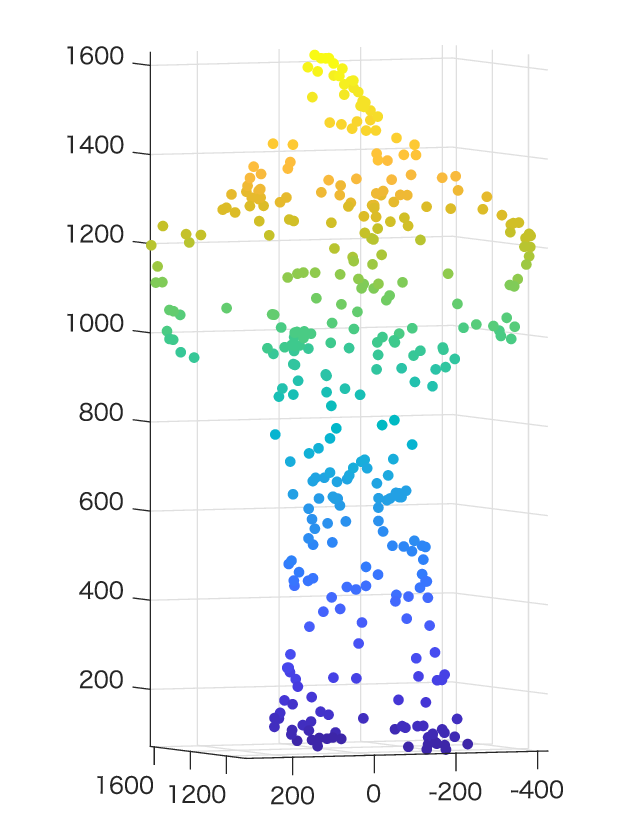}}\
    \subfigure[TGL-TH]{\includegraphics[width=0.23\linewidth]{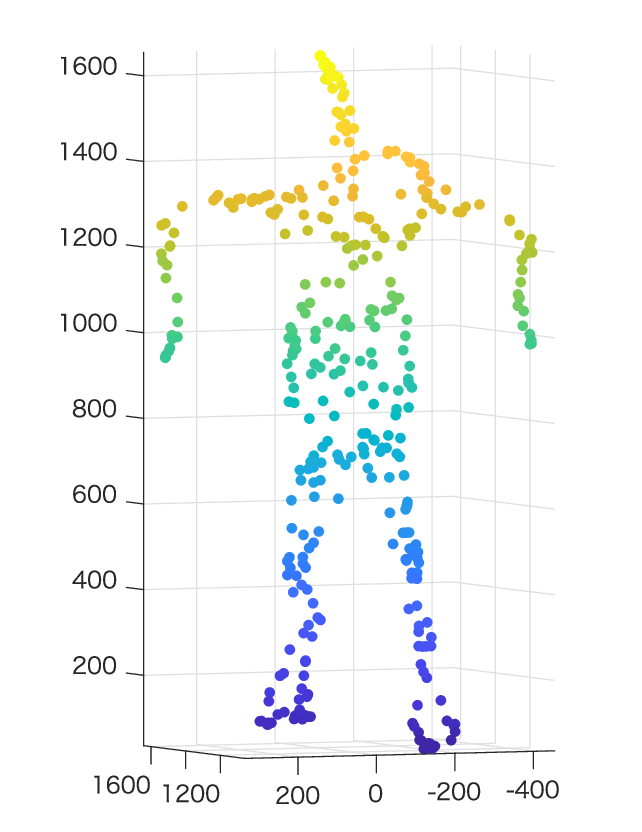}}\\
  \caption{ The visualization of denoising result of {\it wheel} at a $t=10$ time.}\vspace{-5pt}
  \label{fig:pc_denoise}\vspace{-10pt}
\end{figure}

\begin{figure}[tb]
  \centering
      \subfigure[$k$-NN]{\includegraphics[width=0.24\linewidth]{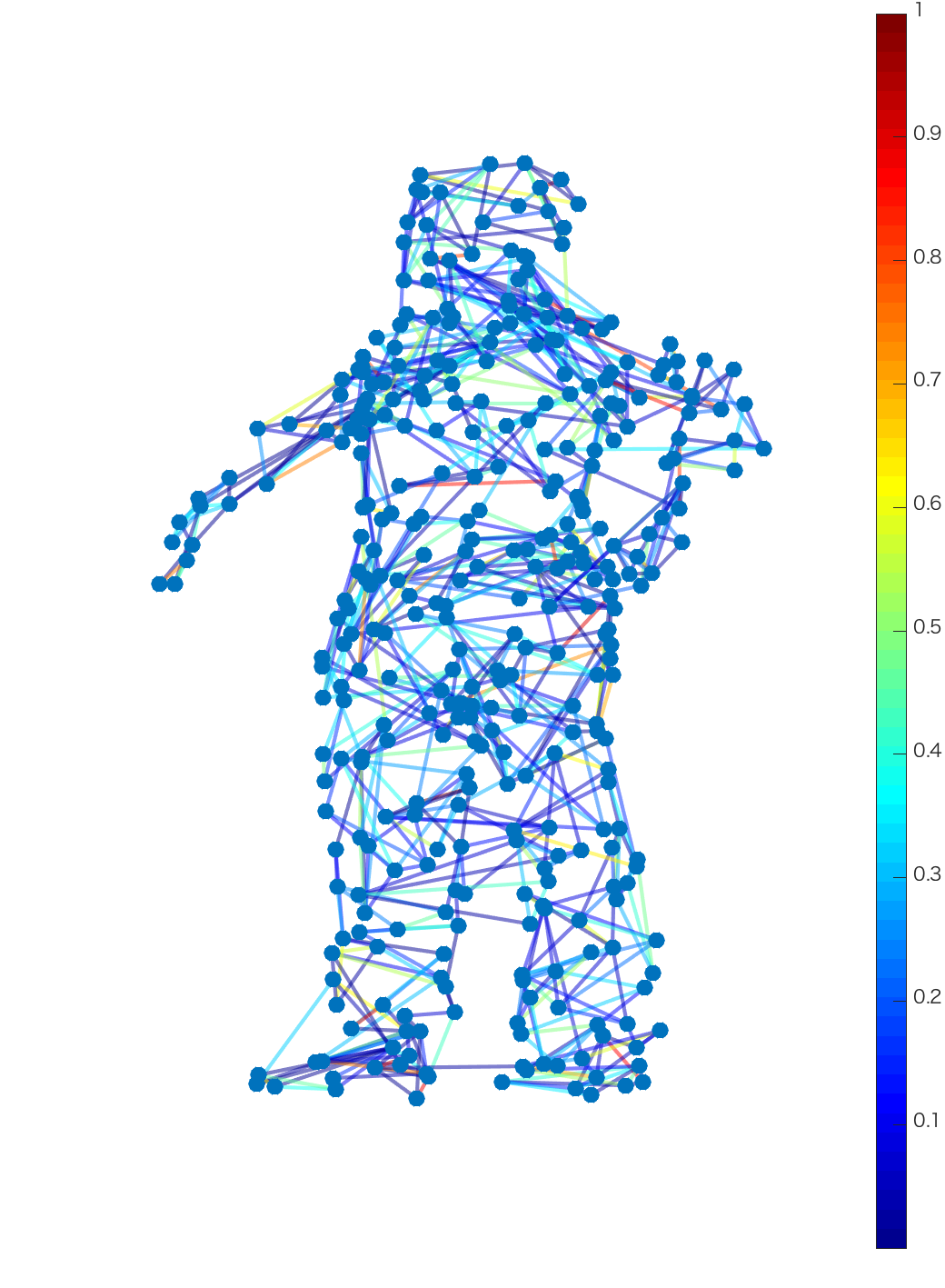}}\
      \subfigure[SGL-Smooth]{\includegraphics[width=0.24\linewidth]{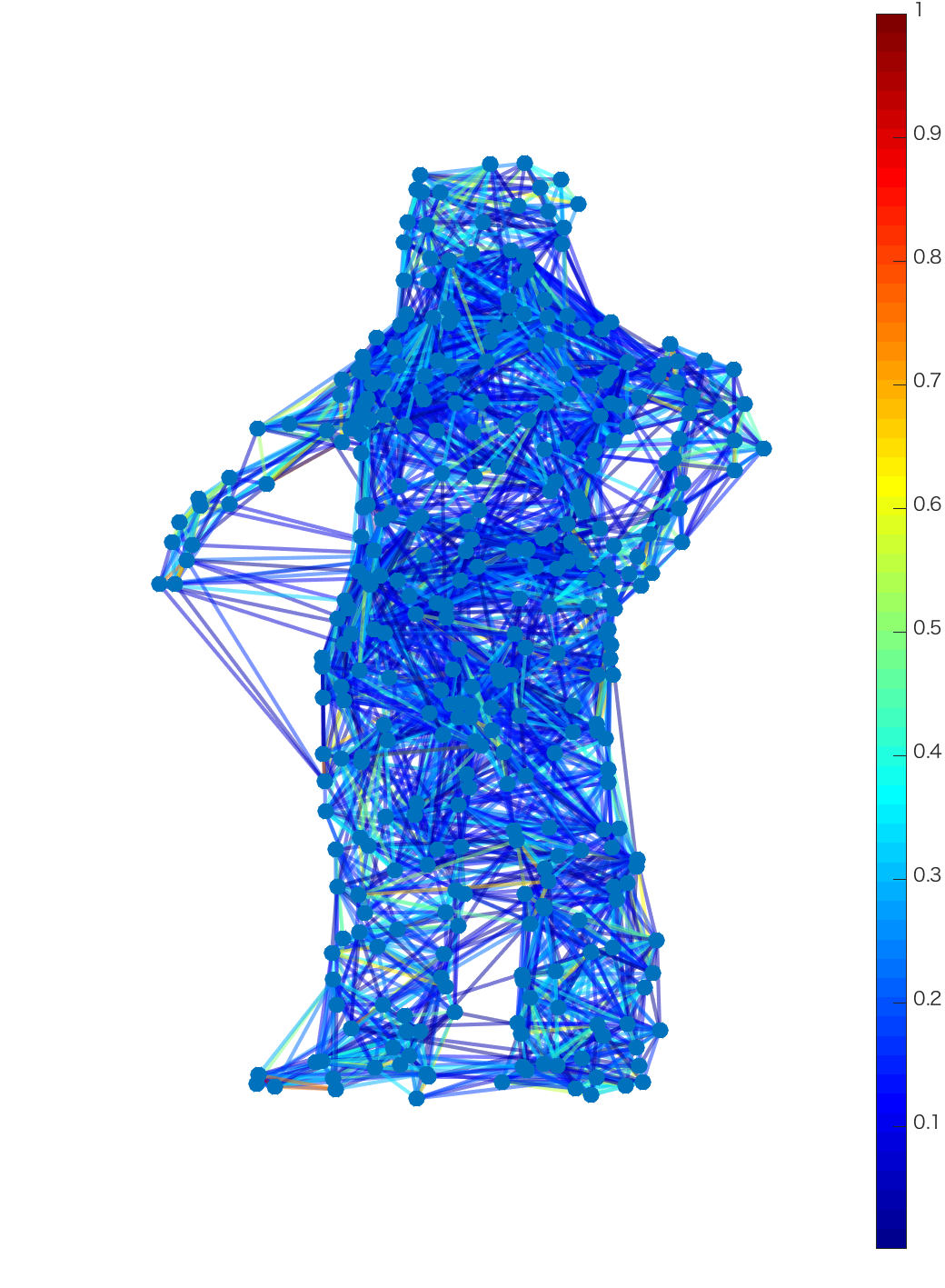}}\
      \subfigure[TGL-Tik]{\includegraphics[width=0.24\linewidth]{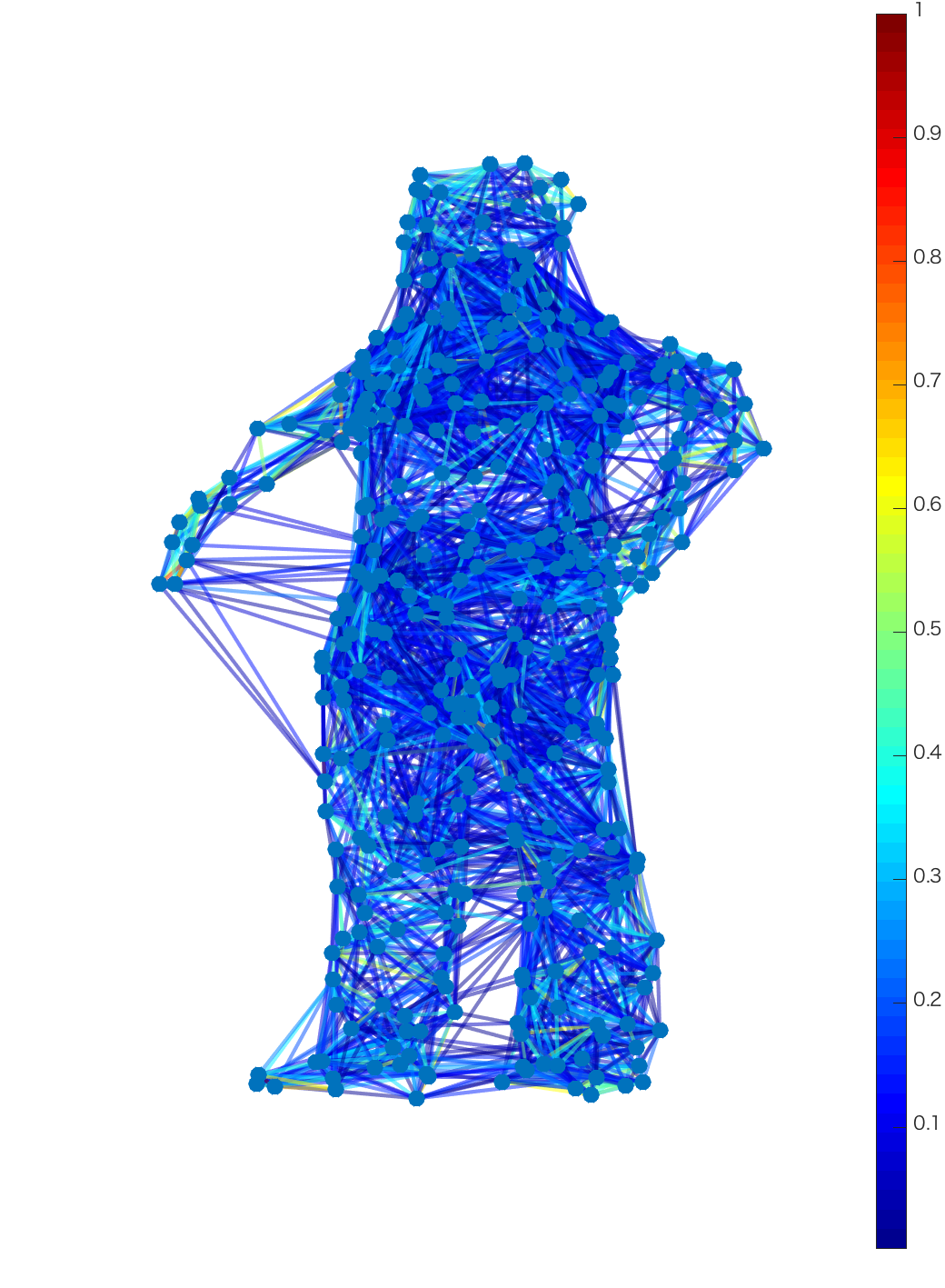}}\
      \subfigure[TGL-TH]{\includegraphics[width=0.24\linewidth]{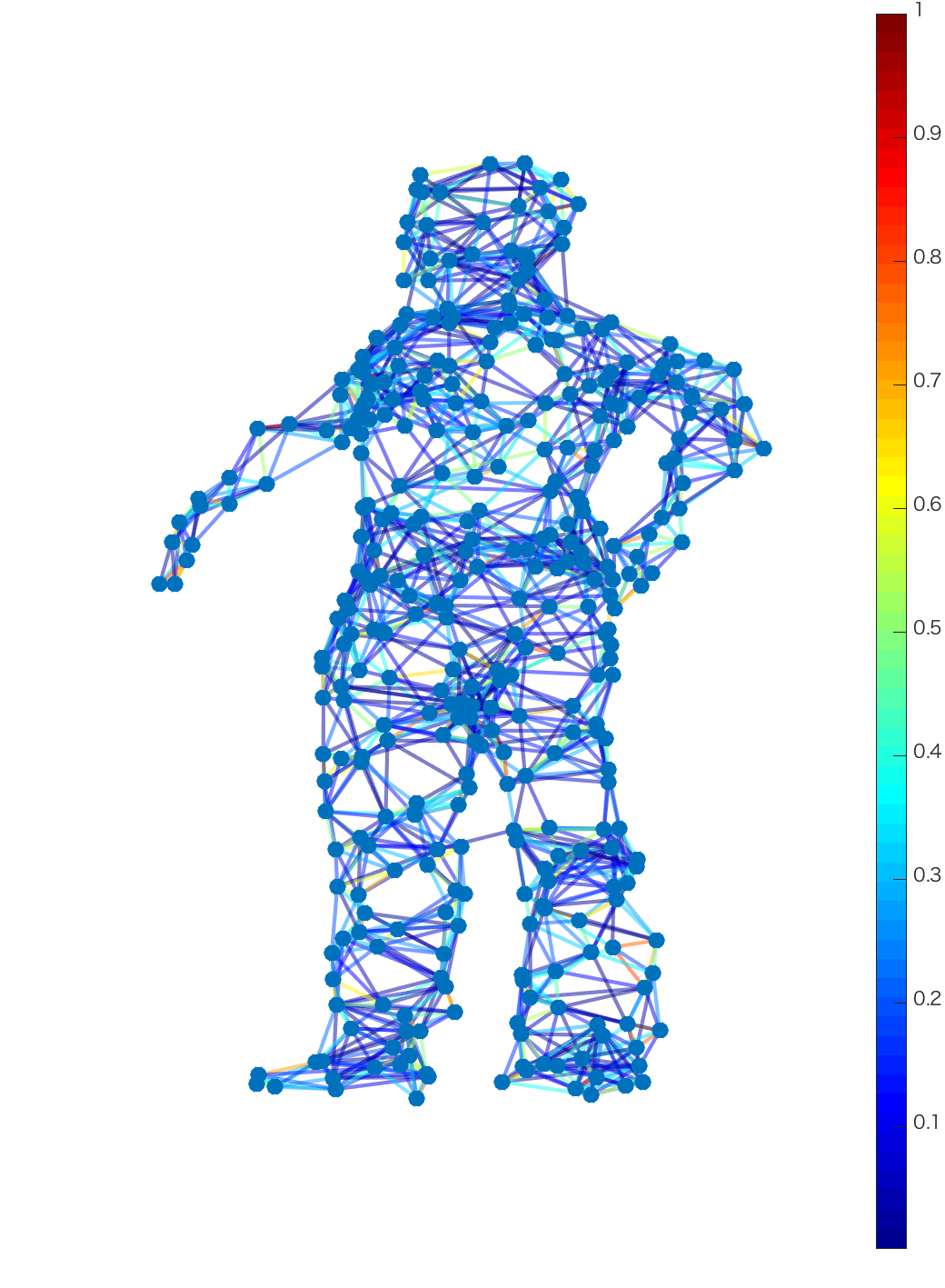}}\
      \caption{The visualization of a graph at a certain time in the TV graph learned from the noisy point cloud data.}
      \label{fig:pc_graph}\vspace{-10pt}
\end{figure}

We apply our method for dynamic point cloud (DPC) data denoising.
DPC data contain 3D coordinates of dynamically evolving points. 
Graphs for point cloud data are typically constructed using static methods such as kNN graphs.
However, when point cloud data are acquired, it is often contaminated with noise due to measurement errors, which leads to displacements of the geometric coordinates of the point clouds.
Therefore, considering the temporal difference of graphs is expected to enhance noise removal performance.
% In this experiment, denoising is performed using 
We use graph heat kernel filtering \cite{zhang2008} for denoising.
% Here, we use a graph-based denoising approach to focus on comparing the connectivity learned using different methods.
The TV graph used in the denoising is estimated from noisy point cloud data.

We use the DPC dataset in \cite{gall2009}, which contains five DPCs: {\it dance, dog, handstand, skirt}, and {\it wheel}.
As this dataset is clean data, with positions ranging from $-260$ to $1932$, we added Gaussian noise with $\sigma = 90$, a very high noise level.
The TV graph is learned from the dataset downsampled to 357 points and evolving over 240 time slots. The dataset for \textit{dog} only contained 59 frames, so we tested on the available frames $T=59$.
In this experiment, we use fixed parameters for each method, which are determined by a grid search with {\it dance} data, and evaluate the denoising performance with the other four data. 

\autoref{tab:denoise} summarizes our DPC denoising results.
We observe that the TV graph learning methods perform better than SGL-Smooth and $k$-nearest neighbor ($k$-NN), indicating that the $k$-NN fails to construct a meaningful graph from noisy data. %cannot construct a meaningful graph from noisy data.
Additionally, TGL-TH outperformed TGL-Tik by up to 6 dB. 

\autoref{fig:pc_denoise} shows the visualization of denoising results of the {\it wheel} at a certain time.
Similar to the numerical performance, the $k$-NN cannot capture the structure of the human body. 
SGL-Smooth and TGL-Tik yield slightly better outputs than those by the $k$-NN; however, the arms and legs are still problematic.
On the other hand, TGL-TH can better capture the structure of the human body than the other methods.

\autoref{fig:pc_graph} visualizes a graph at time slot $t=10$ in the TV graph estimated from the noisy {\it wheel} data in the dynamic point cloud dataset.
In this figure, the nodes in the graphs are plotted in the correct position for visualization.
As shown in the figure, the $k$-NN method yields a sparse graph, but the edge weights and the connectivity are directly affected by noise. In \autoref{fig:pc_graph}(a), we can observe heavily (lightly) weighted edges between nodes that are far apart (close by), especially between the left and the right legs.
SGL-Smooth and TGL-Tik yielded very dense edges and connected different parts of the body.
In contrast, TGL-TH yields graphs whose connectivity preserves the human body structure.

\subsection{Temperature Data}
\begin{figure}[tb]
	\centering
	\subfigure[]{\includegraphics[width=0.4\linewidth]{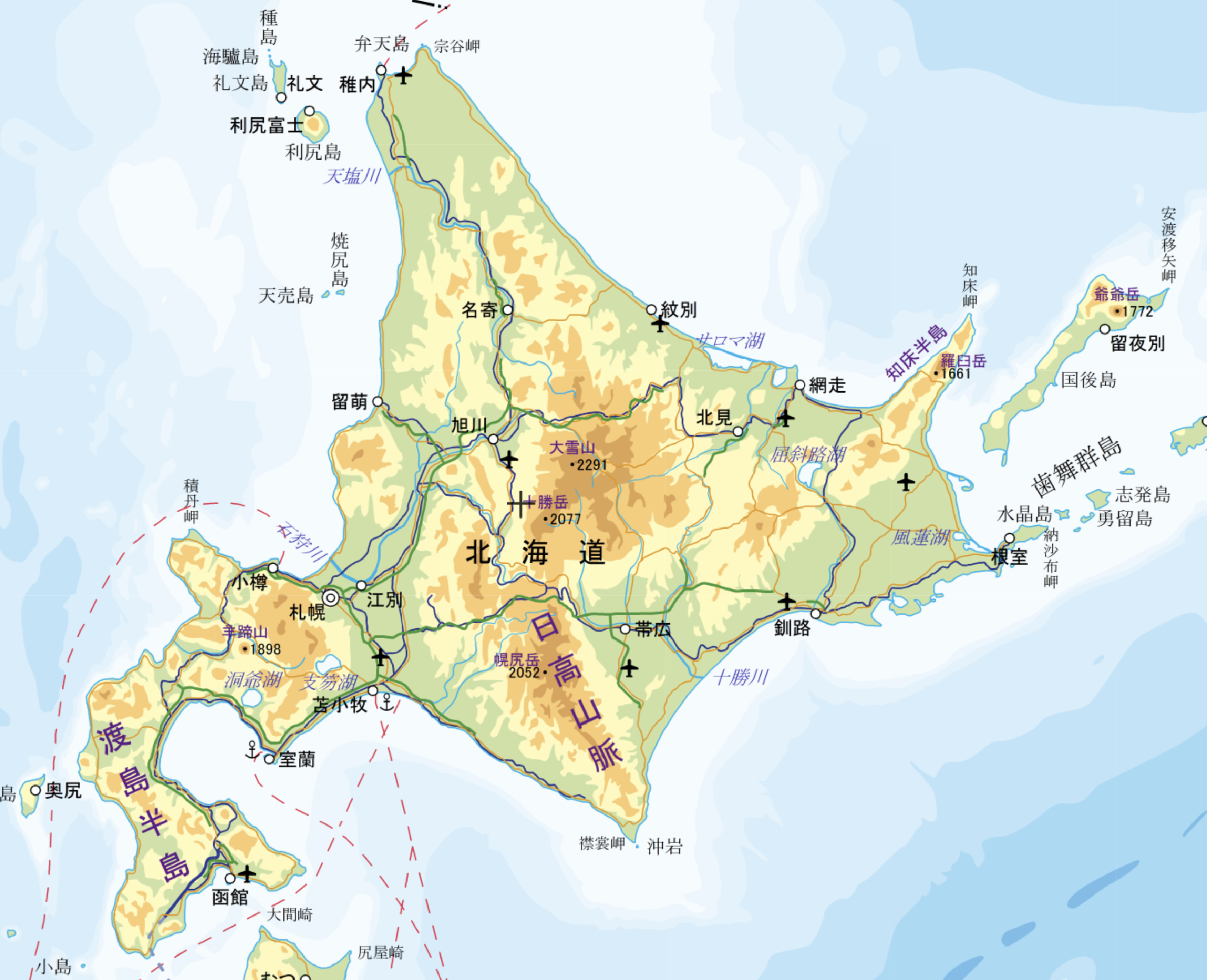} \label{fig:tmp_map}}\
	\subfigure[]{\includegraphics[width=0.4\linewidth]{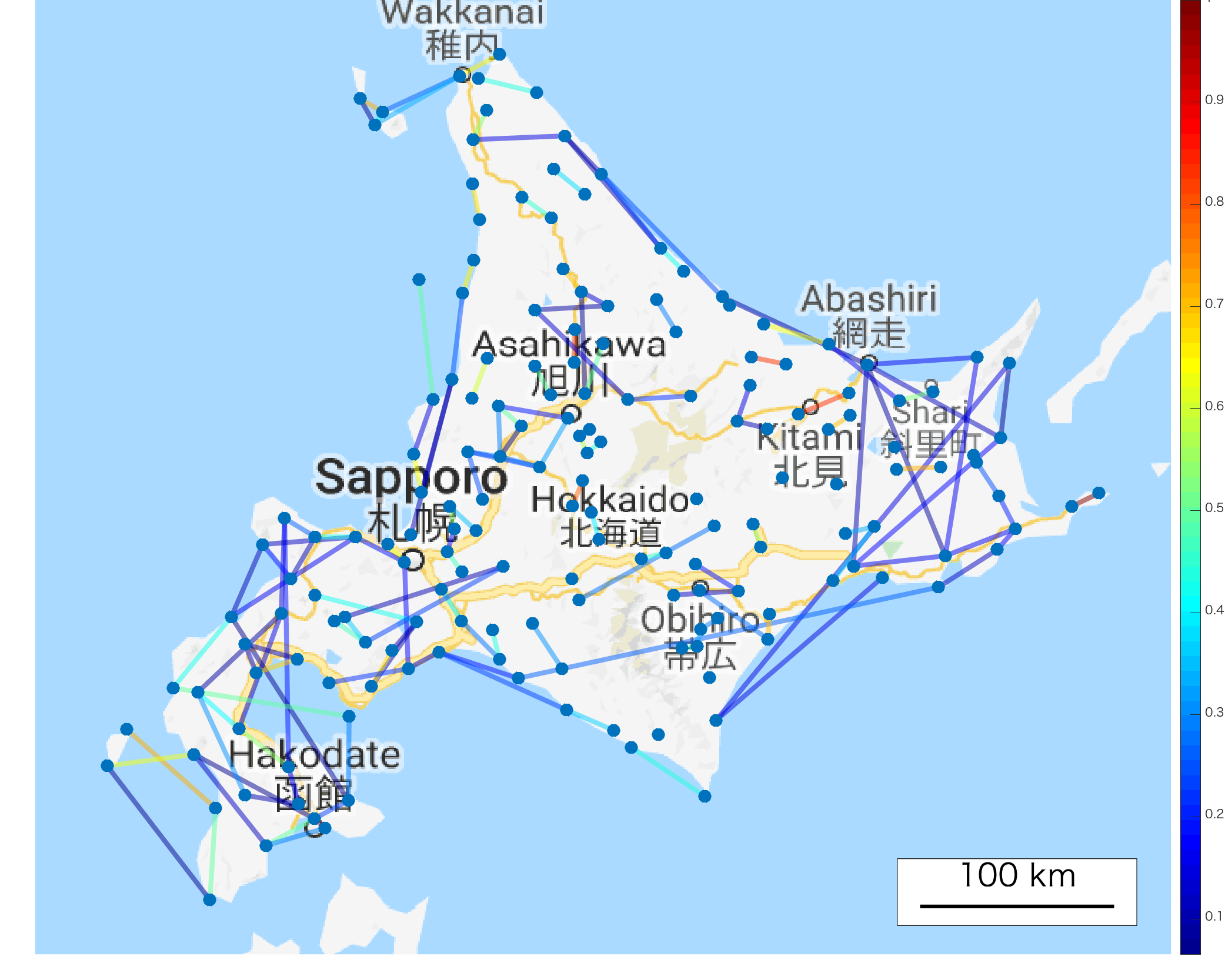} \label{fig:tmp_winter}}\\
	\subfigure[]{\includegraphics[width=0.4\linewidth]{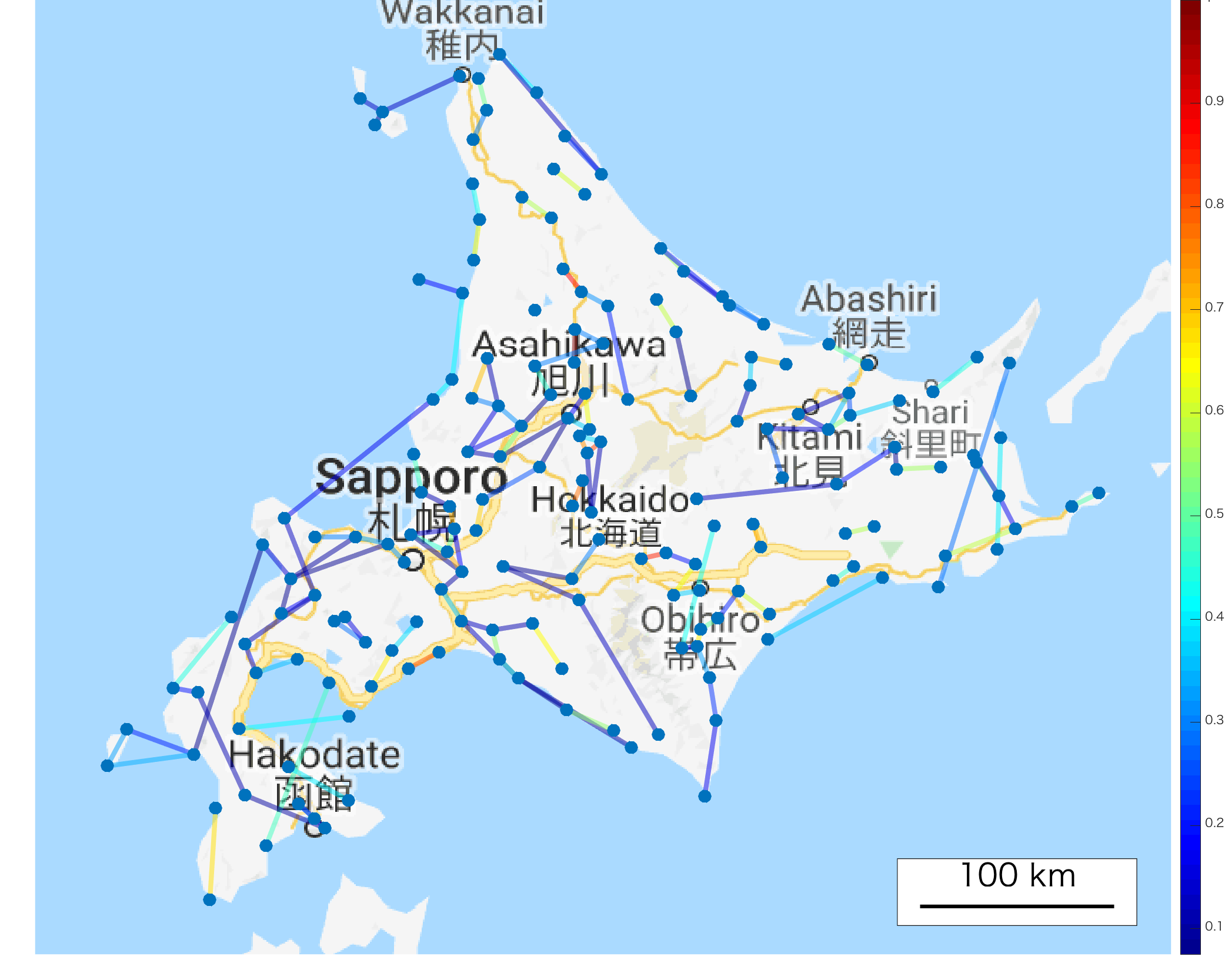} \label{fig:tmp_summer}}
	\subfigure[]{\includegraphics[width=0.4\linewidth]{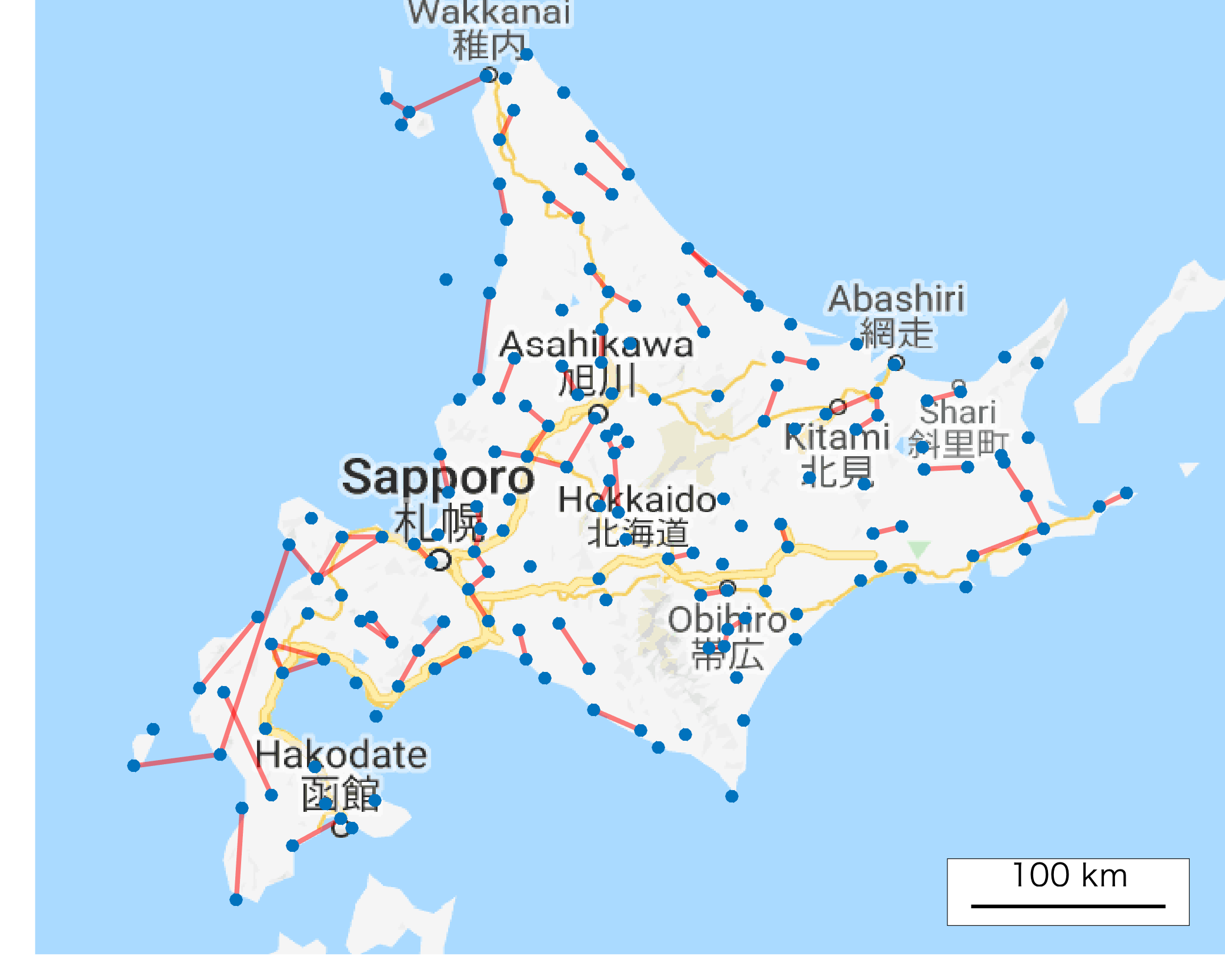} \label{fig:tmp_common}}\\
	\subfigure[]{\includegraphics[width=0.4\linewidth]{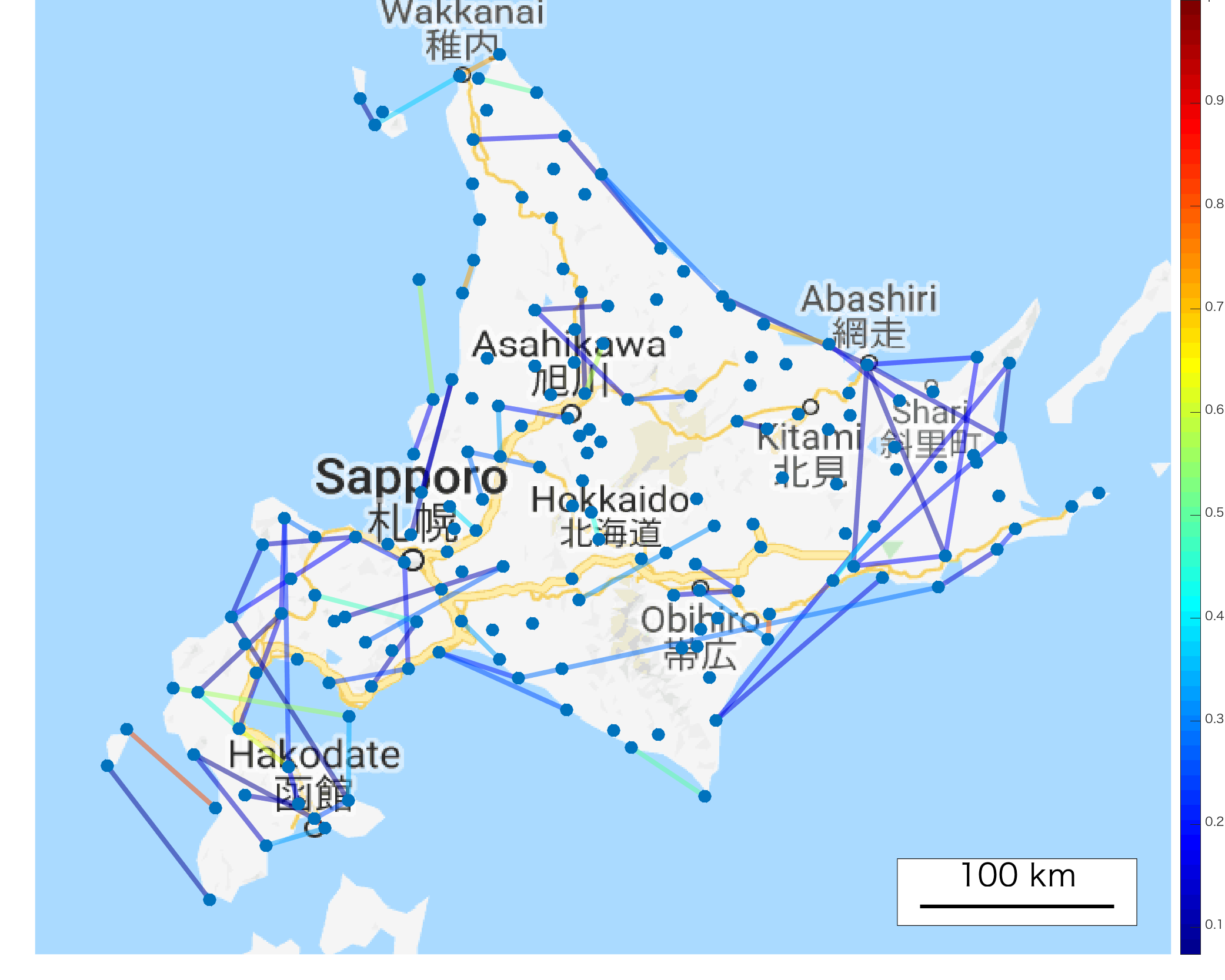} \label{fig:tmp_ws}}
	\subfigure[]{\includegraphics[width=0.4\linewidth]{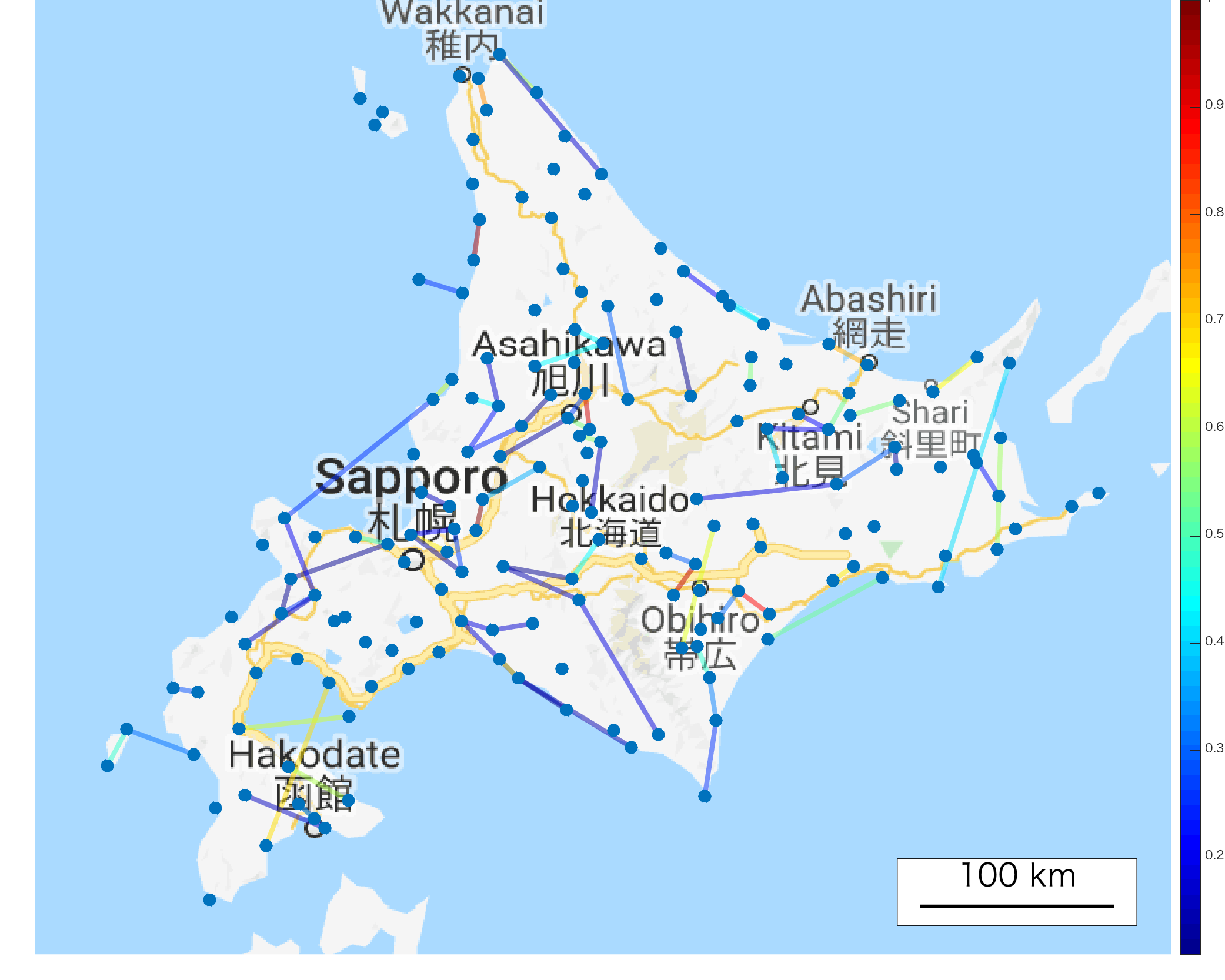} \label{fig:tmp_ss}}\
    \caption{Visualization of the graph learned from the temperature data. (a) Map of Hokkaido with altitude. (b) Graph learned for January 8, 2015 (winter graph). (c) Graph learned for August 9, 2015 (summer graph). (d) Edges that are common in winter and summer. (e) The winter graph without common edges (winter-specific graph). (f) The summer graph without common edges (summer-specific graph). }
    \label{fig:temp_graph}\vspace{-5pt}
\end{figure}

\begin{figure}
	\centering
    \subfigure[8 January 2015]{\includegraphics[width=0.48\linewidth]{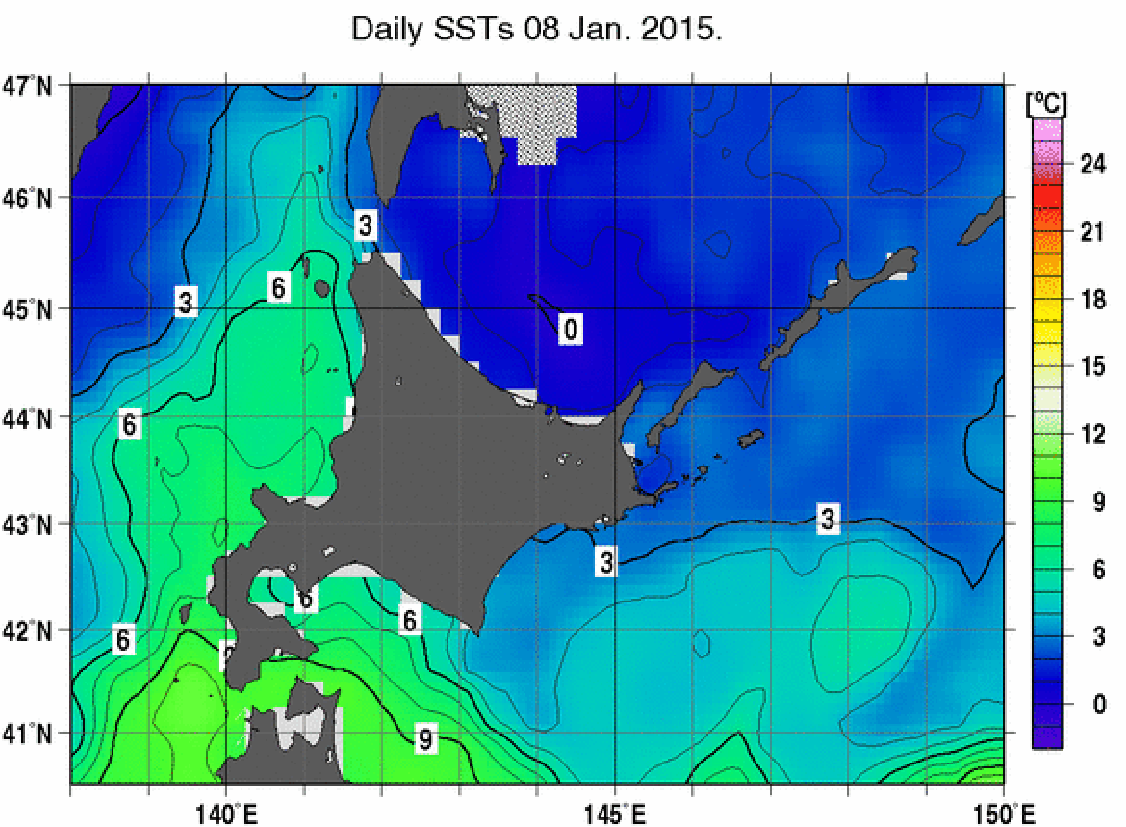}}\
	\subfigure[9 August 2015]{\includegraphics[width=0.48\linewidth]{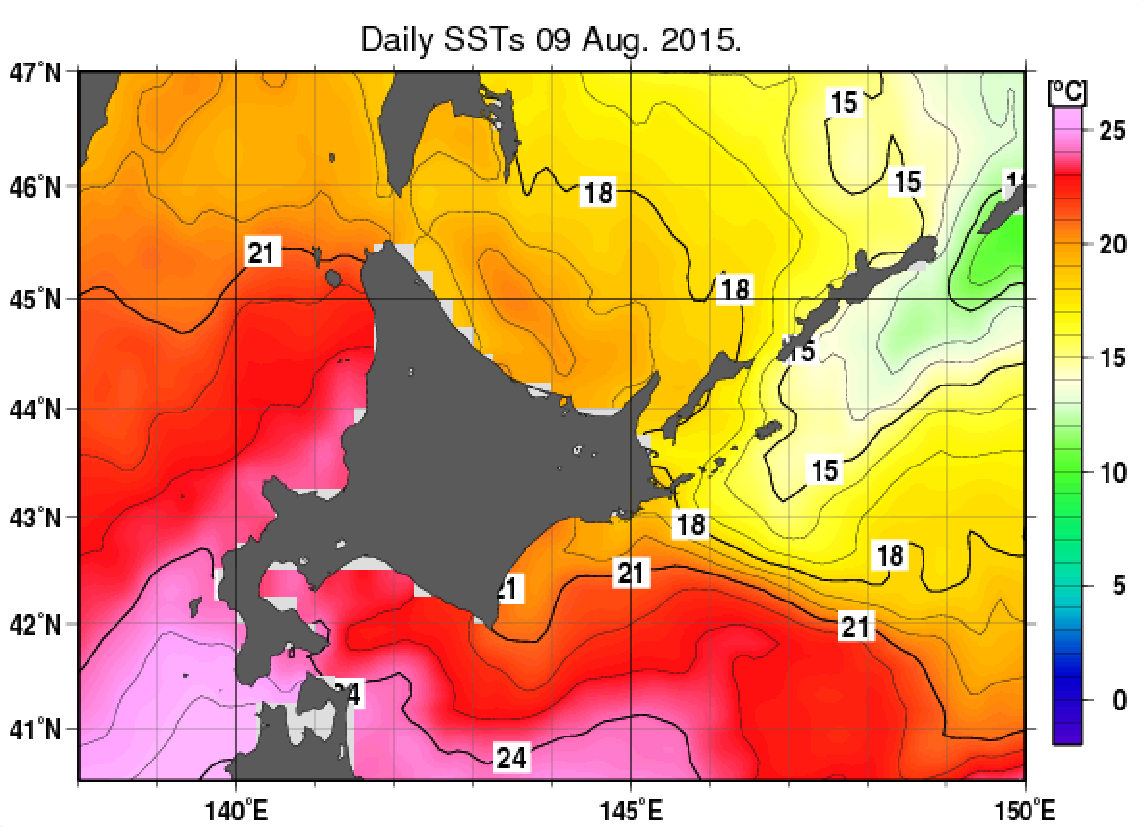}}\
	\caption{Daily sea surface temperature in winter (left) and summer (right).}
	\label{fig:tmp_sst}\vspace{-5pt}
\end{figure}
We also apply the proposed method to estimate a TV graph from temperature data on the island of Hokkaido, the northernmost region in Japan.
The goal of this experiment is to learn a TV graph to explore the relationship between geographical regions over time.
In this experiment, we use the average daily temperature data\footnote{The Japan Meteorological Agency provided the daily temperature data from their website at https://www.jma.go.jp/jma/index.html} collected from 172 locations in Hokkaido in 2015, i.e., the overall dataset has 365 time slots with $N = 172$ and $K=24$.
From this data, we estimate a TV graph by using TGL-TH.

\autoref{fig:temp_graph} shows the graph learned from temperature data\footnote{The Geospatial Information Authority of Japan provided the map in Hokkaido with altitude from their website at http://www.gsi.go.jp/}.
For visualization, we remove edges having very small weight $( < 1.0 \times 10^{-4})$.
The learned graph shows that geographically close nodes are generally connected due to similar temperatures, unless separated by features like mountain ranges. Nodes also form connections along the coast—even over long distances—and separately within inland areas. As shown in Fig.~\ref{fig:tmp_common}, these local structures remain stable over time.
% From the figure, we can infer the following for the learned graph. 
% First, nodes close to each other are basically connected. This is reasonable because the temperature is similar in the same geographical area. Note that if the recording points are separated by mountain ranges, nodes may not be connected even if the distances are short.
% Second, nodes along the coast are often connected, and the inland nodes are also connected. Coastal nodes, in particular, may connect despite being far away.
% Third, locally connected structures in the learned graph mostly remain unchanged over time as shown in Fig.~\ref{fig:tmp_common}.
 
Figs.~\ref{fig:tmp_ws} and \ref{fig:tmp_ss} show the season-specific graphs.
We have found that a coastal node often connects to another coastal node, even if they are not close to each other.
This can be justified by the warming or cooling effect of sea currents that occur seasonally in this area and affect coastal areas in similar ways. 
\autoref{fig:tmp_sst} represents daily sea surface temperatures (SST) on January 8 and August 9, 2015 \footnote{The daily SST was provided by Japan Meteorological Agency, from their website at https://www.jma.go.jp/jma/index.html}.
As can be seen in Figs.~\ref{fig:tmp_ws}, \ref{fig:tmp_ss}, and \ref{fig:tmp_sst}, nodes in similar SST areas are connected to each other in the learned graph. It is important to emphasize that there was no prior information about the SST areas used in the graph learning process. 

\subsection{EEG Data}

As the last example, we analyzed EEG recordings of epileptic seizures, where brain networks are known to undergo significant topological changes \cite{kramer2010, vandiessen2013, schindler2008}. We used the CHB-MIT Scalp EEG Dataset \cite{chbmit_database}, focusing on recordings from 125 seconds before seizure onset until its end. Using 5-second non-overlapping windows, we learned TV graphs with TGL-SB and TGL-LTC to track the network evolution during the pre-ictal, ictal, and post-ictal periods. We set the parameters so that the resulting graph is maximally sparse but connected.

\begin{figure}[tb]
  \centering
	\subfigure[TGL-SB]{\includegraphics[width=0.22\textwidth]{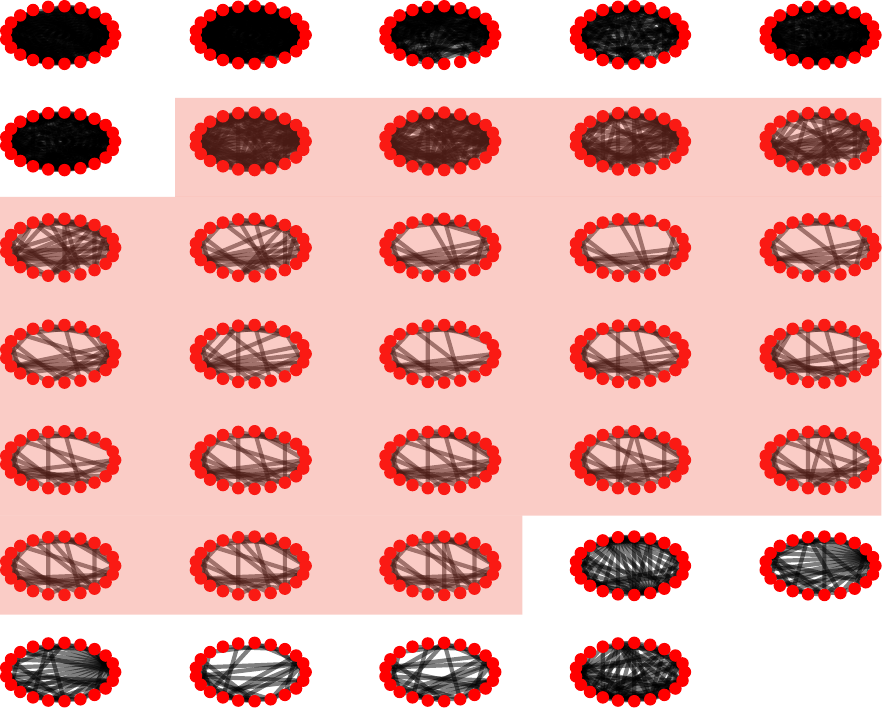}
	\label{fig:eeg_result_GroupL}}
	\subfigure[TGL-LTC]{\includegraphics[width=0.22\textwidth]{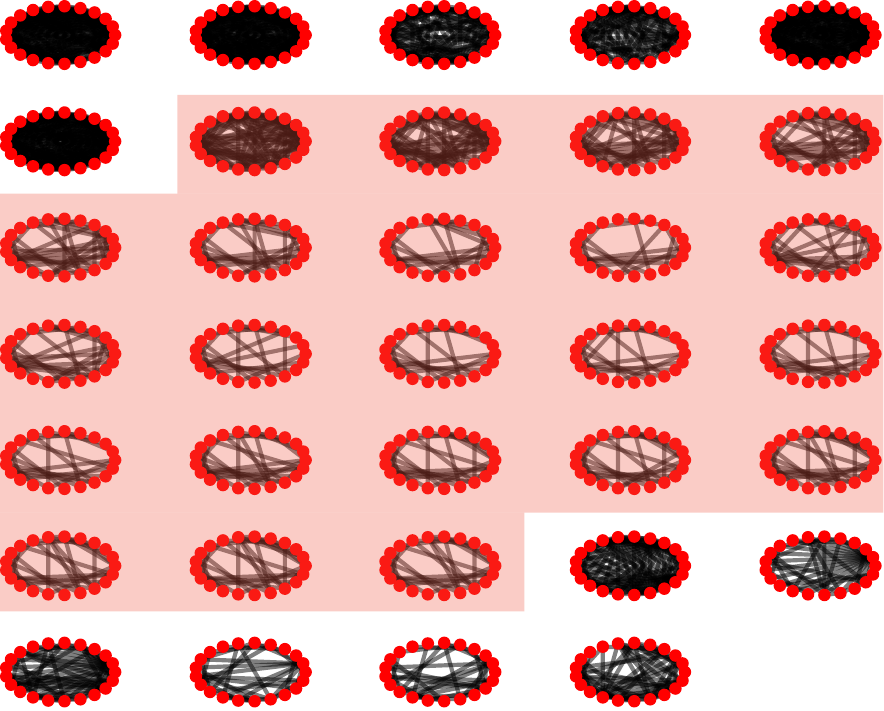} \label{fig:eeg_result_NG}}\
  \caption{Brain functional network estimated from bipolar EEG data. The graphs are placed chronologically from left to right, top to bottom. The red areas indicate the graphs during the seizure period. In (a) and (b),  we can observe that the network becomes sparse in the middle of the seizure and becomes dense as the seizure ends.}
  \label{fig:resulteeg}
\end{figure}

\begin{figure}[tb]
  \centering
  \includegraphics[width=0.32\textwidth]{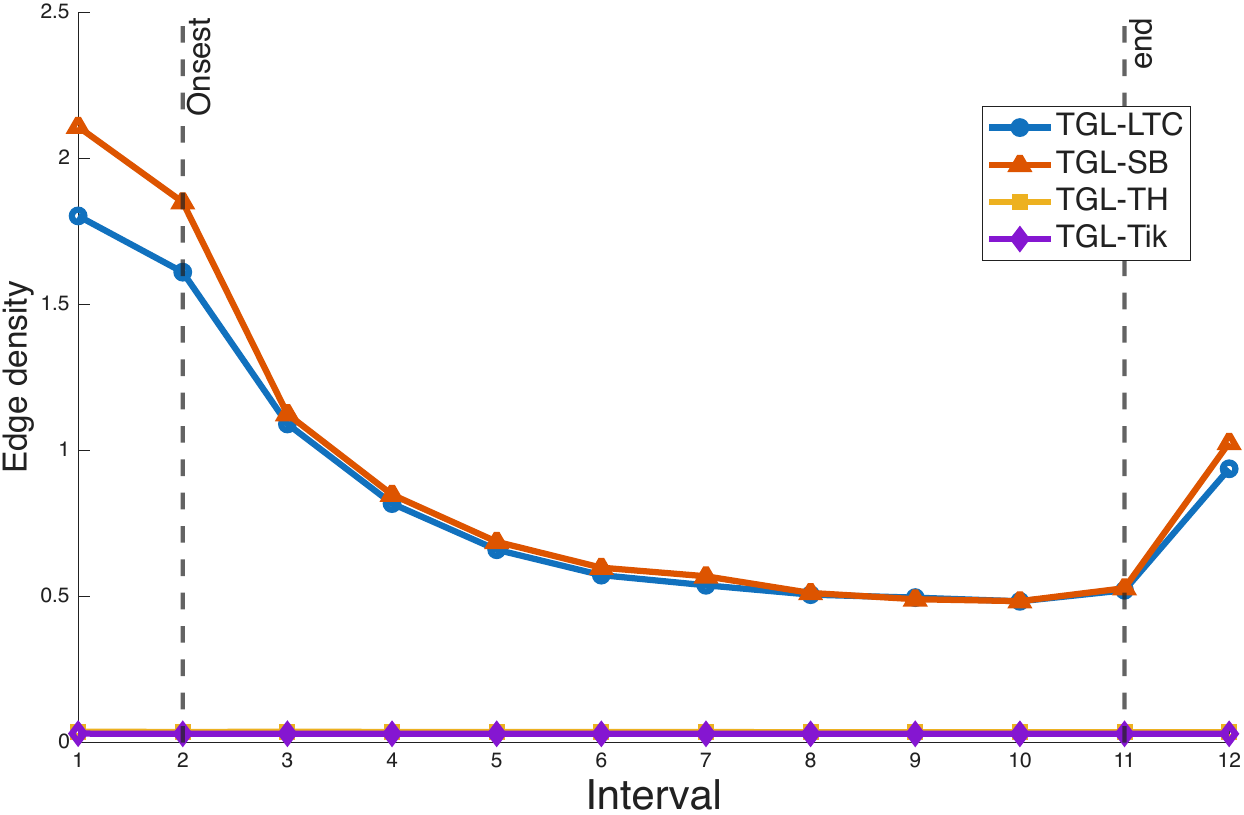}
  \caption{Average density of graphs for 22 different seizures, estimated with TGL methods. The duration of all seizures is normalized into 12 equal intervals. The interval between two vertical dotted lines reflects the seizure period. The learned graphs of TGL-TH and TGL-Tik are fully connected graphs, but with extremely small weights, leading to a near-zero density.}
  \label{fig:mean_aved}
\end{figure}

We first analyze the temporal variation of connectivity by calculating the average edge density for each time window.
As shown in \autoref{fig:resulteeg} for a representative patient, the learned graphs become sparse during a seizure before gradually recovering as the seizure ends. To confirm this trend, we analyzed 22 seizures from five patients by normalizing the seizure durations into 12 uniform intervals. 
The average graph density, calculated as $1/N\sum_{i=1}^N d_i$, $d_i$ being the degree of node $i$, shown in \autoref{fig:mean_aved}, decreases during the seizure and slightly increases just before it ends for TGL-LTC and TGL-SB. These findings are consistent with previous studies \cite{kramer2010, vandiessen2013}. TGL-TH and TGL-Tik have near-zero density across all intervals because the learned graphs are fully connected, yet the edge weights are very small. This demonstrates their inability to capture the non-smooth and non-sparse topological variation of seizure EEGs.

We further evaluate our results using two small-world network metrics for each $t$ \cite{schindler2008}:
\begin{itemize}
    \item Global clustering coefficient ($C_t$): a measure for the tendency of nodes to form a cluster, calculated by $C_t=\frac{1}{N} \sum_{i} c_{i,t}$ where $c_{i,t}$ is the local clustering coefficient defined as $c_{i,t} := 2(|\mathcal{E}_{i}|)/(k_i(k_i-1))$ in which $\mathcal{E}_{i}$ is the number of existing edges in a subgraph of direct neighbors of the $i$the node and $k_i$ is the number of possible edges within the subgraph.
    \item Shortest path length ($L_t$): $L_t$ is the average shortest path between two nodes $v_i$ and $v_j$, given by $L=\frac{1}{N(N-1)} \sum_{i \neq j} s(v_i,v_j)$ where $s(v_i,v_j)$ is the sum of edge weights in the path.
\end{itemize}

\begin{figure}
  \centering
    \subfigure[$C_t/C_{r,t}$]{\includegraphics[width=0.7\linewidth]{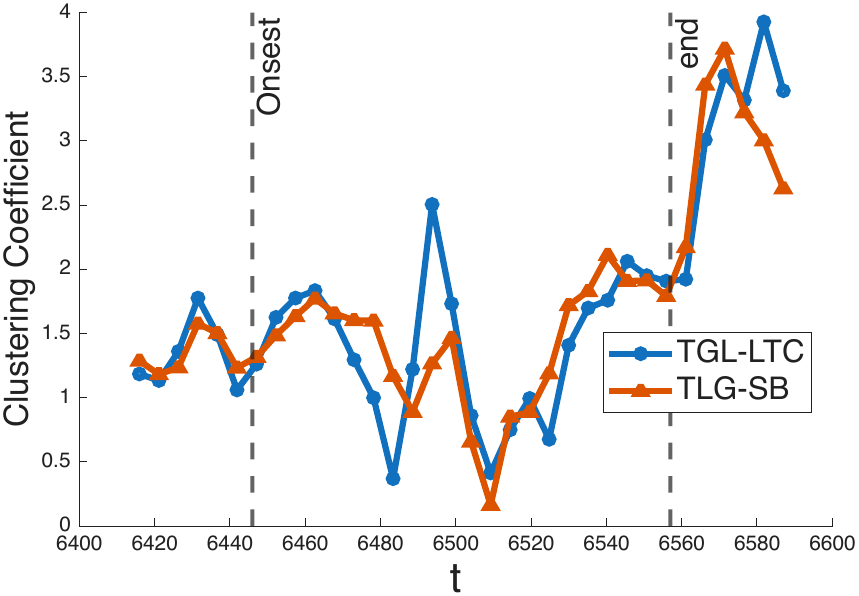}
    \label{fig:clustering_coef}}\\
    \subfigure[$L_t/L_{r,t}$]{\includegraphics[width=0.7\linewidth]{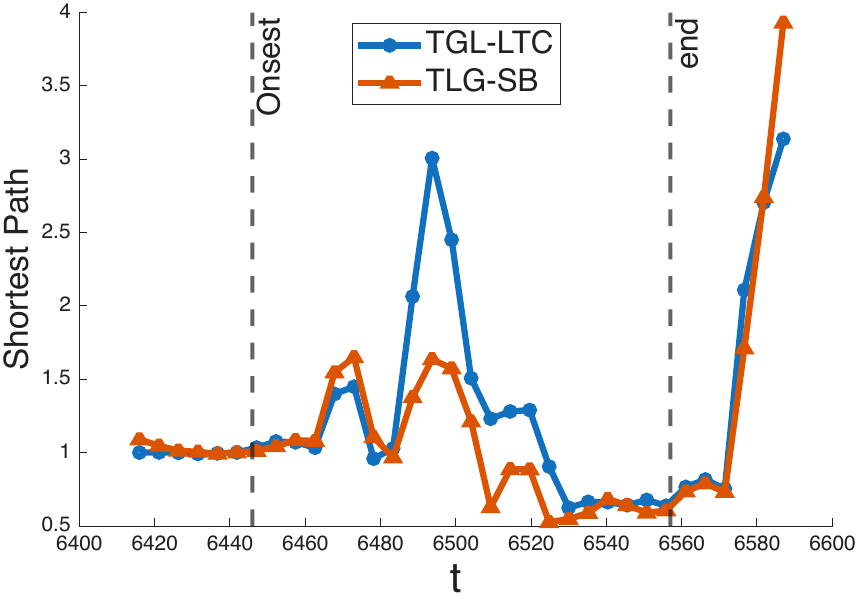}
    \label{fig:shortest_path}}
  \caption{(a) Relative global clustering coefficient and (b) relative shortest path length calculated for each graph at time $t$ of a representative seizure. The values are smoothed with a 2-point moving average. Dotted lines indicate the onset and end of the seizure.}
  \label{fig:sw-metrics}\vspace{-5pt}
\end{figure}

We calculated the ratios of the clustering coefficient ($C_t/C_{r,t}$) and shortest path length ($L_t/L_{r,t}$) relative to a random ER-graph generated with $p_t = E_t/(N(N-1))$ where $E_t$ is the number of edges in the learned graph at $t$.
\autoref{fig:sw-metrics} shows these ratios for a representative seizure. We can see that both values
tend to peak during a seizure, similar to results in \cite{schindler2008}, further validating that our methods can capture the functional brain connectivity during epileptic seizures.

From \autoref{fig:resulteeg}, we can see that TGL-SB and TGL-LTC identify consistent seizure-related network evolution, including pronounced sparsification during the ictal period and rapid densification near seizure end. In \autoref{fig:sw-metrics}, TGL-LTC exhibits larger transient variations in clustering coefficient and shortest path length, reflecting its ability to capture localized changes in node neighborhoods. In contrast, TGL-SB yields smoother network-level transitions because it promotes synchronized changes across edges. Thus, the methods provide complementary global and local descriptions rather than competing conclusions.


\section{Conclusion}
\label{conclusion}
In this paper, we have presented a framework for learning TV graphs suitable for analyzing spatiotemporal data and change detection.
The framework formulates a TV graph learning problem as a convex optimization problem with constraints on the temporal variation in the graph. 
Specifically, the framework introduces fused Lasso regularization and group Lasso regularization of the temporal variation to learn the graph having temporal homogeneity, the graph reflecting a sudden change in networks, and one with local topological changes, respectively.
We also propose algorithms for solving the optimization problems based on the primal-dual splitting algorithm and extend the algorithm to accommodate multiple memory settings.
In synthetic and real data applications, we demonstrated that our proposed model can successfully learn the TV graphs, especially when the number of observations is small.
As an offline batch method, our approach (along with offline baselines) requires computational and memory costs that scale linearly with $T$, which can be significant for long time horizons; online algorithms avoid this factor of $T$. A theoretical analysis, e.g., sample-complexity bounds characterizing the Frobenius error between the estimated and true Laplacian sequence in terms of $K$, $N$, and the regularization parameters, remains an open problem for the three non-smooth regularizers (TGL-TH, TGL-SB, TGL-LTC). Error bounds for related dynamic graph estimation problems exist in the literature~\cite{zhou2010Time,chen2025estimating}; extending these to our setting is left as future work.
% Our future work includes implementing an automatic parameter-tuning method and extending the algorithm for online graph learning.
\ifCLASSOPTIONcaptionsoff
  \newpage
\fi

% \footnotesize
\bibliography{./IEEEabrv.bib, ./refs2.bib}

\begin{thebibliography}{10}
\providecommand{\url}[1]{#1}
\csname url@samestyle\endcsname
\providecommand{\newblock}{\relax}
\providecommand{\bibinfo}[2]{#2}
\providecommand{\BIBentrySTDinterwordspacing}{\spaceskip=0pt\relax}
\providecommand{\BIBentryALTinterwordstretchfactor}{4}
\providecommand{\BIBentryALTinterwordspacing}{\spaceskip=\fontdimen2\font plus
\BIBentryALTinterwordstretchfactor\fontdimen3\font minus \fontdimen4\font\relax}
\providecommand{\BIBforeignlanguage}[2]{{%
\expandafter\ifx\csname l@#1\endcsname\relax
\typeout{** WARNING: IEEEtran.bst: No hyphenation pattern has been}%
\typeout{** loaded for the language `#1'. Using the pattern for}%
\typeout{** the default language instead.}%
\else
\language=\csname l@#1\endcsname
\fi
#2}}
\providecommand{\BIBdecl}{\relax}
\BIBdecl

\bibitem{nowzari2016}
C.~Nowzari, V.~M. Preciado, and G.~J. Pappas, ``Analysis and control of epidemics: {{A}} survey of spreading processes on complex networks,'' \emph{IEEE Control Syst. Mag.}, vol.~36, no.~1, pp. 26--46, 2016.

\bibitem{ganesh2005}
A.~Ganesh, L.~Massoulie, and D.~Towsley, ``The effect of network topology on the spread of epidemics,'' in \emph{Proc. IEEE INFOCOM}, vol.~2, 2005, pp. 1455--1466.

\bibitem{colizza2006}
V.~Colizza, A.~Barrat, M.~Barthélemy, and A.~Vespignani, ``The role of the airline transportation network in the prediction and predictability of global epidemics,'' \emph{Proc. Natl. Acad. Sci. USA}, vol. 103, no.~7, pp. 2015--2020, 2006.

\bibitem{zhang2014}
J.~Zhang and J.~M.~F. Moura, ``Diffusion in social networks as {{SIS}} epidemics: {{Beyond}} full mixing and complete graphs,'' \emph{IEEE J. Sel. Top. Signal Process.}, vol.~8, no.~4, pp. 537--551, 2014.

\bibitem{shahid2016}
N.~Shahid, N.~Perraudin, V.~Kalofolias, G.~Puy, and P.~Vandergheynst, ``Fast robust {{PCA}} on graphs,'' \emph{IEEE J. Sel. Top. Signal Process.}, vol.~10, no.~4, pp. 740--756, 2016.

\bibitem{shang2012}
F.~Shang, L.~C. Jiao, and F.~Wang, ``Graph dual regularization non-negative matrix factorization for co-clustering,'' \emph{Pattern Recognition}, vol.~45, no.~6, pp. 2237--2250, 2012.

\bibitem{zhang2013}
Z.~Zhang and K.~Zhao, ``Low-rank matrix approximation with manifold regularization,'' \emph{IEEE Trans. Pattern Anal. Mach. Intell.}, vol.~35, no.~7, pp. 1717--1729, 2013.

\bibitem{gadde2013}
A.~Gadde, S.~K. Narang, and A.~Ortega, ``Bilateral filter: {{Graph}} spectral interpretation and extensions,'' in \emph{Proc. {IEEE ICIP}}, 2013, pp. 1222--1226.

\bibitem{wang2014}
Y.~Wang, A.~Ortega, D.~Tian, and A.~Vetro, ``A graph-based joint bilateral approach for depth enhancement,'' in \emph{Proc. IEEE ICASSP}, 2014, pp. 885--889.

\bibitem{onuki2016}
M.~Onuki, S.~Ono, M.~Yamagishi, and Y.~Tanaka, ``Graph signal denoising via trilateral filter on graph spectral domain,'' \emph{IEEE Trans. Signal Inf. Process. Netw.}, vol.~2, no.~2, pp. 137--148, 2016.

\bibitem{elmoataz2008}
A.~Elmoataz, O.~Lezoray, and S.~Bougleux, ``Nonlocal discrete regularization on weighted graphs: {{A}} framework for image and manifold processing,'' \emph{IEEE Trans. Image Process.}, vol.~17, no.~7, pp. 1047--1060, 2008.

\bibitem{couprie2013}
C.~Couprie, L.~Grady, L.~Najman, J.~Pesquet, and H.~Talbot, ``Dual constrained {{TV-based}} regularization on graphs,'' \emph{SIAM J. Imaging Sci.}, vol.~6, no.~3, pp. 1246--1273, 2013.

\bibitem{ono2015}
S.~Ono, I.~Yamada, and I.~Kumazawa, ``Total generalized variation for graph signals,'' in \emph{Proc. IEEE ICASSP}, 2015, pp. 5456--5460.

\bibitem{dong2019}
X.~Dong, D.~Thanou, M.~Rabbat, and P.~Frossard, ``Learning graphs from data: {{A}} signal representation perspective,'' \emph{IEEE Signal Process. Mag.}, vol.~36, no.~3, pp. 44--63, 2019.

\bibitem{mateos2019}
G.~Mateos, S.~Segarra, A.~G. Marques, and A.~Ribeiro, ``Connecting the dots: {{Identifying}} network structure via graph signal processing,'' \emph{IEEE Signal Process. Mag.}, vol.~36, no.~3, pp. 16--43, 2019.

\bibitem{giannakis2018}
G.~B. Giannakis, Y.~Shen, and G.~V. Karanikolas, ``Topology identification and learning over graphs: {{Accounting}} for nonlinearities and dynamics,'' \emph{Proc. IEEE}, vol. 106, no.~5, pp. 787--807, 2018.

\bibitem{kadambari2020a}
S.~K. Kadambari and S.~Prabhakar~Chepuri, ``Learning product graphs from multidomain signals,'' in \emph{Proc. IEEE ICASSP}, 2020, pp. 5665--5669.

\bibitem{liu2019b}
Y.~Liu, L.~Yang, K.~You, W.~Guo, and W.~Wang, ``Graph learning based on spatiotemporal smoothness for time-varying graph signal,'' \emph{IEEE Access}, vol.~7, pp. 62\,372--62\,386, 2019.

\bibitem{lodhi2020}
M.~A. Lodhi and W.~U. Bajwa, ``Learning product graphs underlying smooth graph signals,'' \emph{ArXiv}, 2020.

\bibitem{chen2010}
X.~Chen, Y.~Liu, H.~Liu, and J.~G. Carbonell, ``Learning spatial-temporal varying graphs with applications to climate data analysis,'' in \emph{Proc. {{AAAI Conf}}. {{Artif}}. {{Intell}}.}\hskip 1em plus 0.5em minus 0.4em\relax AAAI Press, 2010, pp. 425--430.

\bibitem{kalofolias2016}
V.~Kalofolias, ``\BIBforeignlanguage{en}{How to learn a graph from smooth signals},'' in \emph{\BIBforeignlanguage{en}{Proc. {{Int}}. {{Conf}}. {{Artif}}. {{Intell}}. {{Statist}}.}}, 2016, pp. 920--929.

\bibitem{dong2016}
X.~Dong, D.~Thanou, P.~Frossard, and P.~Vandergheynst, ``Learning {{Laplacian}} matrix in smooth graph signal representations,'' \emph{IEEE Trans. Signal Process.}, vol.~64, no.~23, pp. 6160--6173, 2016.

\bibitem{egilmez2017}
H.~E. Egilmez, E.~Pavez, and A.~Ortega, ``Graph learning from data under {{Laplacian}} and structural constraints,'' \emph{IEEE J. Sel. Top. Signal Process.}, vol.~11, no.~6, pp. 825--841, 2017.

\bibitem{pasdeloup2018}
B.~Pasdeloup, V.~Gripon, G.~Mercier, D.~Pastor, and M.~G. Rabbat, ``Characterization and inference of graph diffusion processes from observations of stationary signals,'' \emph{IEEE Trans. Signal Inf. Process. Netw.}, vol.~4, no.~3, pp. 481--496, 2018.

\bibitem{thanou2017}
D.~Thanou, X.~Dong, D.~Kressner, and P.~Frossard, ``Learning heat diffusion graphs,'' \emph{IEEE Trans. Signal Inf. Process. Netw.}, vol.~3, no.~3, pp. 484--499, 2017.

\bibitem{mei2017}
J.~Mei and J.~M.~F. Moura, ``Signal processing on graphs: {{Causal}} modeling of unstructured data,'' \emph{IEEE Trans. Signal Process.}, vol.~65, no.~8, pp. 2077--2092, 2017.

\bibitem{chepuri2017}
S.~P. Chepuri, S.~Liu, G.~Leus, and A.~O. Hero, ``Learning sparse graphs under smoothness prior,'' in \emph{Proc. IEEE ICASSP}, 2017, pp. 6508--6512.

\bibitem{pavez2018}
E.~Pavez, H.~E. Egilmez, and A.~Ortega, ``Learning graphs with monotone topology properties and multiple connected components,'' \emph{IEEE Trans. Signal Process.}, vol.~66, no.~9, pp. 2399--2413, 2018.

\bibitem{egilmez2018}
H.~E. Egilmez, E.~Pavez, and A.~Ortega, ``Graph learning from filtered signals: Graph system and diffusion kernel identification,'' \emph{IEEE Trans. Signal Inf. Process. Netw.}, vol.~5, no.~2, pp. 360--374, 2019.

\bibitem{pu2021}
X.~Pu, T.~Cao, X.~Zhang, X.~Dong, and S.~Chen, ``Learning to learn graph topologies,'' in \emph{Adv. Neural Inf. Process. Syst.}, M.~Ranzato, A.~Beygelzimer, Y.~Dauphin, P.~Liang, and J.~W. Vaughan, Eds., vol.~34.\hskip 1em plus 0.5em minus 0.4em\relax Curran Associates, Inc., 2021, pp. 4249--4262.

\bibitem{li2023}
R.~Li, J.~Lin, H.~Qiu, W.~Zhang, and J.~Wang, ``Graph learning for latent-variable {{Gaussian}} graphical models under laplacian constraints,'' \emph{Neurocomputing}, vol. 532, pp. 67--76, 2023.

\bibitem{tamaru2024}
A.~Tamaru, J.~Hara, H.~Higashi, Y.~Tanaka, and A.~Ortega, ``Optimizing k in {{kNN}} graphs with graph learning perspective,'' in \emph{Proc. IEEE ICASSP}, 2024, pp. 9441--9445.

\bibitem{preti2017}
M.~G. Preti, T.~A. Bolton, and D.~Van De~Ville, ``The dynamic functional connectome: {{State-of-the-art}} and perspectives,'' \emph{NeuroImage}, vol. 160, pp. 41--54, 2017.

\bibitem{kim2014}
Y.~Kim, S.~Han, S.~Choi, and D.~Hwang, ``\BIBforeignlanguage{eng}{Inference of dynamic networks using time-course {{Data}}},'' \emph{\BIBforeignlanguage{eng}{Brief. Bioinformatics}}, vol.~15, no.~2, pp. 212--228, 2014.

\bibitem{hallac2017}
D.~Hallac, Y.~Park, S.~Boyd, and J.~Leskovec, ``Network inference via the time-varying graphical lasso,'' in \emph{Proc. {{KDD}} '17}.\hskip 1em plus 0.5em minus 0.4em\relax Association for Computing Machinery, 2017, pp. 205--213.

\bibitem{yamada2019}
K.~Yamada, Y.~Tanaka, and A.~Ortega, ``Time-varying graph learning based on sparseness of temporal variation,'' in \emph{Proc. IEEE ICASSP}, 2019, pp. 5411--5415.

\bibitem{kalofolias2017}
V.~Kalofolias, A.~Loukas, D.~Thanou, and P.~Frossard, ``Learning time varying graphs,'' in \emph{Proc. IEEE ICASSP}, 2017, pp. 2826--2830.

\bibitem{zhang2022}
X.~Zhang and Q.~Wang, ``Time-varying graph learning under structured temporal priors,'' in \emph{Proc. EUSIPCO}, 2022, pp. 2141--2145.

\bibitem{ye2024}
R.~Ye, X.-Q. Jiang, H.~Feng, J.~Wang, R.~Qiu, and X.~Hou, ``Time-varying graph learning from smooth and stationary graph signals with hidden nodes,'' \emph{EURASIP J. Adv. in Signal Process.}, vol. 2024, no.~1, p.~33, 2024.

\bibitem{cardoso2020learning}
J.~V.~M. Cardoso and D.~P. Palomar, ``Learning undirected graphs in financial markets,'' in \emph{2020 54th ACSSC}, 2020, pp. 741--745.

\bibitem{Natali2022}
A.~Natali, E.~Isufi, M.~Coutino, and G.~Leus, ``Learning time-varying graphs from online data,'' \emph{IEEE Open J. Signal Process.}, vol.~3, pp. 212--228, 2022.

\bibitem{Bagheri2024}
S.~Bagheri, G.~Cheung, T.~Eadie, and A.~Ortega, ``Joint signal interpolation / time-varying graph estimation via smoothness and low-rank priors,'' in \emph{2024 {{IEEE International Conference}} on {{Acoustics}}, {{Speech}} and {{Signal Processing}} ({{ICASSP}})}, 2024, pp. 9646--9650.

\bibitem{shen2017tensor}
Y.~Shen, B.~Baingana, and G.~B. Giannakis, ``Tensor decompositions for identifying directed graph topologies and tracking dynamic networks,'' \emph{IEEE Trans. Signal Process.}, vol.~65, no.~14, pp. 3675--3687, 2017.

\bibitem{Saboksayr2021}
S.~S. Saboksayr, G.~Mateos, and M.~Cetin, ``Online discriminative graph learning from multi-class smooth signals,'' \emph{Signal Processing}, vol. 186, p. 108101, 2021.

\bibitem{rey2022}
S.~Rey, A.~Buciulea, M.~Navarro, S.~Segarra, and A.~G. Marques, ``Joint inference of multiple graphs with hidden variables from stationary graph signals,'' in \emph{Proc. IEEE ICASSP}, 2022, pp. 5817--5821.

\bibitem{monti2014}
R.~P. Monti, P.~Hellyer, D.~Sharp, R.~Leech, C.~Anagnostopoulos, and G.~Montana, ``Estimating time-varying brain connectivity networks from functional {{MRI}} time series,'' \emph{NeuroImage}, vol. 103, pp. 427--443, 2014.

\bibitem{wang2014a}
J.~Wang, S.~Qiu, Y.~Xu, Z.~Liu, X.~Wen, X.~Hu, R.~Zhang, M.~Li, W.~Wang, and R.~Huang, ``\BIBforeignlanguage{en}{Graph theoretical analysis reveals disrupted topological properties of whole brain functional networks in temporal lobe epilepsy},'' \emph{\BIBforeignlanguage{en}{Clinical Neurophysiology}}, vol. 125, no.~9, pp. 1744--1756, 2014.

\bibitem{velmani2015}
R.~Velmani and B.~Kaarthick, ``An efficient cluster-tree based data collection scheme for large mobile wireless sensor networks,'' \emph{IEEE Sens. J.}, vol.~15, no.~4, pp. 2377--2390, 2015.

\bibitem{condat2013}
L.~Condat, ``\BIBforeignlanguage{en}{A primal–dual splitting method for convex optimization involving {{Lipschitzian}}, proximable and linear composite terms},'' \emph{\BIBforeignlanguage{en}{J. Optim. Theory Appl.}}, vol. 158, no.~2, pp. 460--479, 2013.

\bibitem{yamada2020arxiv}
K.~Yamada, Y.~Tanaka, and A.~Ortega, ``Time-varying graph learning with constraints on graph temporal variation,'' \emph{ArXiv}, 2020.

\bibitem{shuman2013}
D.~I. Shuman, S.~K. Narang, P.~Frossard, A.~Ortega, and P.~Vandergheynst, ``The emerging field of signal processing on graphs: {{Extending}} high-dimensional data analysis to networks and other irregular domains,'' \emph{IEEE Signal Process. Mag.}, vol.~30, no.~3, pp. 83--98, 2013.

\bibitem{ortega2022introduction}
A.~Ortega, \emph{Introduction to Graph Signal Processing}.\hskip 1em plus 0.5em minus 0.4em\relax Cambridge University Press, 2022.

\bibitem{parikh2014}
N.~Parikh and S.~Boyd, ``Proximal algorithms,'' \emph{Found. Trends Optim.}, vol.~1, no.~3, pp. 127--239, 2014.

\bibitem{perraudin2017}
N.~Perraudin and P.~Vandergheynst, ``Stationary signal processing on graphs,'' \emph{IEEE Trans. Signal Process.}, vol.~65, no.~13, pp. 3462--3477, 2017.

\bibitem{tibshirani2005}
R.~Tibshirani, M.~Saunders, S.~Rosset, J.~Zhu, and K.~Knight, ``\BIBforeignlanguage{en}{Sparsity and smoothness via the fused {{Lasso}}},'' \emph{\BIBforeignlanguage{en}{J. R. Stat. Soc. Ser. B Stat. Methodol.}}, vol.~67, no.~1, pp. 91--108, 2005.

\bibitem{yang2011ell}
Y.~Yang, H.~T. Shen, Z.~Ma, Z.~Huang, and X.~Zhou, ``$\ell_{2,1}$-norm regularized discriminative feature selection for unsupervised learning,'' in \emph{Proc. IJCAI}, 2011.

\bibitem{meier2008}
L.~Meier, S.~Van De~Geer, and P.~B{\"u}hlmann, ``\BIBforeignlanguage{en}{The group {{Lasso}} for logistic regression},'' \emph{\BIBforeignlanguage{en}{J. R. Stat. Soc. Ser. B Stat. Methodol.}}, vol.~70, no.~1, pp. 53--71, 2008.

\bibitem{zou2006elastic}
H.~Zou and T.~Hastie, ``Regularization and variable selection via the elastic net,'' \emph{J. R. Stat. Soc. Ser. B Stat. Methodol.}, vol.~67, no.~2, pp. 301--320, 2005.

\bibitem{komodakis2015}
N.~Komodakis and J.~Pesquet, ``Playing with duality: {{An}} overview of recent primal-dual approaches for solving large-scale optimization problems,'' \emph{IEEE Signal Process. Mag.}, vol.~32, no.~6, pp. 31--54, 2015.

\bibitem{daubechies2004}
I.~Daubechies, M.~Defrise, and C.~De~Mol, ``\BIBforeignlanguage{en}{An iterative thresholding algorithm for linear inverse problems with a sparsity constraint},'' \emph{\BIBforeignlanguage{en}{Commun. Pure Appl. Math.}}, vol.~57, no.~11, pp. 1413--1457, 2004.

\bibitem{combettes2012}
P.~Combettes and J.~Pesquet, ``\BIBforeignlanguage{en}{Primal-dual splitting algorithm for solving inclusions with mixtures of composite, {{Lipschitzian}}, and parallel-sum type monotone operators},'' \emph{\BIBforeignlanguage{en}{Set-Valued and Variational Analysis}}, vol.~20, no.~2, pp. 307--330, 2012.

\bibitem{Lambiotte2014}
R.~Lambiotte, J.-C. Delvenne, and M.~Barahona, ``Random walks, markov processes and the multiscale modular organization of complex networks,'' \emph{IEEE Trans. Netw. Sci. Eng.}, vol.~1, no.~2, pp. 76--90, 2014.

\bibitem{johnson1996}
D.~B. Johnson and D.~A. Maltz, ``\BIBforeignlanguage{en}{Dynamic source routing in ad hoc wireless networks},'' in \emph{\BIBforeignlanguage{en}{Mobile {{Computing}}}}, 1996, pp. 153--181.

\bibitem{zhang2008}
F.~Zhang and E.~R. Hancock, ``Graph spectral image smoothing using the heat kernel,'' \emph{Pattern Recognition}, vol.~41, no.~11, pp. 3328--3342, 2008.

\bibitem{gall2009}
J.~Gall, C.~Stoll, E.~de~Aguiar, C.~Theobalt, B.~Rosenhahn, and H.-P. Seidel, ``Motion capture using joint skeleton tracking and surface estimation,'' in \emph{{Proc. IEEE CVPR}}, 2009, pp. 1746--1753.

\bibitem{kramer2010}
M.~A. Kramer, U.~T. Eden, E.~D. Kolaczyk, R.~Zepeda, E.~N. Eskandar, and S.~S. Cash, ``Coalescence and fragmentation of cortical networks during focal seizures,'' \emph{J. Neurosci.}, vol.~30, no.~30, pp. 10\,076--10\,085, 2010.

\bibitem{vandiessen2013}
V.~E. Diessen, S.~J.~H. Diederen, K.~P.~J. Braun, F.~E. Jansen, and C.~J. Stam, ``Functional and structural brain networks in epilepsy: {{What}} have we learned?'' \emph{Epilepsia}, vol.~54, no.~11, pp. 1855--1865, 2013.

\bibitem{schindler2008}
K.~A. Schindler, S.~Bialonski, M.-T. Horstmann, C.~E. Elger, and K.~Lehnertz, ``Evolving functional network properties and synchronizability during human epileptic seizures,'' \emph{Chaos}, vol.~18, no.~3, p. 033119, 2008.

\bibitem{chbmit_database}
J.~Guttag, ``{CHB-MIT Scalp EEG Database},'' \emph{{PhysioNet}}, Jun. 2010, version 1.0.0.

\bibitem{zhou2010Time}
S.~Zhou, J.~Lafferty, and L.~Wasserman, ``Time varying undirected graphs,'' \emph{Machine Learning}, vol.~80, no.~2, pp. 295--319, 2010.

\bibitem{chen2025estimating}
J.~Chen, D.~Li, Y.~Li, and O.~Linton, ``Estimating time-varying networks for high-dimensional time series,'' \emph{J. Econometrics}, vol. 249, May 2025.

\end{thebibliography}
\bibliographystyle{./IEEEtran}

\begin{IEEEbiography}
[{\includegraphics[width=1in,height=1.25in,clip,keepaspectratio]{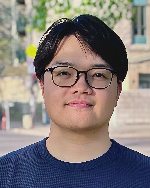}}]
{Haruki Yokota}(Student Member, IEEE) received B.E. and M.E. degrees from Tokyo University of Agriculture and Technology, Tokyo, Japan, in 2020 and 2022, respectively. He is a full-time student at the Graduate School of Engineering, University of Osaka, Osaka, Japan. He is currently supported by the Japan Society for the Promotion of Science (JSPS). His research interests are graph signal processing and biomedical signal processing. He received the IEEE SPS Tokyo Joint Chapter Student Award in 2022.
\end{IEEEbiography}
\begin{IEEEbiography}
[{\includegraphics[width=1in,height=1.25in,clip,keepaspectratio]{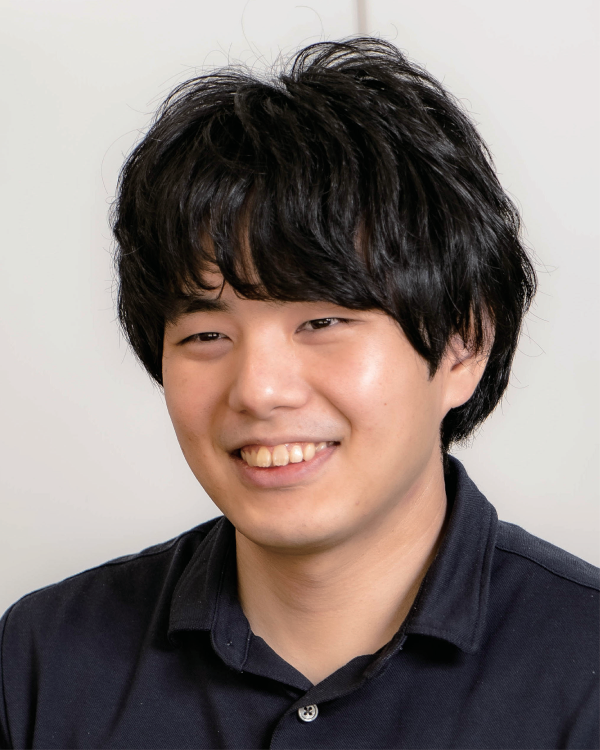}}]{Koki Yamada}(Member, IEEE) received the B.S. degree in physics and informatics and the M.S. degree in science and engineering from Yamaguchi University, Yamaguchi, Japan, in 2015 and 2017, respectively, and the Ph.D. degree in engineering from the Tokyo University of Agriculture and Technology, Fuchu, Japan, in 2022. He has been an Associate Professor with the Department of Electrical Engineering and Computer Science, Tokyo University of Agriculture and Technology.
\end{IEEEbiography}
\begin{IEEEbiography}[{\includegraphics[width=1in,height=1.25in,clip,keepaspectratio]{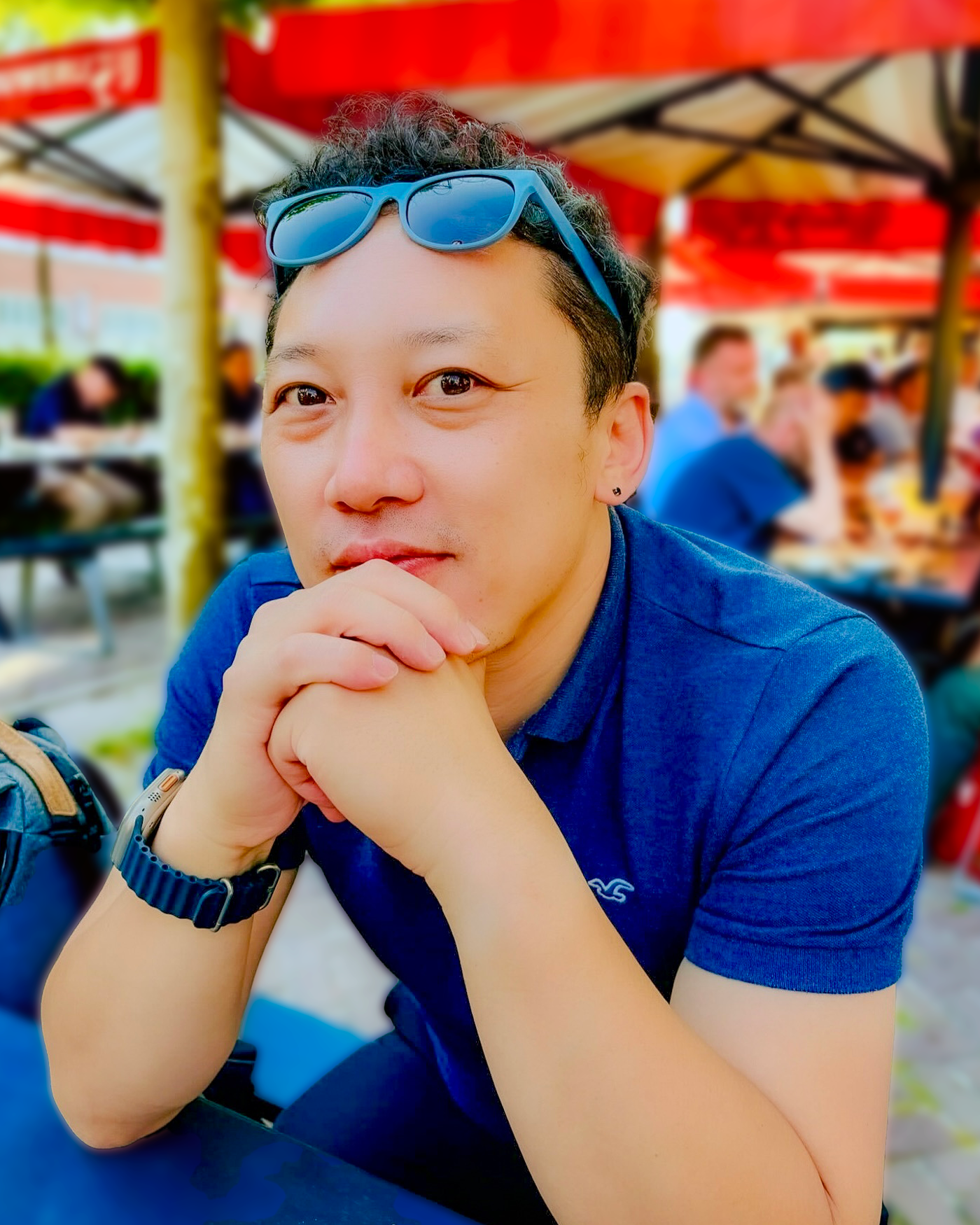}}]{Yuichi Tanaka}(Senior Member, IEEE) received the B.E., M.E., and Ph.D. degrees in electrical engineering from Keio University, Yokohama, Japan, in 2003, 2005, and 2007, respectively. From 2007 to 2008, he was a Postdoctoral Scholar with Keio University, and was supported by the Japan Society for the Promotion of Science (JSPS). From 2006 to 2008, he was a Visiting Scholar with the University of California, San Diego, CA, USA. From 2008 to 2012, he was an Assistant Professor with the Department of Information Science, Utsunomiya University, Tochigi, Japan. From 2012 to 2022, he was an Associate Professor with the Department of Electrical Engineering and Computer Science, Tokyo University of Agriculture and Technology, Tokyo, Japan. Since 2022, he has been a Professor with the Graduate School of Engineering, The University of Osaka, Osaka, Japan. 
His research interests include the field of high-dimensional signal processing and machine learning which includes: graph signal processing, green IoT, geometric deep learning, sensor networks, image/video processing in extreme situations, biomedical signal processing, and remote sensing. Since 2022, he was an Associate Editor for IEEE Signal Processing Letters and APSIPA Transactions on Signal and Information Processing. From 2016 to 2020, he was also an Associate Editor for IEEE Transactions on Signal Processing and IEICE Transactions on Fundamentals from 2013 to 2017. He received Best Paper Awards in APSIPA ASC 2014 and 2015 and IEEE Signal Processing Society Japan Best Paper Award in 2016. He was also the recipient of Yasujiro Niwa Outstanding Paper Award in 2010, TELECOM System Technology Award in 2011, Ando Incentive Prize for the Study of Electronics in 2015, and DOCOMO Mobile Science Award 2025. He is a Board-of-Governors Member-at-Large of APSIPA and Vice Chair of APSIPA IVM (Image, Video and Multimedia) Technical Committee.
\end{IEEEbiography}
\begin{IEEEbiography}[{\includegraphics[width=1in,height=1.25in,clip,keepaspectratio]{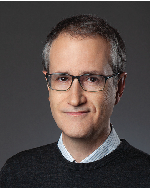}}]{Antonio Ortega}(Fellow, IEEE) is Dean's Professor of Electrical and Computer Engineering at the University of Southern California (USC).  He received his undergraduate and doctoral degrees from the Universidad Politécnica de Madrid, Madrid, Spain, and Columbia University, New York, NY, respectively. He is a Fellow of the IEEE and EURASIP, and he currently serves as the VP for Publications of the IEEE Signal Processing Society. He has received several paper awards, including the 2016 Signal Processing Magazine award. His recent research work focuses on graph signal processing, machine learning, and multimedia compression. He is the author of the book, "Introduction to Graph Signal Processing", published by Cambridge University Press in 2022.  
\end{IEEEbiography}

\end{document}